%% file: paper.tex
\documentclass[12pt]{article} 
\input{macros}

\usepackage[sort, comma, square]{natbib}
\usepackage{url}

\title{More powerful selective inference \\
for the graph fused lasso}
\author[1]{Yiqun Chen\thanks{yiqunc@uw.edu}}
\author[2]{Sean Jewell}
\author[1,2]{Daniela Witten}
\affil[1]{Department of Biostatistics, University of Washington, Seattle, WA}
\affil[2]{Department of Statistics, University of Washington, Seattle, WA, USA}

\newcites{appendix}{References}

\begin{document} 

\def\spacingset#1{\renewcommand{\baselinestretch}%
{#1}\small\normalsize} \spacingset{1}

\maketitle
\begin{abstract}
    \input{sections-revision-v1/abstract}
\end{abstract}

\noindent%
\vfill

\newpage
\spacingset{1.5} 

\section{Introduction}
\label{section:intro}
\input{sections-revision-v1/intro}

\section{Background on the generalized lasso}
\label{section:related_work}
\input{sections-revision-v1/existing_work}

\section{Proposed approach}
\label{section:method}
\input{sections-revision-v1/method}

\section{Extensions}
\label{section:extension}
\input{sections-revision-v1/extension}

\section{Simulation study}
\label{section:sim}
\input{sections-revision-v1/sim}
\section{Data applications}
\label{section:real_data}
\input{sections-revision-v1/real_data}

\section{Discussion}
\label{section:discussion}
\input{sections-revision-v1/discussion}

\medskip

\bibliographystyle{apalike}
\bibliography{ref.bib}

\newpage
\appendix
\section{Appendix}
\input{sections-revision-v1/appendix_a1}

\newpage
\input{sections-revision-v1/appendix_a1_b}

\newpage
\input{sections-revision-v1/appendix_a2}

\newpage
\input{sections-revision-v1/appendix_a3}
\newpage
\input{sections-revision-v1/appendix_a4}

\newpage
\input{sections-revision-v1/appendix_a5}

\newpage
\input{sections-revision-v1/appendix_a6}

\bibliographystyleappendix{apalike}
\bibliographyappendix{ref.bib}
\end{document}

%% file: macros.tex
\usepackage{amsmath, amsthm, amssymb, verbatim, bbm, color, graphics, geometry,physics}
\usepackage{booktabs,color}
\usepackage[norelsize,ruled]{algorithm2e}
\SetAlFnt{\small}
\usepackage{setspace}
\usepackage{multibib}
\usepackage{mathtools}
\usepackage[export]{adjustbox}

\usepackage{bbm}

\usepackage{tabularx}

\usepackage{tablefootnote}

\usepackage[labelformat=simple]{subcaption}

\usepackage{array}
\newcolumntype{$}{>{\global\let\currentrowstyle\relax}}
\newcolumntype{^}{>{\currentrowstyle}}
\newcommand{\rowstyle}[1]{\gdef\currentrowstyle{#1}%
  #1\ignorespaces
}

\newtheorem{proposition}{Proposition}
\newtheorem{corollary}{Corollary}

\DeclareMathOperator*{\argmax}{argmax}
\DeclareMathOperator*{\argmin}{argmin}

\usepackage[colorlinks=true,linkcolor=blue,citecolor=blue]{hyperref}%
\usepackage{subcaption}
\DeclareMathAlphabet{\mathcal}{OMS}{cmsy}{m}{n}
\usepackage{authblk}
\newcommand{\pHyun}{\ensuremath{p_{\text{Hyun}}}}

\newcommand{\pB}{\ensuremath{p_{\hat{C}_1,\hat{C}_2}}}
\newcommand{\pPP}{\ensuremath{p_{\text{LeDuy}}}}

\newcommand{\pD}{\ensuremath{p_{\hat{C}_1,\hat{C}_2}^*}}
\newcommand{\pDvar}{\ensuremath{\tilde{p}_{\hat{C}_1,\hat{C}_2}^*}}

\newcommand{\SB}{\ensuremath{\mathcal{S}_{\hat{C}_1,\hat{C}_2}}}
\newcommand{\SBconserv}{\ensuremath{\tilde{\mathcal{S}}_{\hat{C}_1,\hat{C}_2}}}

\newcommand{\SD}{\ensuremath{\mathcal{S}_{\hat{C}_1,\hat{C}_2}^*}}

\newcommand{\CC}{\ensuremath{\mathcal{CC}}}

%% file: sections-revision-v1/abstract.tex

The graph fused lasso --- which includes as a special case the one-dimensional fused lasso --- is widely used to reconstruct signals that are piecewise constant on a graph, meaning that nodes connected by an edge tend to have identical values. We consider testing for a difference in the means of two connected components estimated using the graph fused lasso. A naive procedure such as a z-test for a difference in means will not control the selective Type I error, since the hypothesis that we are testing is itself a function of the data. 
In this work, we propose a new test for this task that controls the selective Type I error, and conditions on less information than existing approaches, leading to substantially higher power. We illustrate our approach in simulation and on datasets of drug overdose death rates and teenage birth rates in the contiguous United States. Our approach yields more discoveries on both datasets.

%% file: sections-revision-v1/intro.tex

We consider a vector $Y \in \mathbb{R}^n$, assumed to be a noisy realization of a 
signal $\beta \in \mathbb{R}^n$, 
{
\begin{equation}
Y_j = \beta_j + \epsilon_j , \quad \epsilon_j \sim_{\text{i.i.d.}} \mathcal{N}(0, \sigma^2), \quad j =1, \ldots, n ,
\label{eq:model}
\end{equation}
}
 with known variance $\sigma^2$.  
We assume that $\beta$ has some underlying structure of interest. For instance, $\beta$ might be \textit{sparse}, with few non-zero elements, or \textit{piecewise constant}, meaning that the elements of $\beta$ are ordered, and adjacent elements tend to take on equal values. 

It is natural to estimate $\beta$ by solving the optimization problem 
\begin{equation}\label{eq:genlasso}
    \hat{\beta} = \text{argmin}_{\beta \in \mathbb{R}^n} \left\{ \frac{1}{2} \Vert y- \beta \Vert_2^2+\lambda \Vert D \beta \Vert_1 \right\}, 
\end{equation}
 where $D$ is an $m \times n$ penalty matrix that encodes the structure of  $\beta$. Problem \eqref{eq:genlasso} is a special case of the generalized lasso with an identity design matrix  \citep{Tibshirani2011-fq,Hastie2015-tn,Arnold2016-ue}. While the ideas in this paper apply for a general design matrix, we make use of an identity design matrix to simplify the discussion.  
 Many well-known regression problems involving $\ell_1$ penalties can be viewed as special cases of the generalized lasso; examples include the lasso \citep{Tibshirani1996-he}, the fused lasso signal approximator \citep{Tibshirani2005-xd,Friedman2007-fl,Rinaldo2009-pw}, the graph fused lasso \citep{Tibshirani2011-fq,Hastie2015-tn}, and trend filtering \citep{Kim2009-ty,Tibshirani2014-gg}. 

 Despite the abundant literature on  algorithms for computing $\hat{\beta}$ in \eqref{eq:genlasso} \citep{Johnson2013-yo,Tibshirani2011-fq,Xin2014-zo,Ramdas2016-pu,Zhu2017-ij,Arnold2016-ue,Friedman2007-fl} and on its theoretical properties \citep{Sadhanala2016-bm,Tibshirani2011-fq,Rinaldo2009-pw,Harchaoui2010-cl}, the topic of 
 \emph{inference} for the generalized lasso remains less developed. In this work, we focus on testing a null hypothesis that was determined after observing  $\hat\beta$ in \eqref{eq:genlasso}.

More precisely, suppose that we perform the graph fused lasso, a special case of \eqref{eq:genlasso},
\begin{equation}
\label{eq:graph_fused_lasso}
    \hat{\beta} = \text{argmin}_{\beta \in \mathbb{R}^n} \left\{ \frac{1}{2}\Vert y-\beta\Vert_2^2 +\lambda \sum_{(j,j')\in E}|\beta_j-\beta_{j'}| \right\}, 
\end{equation} 
 where $G=(V,E)$ is an undirected graph,  $V=\{1,\ldots,n\}$,  and  $(j,j') \in E$ indicates that the $j$th and $j'$th vertices in the graph are connected by an edge \citep{Tibshirani2011-fq}. For sufficiently large values of the non-negative tuning parameter $\lambda$, we will have $\hat\beta_j = \hat\beta_{j'}$ for some $(j,j') \in E$. We can segment $\hat\beta$ into \emph{connected components} --- that is, sets of elements of $\hat\beta$ that are connected in the original graph and share a common value. We might then consider testing the null hypothesis that the true mean of $\beta$ is the same across two \emph{estimated} connected components, i.e.,
  \begin{equation}
  H_0: {\sum_{j \in \hat{C}_1} \beta_j  }/{|\hat{C}_1|  }  =
 { \sum_{j' \in \hat{C}_2} \beta_{j'} }/{   |\hat{C}_2|} \mbox{ versus }  H_1: {\sum_{j \in \hat{C}_1} \beta_j  }/{|\hat{C}_1|  }  \neq
 { \sum_{j' \in \hat{C}_2} \beta_{j'} }/{ |\hat{C}_2|},
 \label{eq:null}
  \end{equation}
  where $\hat{C}_1\subseteq V$ and $\hat{C}_2\subseteq V$ are connected components of $\hat\beta$, with cardinality $|\hat{C}_1|$ and $|\hat{C}_2|$, and $\hat{C}_1\cap \hat{C}_2 = \emptyset$. This is equivalent to testing 
  $H_0: \nu^\top \beta = 0$ versus $H_1: \nu^\top \beta \neq 0$, where 
  {
   \begin{equation}
  \nu_{j} =  1_{( j \in \hat{C}_1)} / |\hat{C}_1| -  1_{( j \in \hat{C}_2)}/|\hat{C}_2|, \quad j=1,\ldots,n.
 \label{eq:nu_c1_c2}
   \end{equation}
   }
  Here, $H_0$ is chosen based on the data, i.e., we selected the contrast vector $\nu$ in \eqref{eq:nu_c1_c2} because $\hat{C}_1$ and $\hat{C}_2$ are estimated connected components. We focus on developing a test of $H_0$ that controls the  \emph{selective Type I error rate} \citep{Fithian2014-ow}, i.e., one for which the probability of rejecting $H_0$ at level $\alpha$, given that $H_0$ holds \emph{and we decided to test $H_0$}, is no greater than $\alpha$: 
  {
\begin{align}
\label{eq:selective_type_1}
\mathbb{P}_{H_0}\qty(\text{reject $H_0$ at level $\alpha$} \;\middle\vert\; H_0 \text{ is tested}   )\leq \alpha, \;\forall \alpha \in (0,1).
\end{align}
}
  It is not hard to see that a standard two-sample z-test of $H_0: \nu^\top \beta=0$, with $p$-value
  $\mathbb{P}_{H_0}\left( | \nu^\top Y |\geq | \nu^\top y| \right)$, 
 fails to account for the fact that we decided to test $H_0$ after looking at the data, and therefore does not control the selective Type I error rate \eqref{eq:selective_type_1}.  
 To address this problem, \citet{Hyun2018-ta} propose an elegant approach for testing $H_0: \nu^\top \beta=0$  
 that makes use of the selective inference framework developed by \citet{Lee2016-te}, \citet{Fithian2014-ow}, and \citet{Tibshirani2016-bx}.  
Their key insight is as follows: the set of $Y$ that yields a particular output for the first $K$ steps of  the dual path algorithm for solving \eqref{eq:genlasso} is a polyhedron, of the form $\qty{Y: A Y \geq 0 }$, for a matrix $A$ that can be explicitly computed. Thus, \emph{conditional on $Y$ belonging to this polyhedral set}, the linear contrast $\nu^{\top} Y$ follows a truncated normal distribution, with parameters that are a function of $A$, $\sigma^2$, and $\nu$, for any $\nu$ that is based on the output of \eqref{eq:genlasso}. It is thus possible to compute valid $p$-values for the null hypothesis in \eqref{eq:null} in the sense of \eqref{eq:selective_type_1}, \emph{by conditioning on the outputs from the first $K$ steps of the dual path algorithm}. 
  
Our paper relies on a simple observation: the proposal considered in \citet{Hyun2018-ta} involves conditioning on much more information than is used to construct the contrast vector $\nu$ in \eqref{eq:nu_c1_c2}. As pointed out by \citet{Fithian2014-ow} and \citet{Liu2018-zx}, conditioning on unnecessary information leads to reduced power. In this paper, we make use of recent ideas from \citet{jewell2019testing} to develop a computationally-efficient test of $H_0: \nu^{\top} \beta = 0$ that conditions on substantially less information than \citet{Hyun2018-ta}, thereby obtaining much higher power while still guaranteeing valid inference in the sense of \eqref{eq:selective_type_1}. 

While this paper was in preparation, \citet{Le_Duy2021-iy} independently developed a test of $H_0: \nu^\top \beta=0$ that has higher power than \citet{Hyun2018-ta}. Compared to that paper, our proposal (i) conditions on less unnecessary information; (ii) enjoys better numerical stability; and (iii) leads to more interpretable $p$-values. Details are provided in Appendix~\ref{appendix:detailed_comp_PP}.

The rest of this paper is organized as follows. In Section~\ref{section:related_work}, we briefly review the dual path algorithm for solving \eqref{eq:genlasso}, and the existing proposals for selective inference for this problem. In Section~\ref{section:method}, we introduce our selective inference procedure, which provides a computationally-efficient approach to condition on less information than \citet{Hyun2018-ta}. Section~\ref{section:extension} outlines some extensions, and Section~\ref{section:sim} compares the performance of our proposal to that of \citet{Hyun2018-ta} in simulation. A real data application is in Section~\ref{section:real_data}, and a discussion of future work is in Section~\ref{section:discussion}. Proofs and other technical details are relegated to the Appendix. 

Throughout this paper, we will use the following notational conventions. The $i$th row of a matrix $A$ is denoted $A_i$. Given a set $S$ of positive integers, $A_S$ is the submatrix with rows in $S$,  $A_{-S}$ is the submatrix with rows not in $S$, and $|S|$ is the cardinality of the set $S$. For a vector $\nu \in \mathbb{R}^n$, $||\nu||_1,||\nu||_2,$ and $||\nu||_\infty$ denote its $\ell_1,\ell_2,$ and $\ell_\infty$ norms, respectively. In addition, $\Pi_\nu^\perp$ denotes the projection matrix onto the orthogonal complement of $\nu$, i.e., $\Pi_\nu^\perp = I_n-\frac{\nu\nu^{\top}}{||\nu||_2^2}$, where $I_n$ is the $n$-dimensional identity matrix. We use $1_{(\cdot)}$ to denote the indicator function. For a positive integer $m$, we define $[m] = \{1,2,\ldots, m\}$.



%% file: sections-revision-v1/existing_work.tex

In this section, we review the selective inference framework of \citet{Hyun2018-ta} for testing hypotheses based upon the generalized lasso estimator \eqref{eq:genlasso}, which includes the graph fused lasso as a special case. Their framework relies on the dual path algorithm of \citet{Tibshirani2011-fq} for solving \eqref{eq:genlasso}. Thus, we begin with a very brief overview of that algorithm. 

\subsection{The dual problem, and the dual path algorithm}
\label{section:dual_path}

 \citet{Tibshirani2011-fq} develop an efficient
  path algorithm for solving the dual problem for \eqref{eq:genlasso}, which takes the form
\begin{equation}
\label{eq:dual}
\hat{u}(\lambda) = \argmin_{u \in \mathbb{R}^m} \quad \left\Vert y-D^{\top}u\right\Vert_2^2 \quad 
\textrm{subject to} \quad  \Vert u\Vert_\infty \leq \lambda,
\end{equation}
and is related to  \eqref{eq:genlasso} through the identity
$\hat{\beta}(\lambda) =   y - D^{\top}\hat{u}(\lambda)$,
where the notation $\hat{\beta}(\lambda)$ and $\hat{u}(\lambda)$ makes explicit that $\hat{\beta}$ and $\hat{u}$ are functions of $\lambda$. This dual path algorithm is detailed in Appendix~\ref{appendix:dual_path}. 
 While the details of the algorithm are not important for the current paper, we briefly summarize the main idea. 
The algorithm begins with $\lambda=\infty$, and then proceeds through a series of steps, corresponding to decreasing values of $\lambda$. 
The $k$th step involves computing a boundary set $B_k \subseteq [m]$, which consists of the subset of indices of the vector $u$ for which the inequality constraint in \eqref{eq:dual} is tight. The signs of the elements of $u$ associated with this boundary set, $s_{B_k}$, are also computed.
These quantities satisfy
\begin{equation}
\label{eq:primaldual2}
\hat{\beta}(\lambda) =   P_{\text{Null}(D_{-B_k})}\qty(y-\lambda \cdot D_{B_k}^{\top} s_{B_k})
\end{equation}
for a range of $\lambda$ values corresponding to the $k$th step~\citep{Tibshirani2011-fq}. 
In \eqref{eq:primaldual2}, $D_{B_k}$ and $D_{-B_k}$ correspond to the submatrices of $D$ with rows  in $B_k$ and not in $B_k$, respectively,  and $P_{\text{Null}(D_{-B_k})}$ is the projection matrix onto the null space of $D_{-B_k}$. 
To summarize, \eqref{eq:primaldual2} indicates that $\hat{\beta}({\lambda})$ can be computed from $(B_k,s_{B_k})$, for an appropriate range of $\lambda$ values. 

The next proposition considers the special case of the graph fused lasso problem \eqref{eq:graph_fused_lasso}.

\begin{proposition}
\label{prop:piecewise_beta} 
Let $B_k$ denote the boundary set that results from the $k$th step of the dual path algorithm for \eqref{eq:graph_fused_lasso}, and let $\hat\beta$ denote the solution to \eqref{eq:graph_fused_lasso}. Let $G_{-B_k}$ denote the subgraph of $G$ with edges in the boundary set $B_k$ removed, and let $C_1,\ldots, C_L$ denote the $L$ connected components of $G_{-B_k}$. Then, under \eqref{eq:model}, with probability 1, $\hat{\beta}_j = \hat{\beta}_{j'}$ if and only if $j,j' \in C_l$ for some $l \in [L]$.
\end{proposition}  

We close with a brief summary of the main points in this section:
\begin{itemize}
\item For the generalized lasso \eqref{eq:genlasso}, $\hat\beta$ can be computed from $\lambda$ and $(B_k,s_{B_k})$ from the dual path algorithm. 
\item In the special case of the graph fused lasso \eqref{eq:graph_fused_lasso},  the connected components of $\hat\beta$ can be computed from $B_k$ from the dual path algorithm. 
\end{itemize}

\subsection{Existing work on selective inference for the generalized lasso}
\label{section:existing_work_gen_lasso}

The main idea behind selective inference is as follows: when testing a null hypothesis that is a function of the data, to control the selective Type I error in the sense of \eqref{eq:selective_type_1}, we must condition on the information used to construct that null hypothesis  \citep{Tibshirani2016-bx,Lee2016-te,Fithian2014-ow}. In particular, to test a null hypothesis of the form $H_0: \nu^{\top} \beta=0$ where $\nu$ is a function of the data, we must condition on the information used to construct $\nu$.

 In a recent elegant line of work, a number of authors have shown that the model selection events of several well-known model selection procedures, including the lasso \citep{Lee2016-te}, stepwise regression \citep{Loftus2014-eq,Tibshirani2016-bx}, and marginal screening \citep{Reid2017-pn}, can be written as polyhedral constraints on $Y$. More precisely, conditioning on the selected model (and in some cases, additional information) is equivalent to conditioning on a polyhedral set $\{Y:AY\leq b\}$, where the matrix $A$ and the vector $b$ can be explicitly computed. Thus, we can test null hypotheses that are a function of the selected model by considering the null distribution of $Y$ truncated to a polyhedral set.


 
Recently, \citet{Hyun2018-ta} extended this line of work to develop an approach for selective inference for the generalized lasso \eqref{eq:genlasso}. Their key insight is as follows: 
\begin{quote}
 \emph{The set of $Y$ that leads to a specified output for the first $K$ steps of the dual path algorithm for \eqref{eq:genlasso} is a polyhedron, i.e., $\{Y: AY \leq 0\}$, for a matrix $A$ that can be explicitly computed.}
\end{quote}

 Proposition \ref{prop:generalized_hyun} details this result. 

\begin{proposition}[Proposition 3.1 in \citet{Hyun2018-ta}]
\label{prop:generalized_hyun} Consider solving \eqref{eq:genlasso} using the dual path algorithm for $y\in\mathbb{R}^n$. For the $k$th step, $k=1,\ldots,K$, define
\begin{equation}
{M}_k(y) \equiv \left(B_k(y), s_{B_k}(y), R_k(y), L_k(y) \right)
\label{eq:M}
\end{equation}
 where the boundary set $B_k(y)$ and the sign vector of the boundary set $s_{B_k}(y)$ are defined in Algorithm~\ref{alg:dual_path} (see Appendix~\ref{appendix:dual_path}), and 
  $R_{k+1}(y) = \qty(\emph{\text{sign}}(a_i): i\notin B_{k}(y))$ and $L_{k+1}(y) = \{i:i\in B_k(y), c_i<0, d_i<0\},$
 for $a_i$, $c_i,$ and $d_i$ specified in Algorithm \ref{alg:dual_path}. 
  
  Then the set $\left\{Y\in\mathbb{R}^n:  \bigcap_{k=1}^K \left\{ M_k\qty(Y) = M_k\qty(y) \right\}  \right\} $ is of the form  $\{Y:AY\leq 0\}$ for some  matrix $A$ that can be constructed explicitly based on $M_1(y),\ldots,M_K(y)$.  

\end{proposition}

Motivated by this result, \citet{Hyun2018-ta} proposed to test $H_0: \nu^{\top} \beta=0$, where $\nu$ is a function of the generalized lasso estimator, via a $p$-value of the form 
\begin{equation}
\pHyun \equiv \mathbb{P}_{H_0}\qty(|\nu^{\top} Y |\geq |\nu^{\top} y| \;\middle\vert\; \bigcap_{k=1}^K \left\{ M_k(Y) = M_k(y) \right\}, \Pi_{\nu}^{\perp} Y= \Pi_{\nu}^{\perp} y  ).
\label{eq:hyun_pval}    
\end{equation}  
In \eqref{eq:hyun_pval}, conditioning on $\Pi_{\nu}^{\perp} Y$ eliminates the nuisance parameter $\Pi_{\nu}^{\perp} \beta$; see Section 3.1 of \citet{Fithian2014-ow}. Now, under \eqref{eq:model}, the conditional distribution of $\nu^{\top} Y$ is normal with mean zero and variance $\sigma^2||\nu||_2^2$, truncated to a set that can be characterized and efficiently computed using Proposition~\ref{prop:generalized_hyun}. This yields the $p$-value in \eqref{eq:hyun_pval}. Furthermore, unlike the z-test based on the naive $p$-value $\mathbb{P}_{H_0}\qty(|\nu^{\top} Y |\geq |\nu^{\top} y| )$, a test that rejects $H_0$ when the $p$-value in \eqref{eq:hyun_pval} is less than some level $\alpha$ controls the selective Type I error rate, in the sense of \eqref{eq:selective_type_1}.

We emphasize that the $p$-value in \eqref{eq:hyun_pval} conditions on the event $\bigcap_{k=1}^K \left\{ M_k(Y) = M_k(y) \right\}$; that is,  on all of the outputs of the first $K$ steps of the dual path algorithm (rather than simply the $K$th step). However, typically the contrast vector $\nu$ in $H_0: \nu^{\top} \beta=0$ is constructed using only (at most) the output of the $K$th step in the dual path algorithm.  In what follows, we will consider conditioning on much less information than \eqref{eq:hyun_pval}. This will result in a test that controls the selective Type I error as in \eqref{eq:selective_type_1}, and that has substantially higher power under the alternative.

%% file: sections-revision-v1/method.tex

\subsection{What \emph{should} we condition on?}
\label{section:what_to_condition}
To control the selective Type I error in \eqref{eq:selective_type_1}, we must condition on the aspect of the data that led us to test the specific null hypothesis $H_0: \nu^\top \beta=0$ \citep{Fithian2014-ow,Hyun2018-ta}. 

If a data analyst wishes to choose the contrast vector $\nu$ in the null hypothesis $H_0: \nu^\top \beta=0$ by inspecting the elements of $\hat\beta$ resulting from the $K$th step of the dual algorithm of the generalized lasso problem \eqref{eq:genlasso}, then there is no reason to  condition on $ \bigcap_{k=1}^{K-1} \left\{ M_k(Y) = M_k(y) \right\}$ (as was done by \citet{Hyun2018-ta}), since the outputs of the first $K-1$ steps of the dual path algorithm are not considered in constructing $\nu$. In fact, according to \eqref{eq:primaldual2}, which states that $B_K$ and $s_{B_K}$ uniquely determine $\hat\beta$, the data analyst need only condition on $B_K$ and $s_{B_K}$, rather than on ${M}_K = \left(B_K, s_{B_K}, R_K, L_K \right)$. 

Furthermore, the data analyst might construct the contrast vector $\nu$ in $H_0: \nu^\top \beta=0$ to take on a constant value within each connected component of $\hat\beta$, as in \eqref{eq:null} and \eqref{eq:nu_c1_c2}. Recall from Proposition~\ref{prop:piecewise_beta} that the connected components of $\hat\beta$ are equivalent to the connected components of the subgraph $G_{-B_K}$. Therefore, it suffices to condition only on the connected components of the subgraph $G_{-B_K}$, or even on just the pair of connected components under investigation in \eqref{eq:null}.

What is the disadvantage of conditioning on $ \bigcap_{k=1}^K \left\{ M_k(Y) = M_k(y) \right\}$, as in \citet{Hyun2018-ta}? Conditioning on too much information leads to a loss of power \citep{Fithian2014-ow,Liu2018-zx,Lee2016-te,jewell2019testing}. We wish to condition on less information to achieve higher power than \citet{Hyun2018-ta}, while controlling the selective Type I error \eqref{eq:selective_type_1}. Of course, this could result in computational challenges, as the conditioning sets described in the use cases above are not polyhedral, so the conditional distribution of $\nu^\top Y$ is no longer a normal truncated to an easily-characterized set. 

In what follows, we focus on the case where the data analyst constructs the contrast vector $\nu$ to take on a constant value in each connected component of $\hat\beta$, and consider a $p$-value of the form 
\begin{equation}
\label{eq:pB}
\pB \equiv  \mathbb{P}_{H_0}\left(|\nu^\top Y |\geq |\nu^\top y|  \;\middle\vert\;  \hat{C}_1(y),  \hat{C}_2(y) \in \CC_K(Y), \Pi_{\nu}^{\perp} Y= \Pi_{\nu}^{\perp} y \right).
\end{equation}
 In \eqref{eq:pB}, $\hat{C}_1(y)$ and $\hat{C}_2(y)$ are two connected components estimated from the data realization $y$ and used to construct the contrast vector $\nu$ in \eqref{eq:nu_c1_c2}, and $\CC_K(Y)$ is the set of connected components obtained from applying $K$ steps of  the dual path algorithm for  \eqref{eq:graph_fused_lasso}  to the random variable $Y$.
%
Roughly speaking, this $p$-value answers the following question: 
\begin{quote}
\emph{Assuming that there is no difference between the \text{population} means of $\hat{C}_1$ and $\hat{C}_2$, then what's the probability of observing such a large difference in the \text{sample} means of $\hat{C}_1$ and $\hat{C}_2$, given that these two connected components were estimated from the data?}
\end{quote}

While our proposed $p$-value $\pB$ conditions on far less information than \cite{Hyun2018-ta}, the recent proposal of \citet{Le_Duy2021-iy} takes an intermediate approach. They condition on the full boundary set $B_K$ at the $K$th step of the dual path algorithm (and thus, implicitly, on \emph{all} of the connected components in $\CC_K(Y)$), whereas we condition only on the two connected components of interest. See Appendix~\ref{appendix:detailed_comp_PP} for further discussion and comparison.

\subsection{Illustrative example}

We now demonstrate that conditioning on less information leads to increased power, using an example involving the graph fused lasso applied to a two-dimensional grid graph.  

\begin{figure}[!ht]
\begin{subfigure}[c]{.25\textwidth}
  \caption{}
  \includegraphics[width=\linewidth]{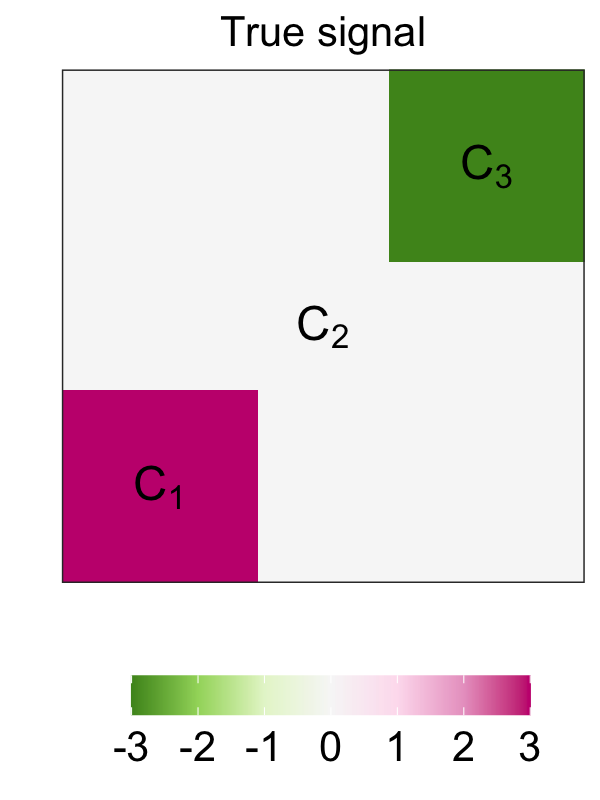}
\end{subfigure}
\begin{subfigure}[c]{.25\textwidth}
    \caption{}
  \includegraphics[width=\linewidth]{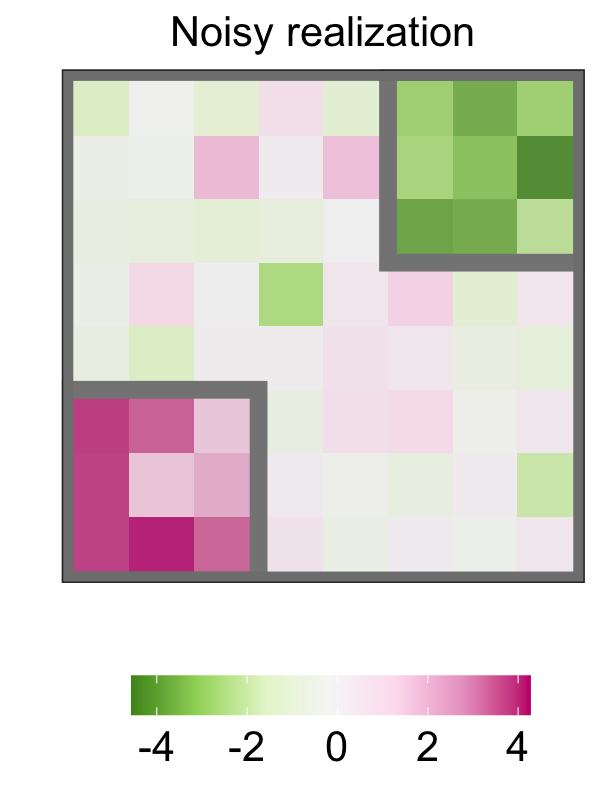}
\end{subfigure}
\begin{subfigure}[c]{0.3\textwidth}
  \centering
  \caption{}
\begin{tabular}{ccc}
\toprule
$H_0$ & \pHyun & \pB \\ 
\toprule
$\bar{\beta}_{\hat{C}_1} = \bar{\beta}_{\hat{C}_2}$     &    0.63    &          $<0.001$         \\ 
$\bar{\beta}_{\hat{C}_2} = \bar{\beta}_{\hat{C}_3}$     &    0.25    &          $<0.001$          \\
$\bar{\beta}_{\hat{C}_1} = \bar{\beta}_{\hat{C}_3}$     &    0.17    &          $<0.001$   \\ 
\bottomrule
\end{tabular}
\end{subfigure}
\begin{subfigure}{\textwidth}
    \caption{}
    \centering
  \includegraphics[width=\linewidth]{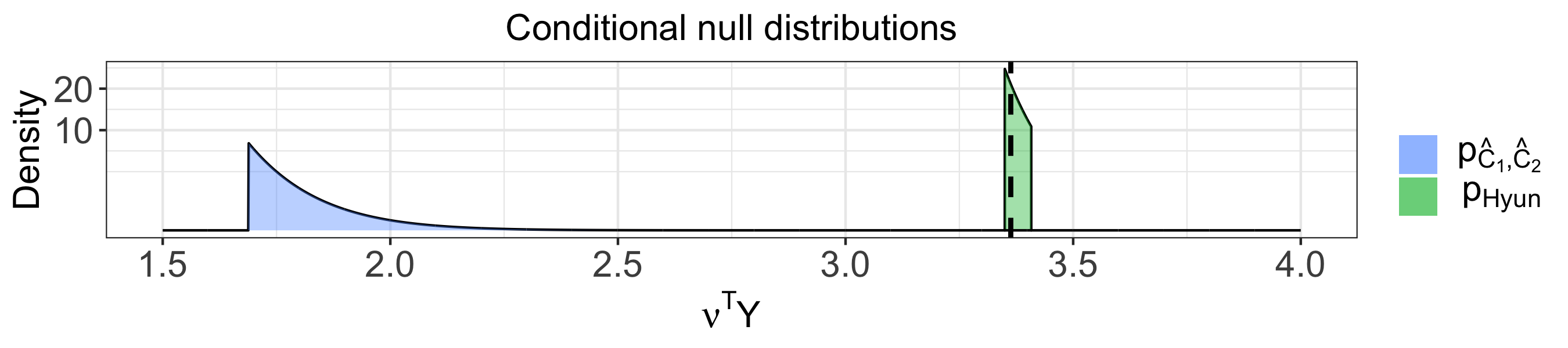}
    \end{subfigure}
     \bigskip


\caption{\textit{(a): }We generated $\beta$ on an $8 \times 8$ grid. There are three true connected components, which take on values of $-3$, $0$, and $3$. \textit{(b): }A noisy realization from the model  $Y \sim \mathcal{N}(\beta, I_{64})$. In this particular example, running $13$ steps of the dual path algorithm for the graph fused lasso results in perfect recovery of the true connected components of $\beta$ (displayed in grey). \textit{(c):} For each pair of estimated connected components, we tested the null hypothesis of equality in means using $\pHyun$ in \eqref{eq:hyun_pval} and $\pB$ in \eqref{eq:pB}. \textit{(d): }The conditional null distributions of $\nu^\top Y$, where $\nu$ is chosen to test for a difference in means between $\hat{C}_1$ and $\hat{C}_2$, conditional on the conditioning sets in the definitions of $\pHyun$ in \eqref{eq:hyun_pval} and $\pB$ in \eqref{eq:pB}. In \textit{(d)}, the test statistic  $|\nu^\top y| = 3.36$ is displayed as a dashed black line; this value is quite large relative to the null distribution of $\pB$, but modest relative to that of $\pHyun$.}
\label{fig:toy_example}
\end{figure}

To begin, we constructed a graph composed of $64$ nodes arranged in an $8 \times 8$ grid, such that each node is connected to its four closest (up, down, left, right) neighbors. 
We generated data on this grid according to \eqref{eq:model}, where $\beta$ has three piecewise constant segments, $C_1$, $C_2$, and $C_3$, with means of  $3$, $0$, and $-3$, respectively. The true values of $\beta$, as well as the data generated from this model, are shown in Figures~\ref{fig:toy_example}(a)--(b). On this particular data set, $K=13$ steps of the dual path algorithm for the graph fused lasso recovered the true connected components exactly. 

For each pair of connected components, we then constructed a contrast vector $\nu$ as in \eqref{eq:null}, so that $H_0: \nu^\top \beta=0$ posits that the two components being tested have the same mean. We tested $H_0$ using the $p$-values $\pHyun$ and $\pB$ given in \eqref{eq:hyun_pval} and \eqref{eq:pB}, respectively. The $p$-values for all pairs of connected components are displayed in Figure~\ref{fig:toy_example}(c). Because $\pHyun$ conditions on unnecessary information, the test based on $\pHyun$ has  extremely low power and it cannot reject any $H_0$. By contrast, the test based on $\pB$ has higher power. In Figure~\ref{fig:toy_example}(d), we display the null distribution of $\nu^\top Y$, conditional on the conditioning sets in \eqref{eq:hyun_pval} and \eqref{eq:pB}. 

\subsection{Properties of $\pB$}
\label{section:method_pb_algo}
The following result establishes key properties of $\pB$ in \eqref{eq:pB}.

\begin{proposition}
\label{prop:pval}
Suppose that $Y \sim \mathcal{N}(\beta,\sigma^2 I_n)$. 
 Define 
\begin{align}
\label{eq:phi}
y'(\phi) = \Pi_\nu^\perp y+\phi\cdot \frac{\nu}{||\nu||_2^2}  = y + \left(\frac{\phi-\nu^{\top}y}{||\nu||_2^2}\right)\nu.
 \end{align}
Let $\phi \sim \mathcal{N}(0,\sigma^2||\nu||_2^2)$. Then, under $H_0:\nu^\top \beta = 0$,
\begin{align}
\label{eq:single_param_null}
\pB = \mathbb{P}\qty(|\phi|\geq |\nu^\top y| \;\middle\vert\; \hat{C}_1(y),\hat{C}_2(y) \in \CC_K(y'(\phi))).
\end{align}
Moreover, the test that rejects $H_0:\nu^\top \beta = 0$ if $\pB\leq \alpha$ controls the selective Type I error.
\end{proposition}
Therefore, to compute the $p$-value in \eqref{eq:pB}, it suffices to characterize the set 
\begin{equation}
\label{eq:SB}
\SB \equiv \left\{\phi\in\mathbb{R}: \hat{C}_{1}(y),\hat{C}_{2}(y)\in \CC_K\qty(y'(\phi)) \right\}.
\end{equation} 

\begin{figure}[!ht]
\begin{subfigure}{0.3\textwidth}
  \centering
  \caption{}
  \includegraphics[width=\linewidth]{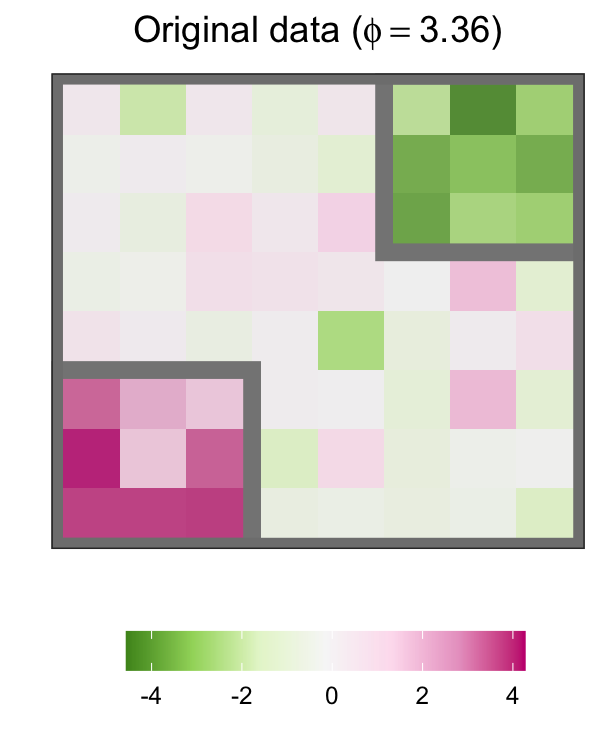}
\end{subfigure}
\begin{subfigure}{0.3\textwidth}
  \centering
  \caption{}
  \includegraphics[width=\linewidth]{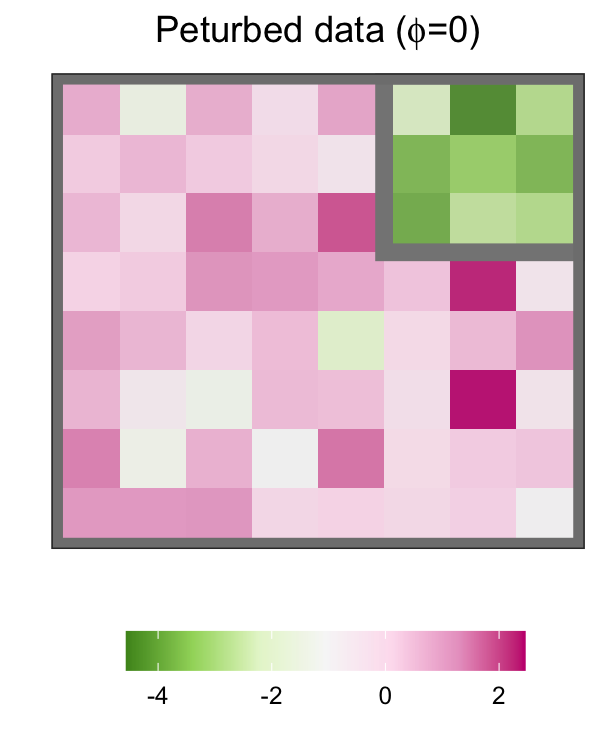}
\end{subfigure}
\begin{subfigure}{0.3\textwidth}
  \centering
  \caption{}
  \includegraphics[width=\linewidth]{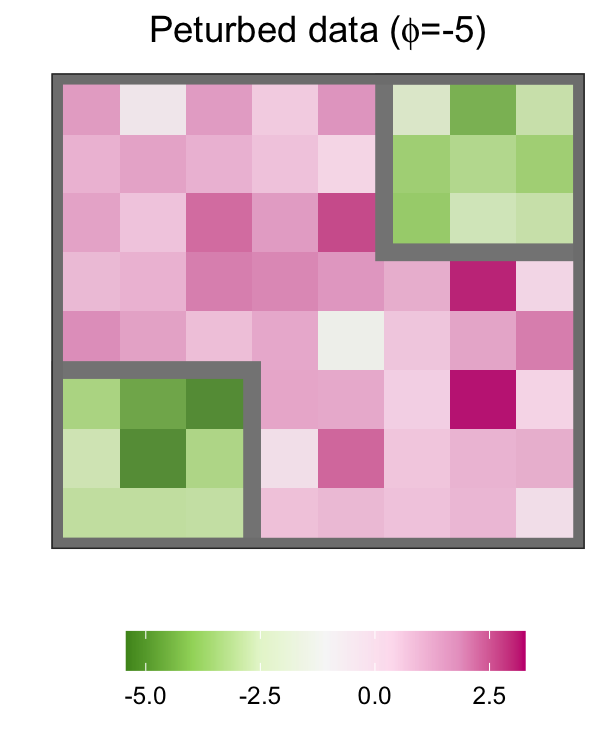}
\end{subfigure}
 \begin{subfigure}{\textwidth}
  \centering
   \caption{}
  \includegraphics[width=0.9\linewidth]{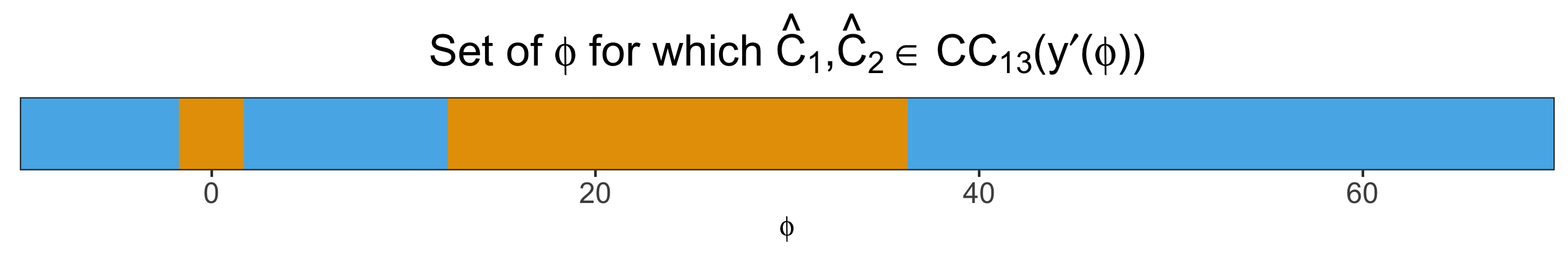}
    \end{subfigure}

\caption{Data generated according to the model in Figure~\ref{fig:toy_example}. \textit{(a): }The data $y$ in Figure~\ref{fig:toy_example}(b) corresponds to $y'(\phi)$ with $\phi = \nu^\top y = 3.36$. Applying the graph fused lasso with $K=13$ steps in the dual path algorithm results in three estimated connected components, displayed in grey boxes. Here, $\nu$ is chosen to test for a difference between the means of $\hat{C}_1(y)$ (lower left) and $\hat{C}_2(y)$ (middle). \textit{(b): }The perturbed dataset $y'(\phi)$ with $\phi = 0$. Applying the graph fused lasso with $K=13$ results in two connected components, displayed in grey boxes. \textit{(c): }The perturbed dataset $y'(\phi)$ with $\phi=-5$. Applying the graph fused lasso with $K=13$ results in $\hat{C}_1(y)$ and $\hat{C}_2(y)$. \textit{(d): }The set of $\phi$ for which $\hat{C}_1(y),\hat{C}_2(y) \in \CC_{13}\qty(y'\qty(\phi))$ is displayed in blue; other values are in orange.}
\label{fig:perturbation}
\end{figure}

We can think of $y'(\phi)$ in \eqref{eq:phi} as a perturbation of the data by a function of $\phi$ along the direction defined by $\nu$. Figure \ref{fig:perturbation} illustrates this intuition in the toy example from Figure \ref{fig:toy_example}, in the context of a test for the difference in the means of $\hat{C}_1$ and $\hat{C}_2$ (see Figure \ref{fig:perturbation}(a)). Panel (a) displays the observed data, for which $\phi=\nu^\top y = 3.36$, where $\nu$ is defined in \eqref{eq:nu_c1_c2}. In panel (b), we perturb the observed data to $\phi=0$. Now the graph fused lasso with $K=13$ no longer detects the three connected components. In panel (c), we perturb the observed data to $\phi=-5$; in this case, the graph fused lasso with $K=13$ estimates all three connected components. Therefore, $\phi = 3.36$ and $-5$ are in the set $\qty{\phi\in\mathbb{R}: \hat{C}_1,\hat{C}_2 \in \mathcal{CC}_{13}(y'(\phi))}$, but $\phi=0$ is not. Panel (d) displays $\qty{ \phi\in\mathbb{R}: \hat{C}_1,\hat{C}_2 \in \mathcal{CC}_{13}(y'(\phi)) } = (-\infty,-1.71)\cup (1.69, 12.3)\cup (36.3, \infty)$.





We now leverage ideas from \citet{jewell2019testing} to develop an efficient approach to compute the set \eqref{eq:SB}. First, we characterize the set $\SB$ in \eqref{eq:SB} in Proposition~\ref{prop:finite_union}. Recall that 
${M}_k(y) = \left(B_k(y), s_{B_k}(y), R_k(y), L_k(y) \right)$ is the output of the $k$th step of Algorithm~\ref{alg:dual_path}. 
We first present a corollary of Proposition~\ref{prop:generalized_hyun}.
\begin{corollary}
\label{cor:hyun_single_union}
The set $\left\{ \phi\in\mathbb{R}:  \bigcap_{k=1}^K \left\{ M_k(y'(\phi)) = M_k(y)\right\} \right\}$ is an interval.
\end{corollary}
\begin{proposition}
\label{prop:finite_union}
Let $\mathcal{I}$ be the set of possible outputs of Algorithm~\ref{alg:dual_path} that yield $\hat{C}_1$ and $\hat{C}_2$ and can be obtained via a perturbation of $y$ defined in \eqref{eq:phi}, i.e.,\begin{align}
\label{eq:I_s_b_set}
\mathcal{I} \equiv \qty{ (m_1,\ldots,m_K): \exists \alpha \in \mathbb{R} \text{ such that } \hat{C}_1(y), \hat{C}_2(y) \in \mathcal{CC}_K(y'(\alpha)),\bigcap_{k=1}^K \qty{{M}_k\qty(y'\qty(\alpha)) = m_k}}.
\end{align}
Then, there exists an index set $\mathcal{J}$ and scalars $\ldots< a_{-2}<a_{-1}<a_{0}<a_1<a_2<\ldots$ such that 
\begin{enumerate}
\item
 the set $\SB$ in \eqref{eq:SB} is the union of $\left|\mathcal{J}\right|$ intervals: 
\begin{align}
\label{eq:s_b_union}
\SB = \left\{\phi\in\mathbb{R}: \hat{C}_{1}(y),\hat{C}_{2}(y)\in \CC_K\qty(y'\qty(\phi)) \right\} = \bigcup_{i \in \mathcal{J}} \qty[a_i, a_{i+1}];
\end{align}
\item
$\left|\mathcal{I}\right| = \left|\mathcal{J}\right|$ (i.e., the sets $\mathcal{I}$ and $\mathcal{J}$ have the same cardinality); and
\item
$\forall i\in \mathcal{J}$, $\exists (m_1,\ldots,m_K) \in \mathcal{I}$ such that $[a_i,a_{i+1}] = \left\{ \phi\in\mathbb{R}:  \bigcap_{k=1}^K \left\{ M_k(y'(\phi)) =  (m_1,\ldots,m_K) \right\} \right\}$.
\end{enumerate}
\end{proposition} 
In words, Proposition~\ref{prop:finite_union} states that the set $\SB$ in \eqref{eq:SB} can be expressed as a union of intervals, each of which can be computed by applying Corollary~\ref{cor:hyun_single_union} on a perturbation of $y$. Next, we use Proposition~\ref{prop:finite_union} to develop an efficient recipe to compute $\SB$ by constructing the index set $\mathcal{J}$ and scalars $\ldots< a_{-2}<a_{-1}<a_{0}<a_1<a_2<\ldots$. To begin, we run the first $K$ steps of the dual path algorithm on the data $y$. We then apply Corollary~\ref{cor:hyun_single_union} to obtain the set $[a_0,a_1] = \left\{\phi\in\mathbb{R}: \bigcap_{k=1}^K \left\{ M_k\qty(y'\qty(\phi)) = M_k(y)\right\} \right\}$. By construction, $[a_0,a_1]\subset \SB$, because $\hat{C}_1$ and $\hat{C}_2$ are connected components estimated from the data $y$. Therefore, we initialize the index set $\mathcal{J}$ as $\{0\}$.
Then, for a small $\eta>0$, we apply Corollary~\ref{cor:hyun_single_union} to obtain the interval $\left\{ \phi\in\mathbb{R}: \bigcap_{k=1}^K \left\{ M_k\qty(y'\qty(\phi)) = M_k\qty(y'\qty(a_1 + \eta))\right\} \right\}$. If the left endpoint of this interval does not equal $a_1$, then we must repeat with a smaller value of $\eta$ until we obtain an interval of the form $[a_1, a_2]$. We can then check whether $\hat{C}_1, \hat{C}_2 \in \mathcal{CC}_K(y'(a_1+\eta))$: if so, then $[a_1, a_2] \subset \SB$  and we update $\mathcal{J}$ to include $\{1\}$. Otherwise, $\mathcal{J}$ remains unchanged. We continue in this vein, along the positive real line, until we reach an interval for which the right endpoint equals $\infty$.

Finally, we proceed along the negative real line: we apply Corollary~\ref{cor:hyun_single_union} to compute the interval $[a_{-1}, a_0] = \left\{ \phi\in\mathbb{R}:  \bigcap_{k=1}^K \left\{ M_k\qty(y'\qty(\phi)) = M_k\qty(y'\qty(a_0 - \eta))\right\} \right\}$. If $\hat{C}_1,\hat{C}_2 \in \mathcal{CC}_K(y'(a_0-\eta))$, then $\mathcal{J}$ is set to $\mathcal{J}\cup\{-1\}$; otherwise, $\mathcal{J}$ remains unchanged. We iterate until the algorithm outputs an interval for which the left endpoint equals $-\infty$. Finally, $\SB = \bigcup_{i \in \mathcal{J}} \qty[a_i, a_{i+1}]$. The procedure is summarized in Algorithm~\ref{alg:SB} of Appendix~\ref{appendix:S_B_algo}. 

In our implementation, we initialize with $\eta = 10^{-4}$, which proves to be an efficient choice in experiments in Section~\ref{section:sim} (see details in Appendix~\ref{appendix:S_B_algo}). In principle, the running time of Algorithm~\ref{alg:SB} can be quite slow, and potentially even exponential in $K$. However, in practice, the runtime of Algorithm~\ref{alg:SB} is nowhere near the worst-case upper bound (see Appendix~\ref{appendix:timing_complexity} for a  detailed empirical study of the timing complexity of Algorithm~\ref{alg:SB}). In addition, in Proposition~\ref{prop:conservative_p}, we describe an ``early stopping'' rule that guarantees a conservative $p$-value and only requires running Algorithm~\ref{alg:SB} until we reach intervals containing $|\nu^\top y|+\delta$ and $-|\nu^\top y|-\delta$ for some $\delta>0$, as opposed to $\infty$ and $-\infty$. Then, the set is appended with $(-\infty,-|\nu^\top y|-\delta]$ and $[|\nu^\top y|+\delta,\infty)$. This ``early stopping'' rule also applies to the extensions in Section~\ref{section:extension}.

\begin{proposition}
\label{prop:conservative_p}

Provided that $\mathbb{P}\qty(\phi \in \SB)>0$, for any $\delta>0$, we have that
\begin{align}
\mathbb{P}\left( |\phi|\geq |\nu^\top y| \;\middle\vert\; \phi \in \SBconserv \right) \geq \mathbb{P}\left( |\phi|\geq |\nu^\top y| \;\middle\vert\; \phi \in \SB \right),
\end{align}
where 
$\SBconserv \equiv (-\infty, -|\nu^\top y|-\delta] \cup \SB \cup [|\nu^\top y|+\delta,\infty)$.
\end{proposition}


%% file: sections-revision-v1/extension.tex

\subsection{Confidence intervals for $\nu^\top \beta$}
\label{sec:conf_int}
 We now construct a $(1-\alpha)$ confidence interval for $\nu^{\top} \beta$, the difference between the population means of two connected components $\hat{C}_1$ and $\hat{C}_2$ resulting from the graph fused lasso.
 
\begin{proposition}
\label{prop:ci}
Suppose that \eqref{eq:model} holds, and let $\hat{C}_1$ and $\hat{C}_2$ be two connected components obtained from performing $K$ steps of the dual path algorithm for the graph fused lasso \eqref{eq:graph_fused_lasso}.  
 For a given value of $\alpha\in(0,1)$, define functions $\theta_l(t)$ and $\theta_u(t)$ such that
 {
\begin{align}
F_{\theta_l(t),\sigma^2||\nu||_2^2}^{\SB}(t) = 1-\frac{\alpha}{2}, \quad  F_{\theta_u(t),\sigma^2||\nu||_2^2}^{\SB}(t) = \frac{\alpha}{2},
\label{eq:LCB_UCB}
\end{align}
 }
  where $F_{\mu,\sigma^2}^{\SB}(t)$ is the cumulative distribution function of a $\mathcal{N}(\mu,\sigma^2)$ random variable, truncated to the set $\SB$ defined in \eqref{eq:SB}. 
Then $[\theta_l(\nu^{\top} Y),\theta_u(\nu^{\top} Y)]$ has $(1-\alpha)$ selective coverage \citep{Lee2016-te,Fithian2014-ow,Tibshirani2016-bx} for $\nu^\top \beta$, in the sense that 
 {
 \begin{align}
 \mathbb{P}\Big(\nu^{\top}\beta \in \left[\theta_l(\nu^{\top} Y),\theta_u(\nu^{\top} Y)\right]  \;\Big\vert\; \hat{C}_1,\hat{C}_2 \in \mathcal{CC}_K(Y), \Pi_{\nu}^{\perp} Y= \Pi_{\nu}^{\perp} y  \Big) = 1-\alpha.
 \end{align}
  }
\end{proposition}
Computing $\theta_l$ and $\theta_u$ in \eqref{eq:LCB_UCB} amounts to a root-finding problem, which can be solved using bisection \citep{Chen2020-rh}. A similar result is used to construct confidence intervals corresponding to $\pHyun$ in \citet{Hyun2018-ta}.

\subsection{An alternative conditioning set}

The conditioning set for $\pB$ involves the connected components of the graph fused lasso solution after $K$ steps of the dual path algorithm. However, in practice, a data analyst might prefer a more ``user-facing" choice of $K$, such as the value that yields $L$ connected components in the solution $\hat\beta$. 

For this reason, we now consider a slight modification of $\pB$,
\begin{equation}
\label{eq:pD}
\pD = \mathbb{P}_{H_0}\qty(|\nu^\top Y |\geq |\nu^\top y | \;\middle\vert\; \hat{C}_1(y), \hat{C}_2(y) \in \CC(Y) ,  \Pi_{\nu}^{\perp} Y= \Pi_{\nu}^{\perp} y ),
\end{equation} 
where the subscript $K$ on $\CC$ has been dropped, indicating that the number of steps of the graph fused lasso algorithm is no longer fixed; instead, the function $\CC$ now represents the graph fused lasso estimator tuned to yield exactly $L$ connected components. Thus, in $\pD$, we condition on datasets for which $\hat{C}_1(y), \hat{C}_2(y)$ are among $L$ connected components estimated using the graph fused lasso. It is not hard to show that Proposition~\ref{prop:finite_union} and Algorithm~\ref{alg:SB} require only minor modifications to enable the computation of the $p$-values $\pD$; details are provided in Section~\ref{section:appendix:pC_PD} of the Appendix.

%% file: sections-revision-v1/sim.tex

We consider testing the null hypothesis $H_0: \nu^\top \beta = 0 \mbox{ versus } H_1: \nu^\top \beta \neq 0$, where, unless otherwise stated, $\nu$ is defined in \eqref{eq:nu_c1_c2} for a randomly-chosen pair of estimated connected components $\hat{C}_1$, $\hat{C}_2$ of the solution to \eqref{eq:graph_fused_lasso}. We consider three $p$-values: $\pHyun$ in \eqref{eq:hyun_pval}, $\pB$ in \eqref{eq:pB}, and the naive $p$-value 
  \begin{equation}
  \label{eq:naive_p}
  p_{\text{Naive}} \equiv \mathbb{P}_{H_0}\left( | \nu^\top Y |\geq | \nu^\top y| \right),
  \end{equation} 
  and compare the selective Type I error \eqref{eq:selective_type_1} and power of the tests that reject $H_0$ when these $p$-values are less than $\alpha=0.05$. 

  In the simulations that follow, comparing the power of the tests requires a bit of care. Because the null hypothesis $H_0:\nu^\top \beta = 0$ involves the contrast vector $\nu$, which is a function of the data, the effect size $|\nu^\top \beta|$ may differ across simulated datasets from the same data-generating distribution. Therefore, in what follows, we consider the power as a function of $|\nu^\top \beta|$. Alternatively, we can separately assess the detection probability (i.e., the probability that $\hat{C}_1$ and $\hat{C}_2$ are true piecewise constant segments) and the  ``conditional power''~\citep{Hyun2018-gx,Gao2020-yt} (i.e., the probability of rejecting $H_0$, given that $\hat{C}_1$, $\hat{C}_2$ are true piecewise constant segments). Details are in Appendix~\ref{appendix:conditional_power}.

\subsection{One-dimensional fused lasso}
\label{section:sim_one_d}

We first consider the special case of the graph fused lasso on a chain graph, in which the observations are ordered, and there is an edge between each pair of adjacent observations. This leads to the one-dimensional fused lasso problem \citep{Tibshirani2011-fq}. 
We simulated from the ``middle mutation'' model of \citet{Hyun2018-gx}, where the signal contains two true changepoints of size $\delta$, and in turn, three connected components:
\begin{align}
 \label{sim:1d_eq}
Y_j \overset{ind.}{\sim} \mathcal{N}(\beta_j,\sigma^2), \quad  \beta_j = \delta \times 1_{(101 \leq j \leq 140)}, \quad j = 1,\ldots, 200.
\end{align}
Figure~\ref{fig:one_d_sim}(a) displays an example of this synthetic data with $\delta=3$ and $\sigma=1$.

\subsubsection{Selective Type I error control under the global null}
\label{section:sim_one_d_type_1}

We simulated $y_1,\ldots, y_{200}$ according to \eqref{sim:1d_eq} with $\delta = 0$ and $\sigma = 1$. Therefore, the null hypothesis $H_0:\nu^\top \beta = 0$ holds for all contrast vectors $\nu$ in $\eqref{eq:nu_c1_c2}$, regardless of the pair of estimated connected components under consideration. 

We solved \eqref{eq:graph_fused_lasso} with $K=2$ steps in the dual path algorithm, which yields exactly three estimated connected components by the properties of the one-dimensional fused lasso. Then, for each simulated dataset, we computed $\pB$ in \eqref{eq:pB}, $\pHyun$ in \eqref{eq:hyun_pval}, and the naive $p$-value in \eqref{eq:naive_p}.

Figure~\ref{fig:one_d_sim}(b) displays the observed $p$-value quantiles versus Uniform$(0,1)$ quantiles, aggregated over 1,000 simulated datasets. We see that (i) the test based on the naive $p$-value in \eqref{eq:naive_p}, which does not account for the fact that the connected components were estimated from the data, is anti-conservative; and (ii) tests based on $\pHyun$ and $\pB$ control the selective Type I error \eqref{eq:selective_type_1}.

\subsubsection{Power as a function of effect size}
\label{section:sim_one_d_power}

Next, we show that the test based on $\pB$ has higher power than that based on $\pHyun$. We generated 1,500 datasets from \eqref{sim:1d_eq} with $\sigma\in\{0.5,1,2\}$, for each of ten evenly-spaced values of $\delta\in [0.5,5]$. For every simulated dataset, we solved \eqref{eq:graph_fused_lasso} with $K=2$. We then rejected $H_0:\nu^\top \beta = 0$ if $\pHyun$ or $\pB$ was less than $\alpha=0.05$. Recalling that $\nu$ in \eqref{eq:nu_c1_c2} is a function of the data, and the effect size $|\nu^\top \beta|$ will vary across simulated datasets drawn from an identical distribution, we created seven evenly-spaced bins of the observed values of $|\nu^\top\beta|$, and then computed the proportion of simulated datasets for which we rejected $H_0$ within each bin. 

\begin{figure}[htbp!]
\hspace{15mm} (a) \hspace{48mm} (b) \hspace{48mm} (c) \\
 \centering
  \includegraphics[width=\linewidth]{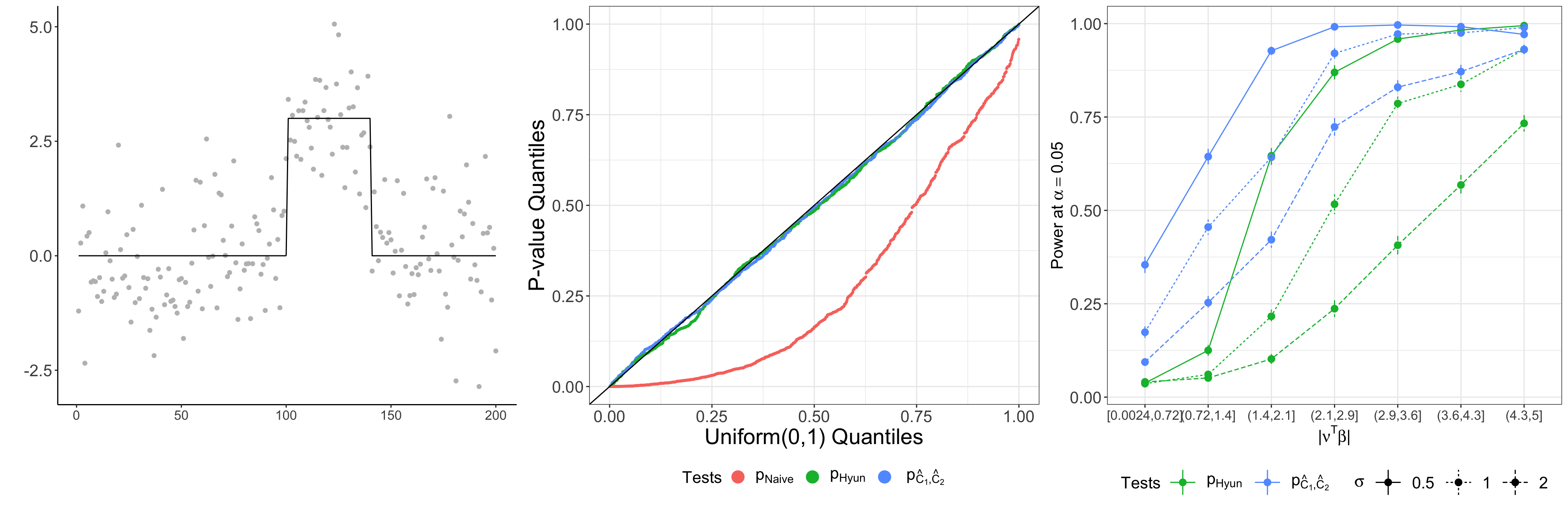}
\caption{\textit{(a):} One realization of $y$ generated according to \eqref{sim:1d_eq} with $\delta=3$ and $\sigma=1$ (grey dots), along with the true signal $\beta$ (black curve). \textit{(b):} When $\delta=0$, tests based on both $\pHyun$ in \eqref{eq:hyun_pval} and $\pB$ in \eqref{eq:pB} control the selective Type I error in the sense of \eqref{eq:selective_type_1}. By contrast, the naive $p$-value in \eqref{eq:naive_p} leads to a test with inflated selective Type I error. \textit{(c):} The power of the tests based on both $\pHyun$ and $\pB$ increases as a function of $|\nu^\top\beta|$. For a given bin of $|\nu^\top\beta|$, the test based on $\pB$ has higher power than the test based on $\pHyun$; the power of each test increases as $\sigma$ decreases.}
\label{fig:one_d_sim}
\end{figure}

Results are in Figure~\ref{fig:one_d_sim}(c). The power of each test increases as the value of $|\nu^\top\beta|$ increases. For a given bin of $|\nu^\top\beta|$, the test based on $\pB$ has higher power than the test based on $\pHyun$. For a given test and bin of $|\nu^\top\beta|$, a smaller value of $\sigma$ results in higher power. As an alternative to binning, we can use regression splines to estimate the power as a smooth function of the effect size; see Appendix~\ref{appendix:conditional_power}. 

\subsection{Two-dimensional fused lasso}
\label{section:sim_two_d}
We consider the graph fused lasso on a grid graph, constructed by connecting each node to its four closest neighbors (up, down, left, right). This leads to the two-dimensional fused lasso problem, also known as total-variation denoising~\citep{Tibshirani2011-fq,Rudin1992-gb}. 

The signal $\beta$ consists of with 64 observations arranged in an $8 \times 8$ grid. It has three piecewise constant segments with means $\delta$, 0, and $-\delta$, displayed in Figure~\ref{fig:two_d_sim}(a):
\begin{align}
 \label{sim:2d_eq}
Y_j \overset{ind.}{\sim} \mathcal{N}(\beta_j,\sigma^2), \quad \beta_j = \delta \times 1_{j\in C_1} + (-\delta) \times 1_{j\in C_3}, \quad j=1,\ldots,64.
\end{align}

\subsubsection{Selective Type I error control under the global null}
\label{section:sim_two_d_type_1}

We simulated $y_1,\ldots,y_{64}$ according to \eqref{sim:2d_eq} with $\delta=0$ and $\sigma=1$. Thus, the null hypothesis $H_0:\nu^\top \beta = 0$ holds for any contrast vector $\nu$ under consideration.

For each simulated dataset, we solved \eqref{eq:graph_fused_lasso} with $K=15$ steps in the dual path algorithm, which typically yields between 2 and 4 estimated connected components. Then, provided that there was more than one connected component in the solution $\hat\beta$, we computed $p_{\text{Naive}}$ in \eqref{eq:naive_p}, $\pHyun$ in \eqref{eq:hyun_pval}, and $\pB$ in \eqref{eq:pB}. We rejected $H_0$ if the $p$-values are less than $\alpha=0.05$. 

 Panel (b) of Figure~\ref{fig:one_d_sim} displays the observed $p$-values quantiles versus the Uniform$(0,1)$ quantiles, over 1,000 simulated datasets. As in Section~\ref{section:sim_one_d_type_1}, the tests based on both $\pHyun$ and $\pB$ control the selective Type I error in \eqref{eq:selective_type_1}, whereas the test based on $p_{\text{Naive}}$ is anti-conservative.

\subsubsection{Power as a function of effect size}
\label{section:sim_two_d_power}

 We generated data according to \eqref{sim:2d_eq} with each of eight evenly-spaced values of $\delta\in [0.5,4]$ and $\sigma\in\{0.5,1,2\}$. For each simulated dataset, we solved \eqref{eq:graph_fused_lasso} with $K=15$ steps in the dual path algorithm. Provided that there were at least two estimated connected components, we then computed $\pHyun$ in \eqref{eq:hyun_pval} and $\pB$ in \eqref{eq:pB}, and rejected $H_0$ if the $p$-values were less than $0.05$. 

\begin{figure}[htbp!]
\hspace{13mm} (a) \hspace{48mm} (b) \hspace{48mm} (c) \\
 \centering
  \includegraphics[width=\linewidth]{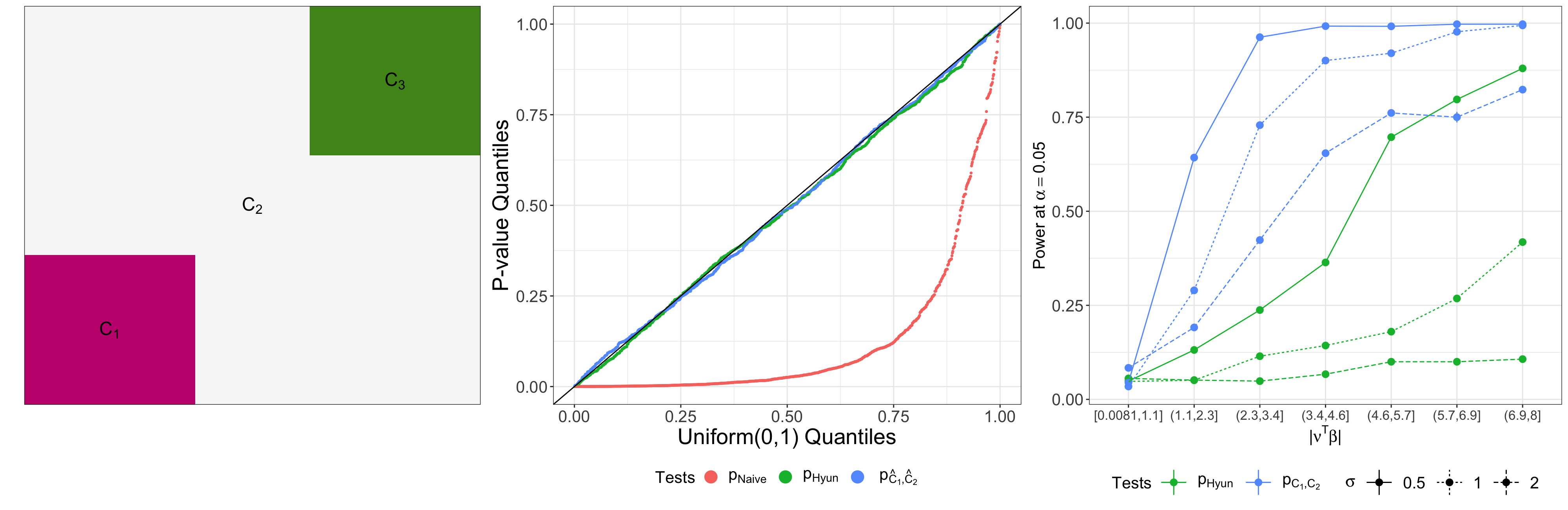}
\caption{\textit{(a):} The piecewise constant segments of $\beta$ in \eqref{sim:2d_eq}. \textit{(b):} When $\delta=0$, tests based on both $\pHyun$ in \eqref{eq:hyun_pval} and $\pB$ in \eqref{eq:pB} control the selective Type I error. By contrast, the test based on $p_{\text{Naive}}$ in \eqref{eq:naive_p} has inflated selective Type I error. \textit{(c):} The power of the tests based on $\pHyun$ and $\pB$ increases as a function of $|\nu^\top\beta|$. For a given bin of $|\nu^\top\beta|$, the test based on $\pB$ has substantially higher power than the test based on $\pHyun$. The power of each test increases as $\sigma$ decreases.}

\label{fig:two_d_sim}
\end{figure}

In Figure~\ref{fig:two_d_sim}(c), we display the proportion of simulated datasets for which we rejected $H_0$ using the two tests, over seven evenly-spaced bins of $|\nu^\top\beta|$. For a given bin, the test based on $\pB$ has substantially higher power than that based on $\pHyun$; the power of each test increases as $\sigma$ decreases. 

\subsubsection{Allowing for unknown variance}
\label{section:unknown_var}

Throughout this section, we have assumed that $\sigma^2$ in \eqref{eq:model} is known. In Appendix~\ref{appendix:estimated_sigma}, we investigate the Type I error control and power of several variance estimators in simulations. 

%% file: sections-revision-v1/real_data.tex

In this section, we apply our proposed $p$-value $\pB$ to a dataset consisting of two measures: (i) drug overdose death rates (deaths per 100,000 persons), and (ii) teenage birth rates (births per 1,000 females aged 15--19), in each of the 48 contiguous states in the United States~\citep{cdc_2020-drug,cdc_2020-teen}. In what follows, we consider the two measures after applying a log transformation. We can think of the data as noisy measurements of the true drug overdose death and teenage birth rate rates in each state, which are known to exhibit geographic trends~\citep{Schieber2019-pq,Ventura2014-ir,Amin2017-kr}. Therefore, we solve the graph fused lasso in \eqref{eq:graph_fused_lasso} with  a custom graph that encodes the geography of the 48 states: each state is a node, and there is an edge between each contiguous pair of states. We then consider testing the equality of measures for pairs of estimated connected components. 

For each pair of connected components, we computed three $p$-values: $\pB$ in \eqref{eq:pB}, $\pHyun$ in \eqref{eq:hyun_pval}, and $p_{\text{Naive}}$ in \eqref{eq:naive_p}. We also computed confidence intervals for $\nu^\top \beta$, the difference between population means of a pair of estimated connected components, using $\pB$ and $\pHyun$, as described in Section~\ref{sec:conf_int}, along with the naive confidence interval $\qty[\nu^\top y - z_{1-\alpha/2}\cdot  \sigma\Vert\nu\Vert_2 ,\nu^\top y + z_{1-\alpha/2} \cdot \sigma\Vert\nu\Vert_2]$, where $z_\alpha$ is the $\alpha$th quantile of the standard normal distribution. For each $p$-value and confidence interval, we used $\hat{\sigma}^2 = \frac{1}{48-L}\sum_{l=1}^{L}\sum_{j\in \hat{C}_l}\qty(y_j-\qty(\sum_{j'\in\hat{C}_l}y_{j'})/|\hat{C}_l| )^2$ to estimate $\sigma^2$ in \eqref{eq:model}, where $\hat{C}_1,\ldots,\hat{C}_L$ are the estimated connected components.




\subsection{Drug overdose death rates in the contiguous U.S. in 2018}
\label{sec:drug_overdose}

\begin{figure}[!htbp]
\hspace{15mm} (a) \hspace{45mm} (b) \hspace{45mm} (c) \\
\centering
\raisebox{-.5\height}{\includegraphics[width=0.32\linewidth]{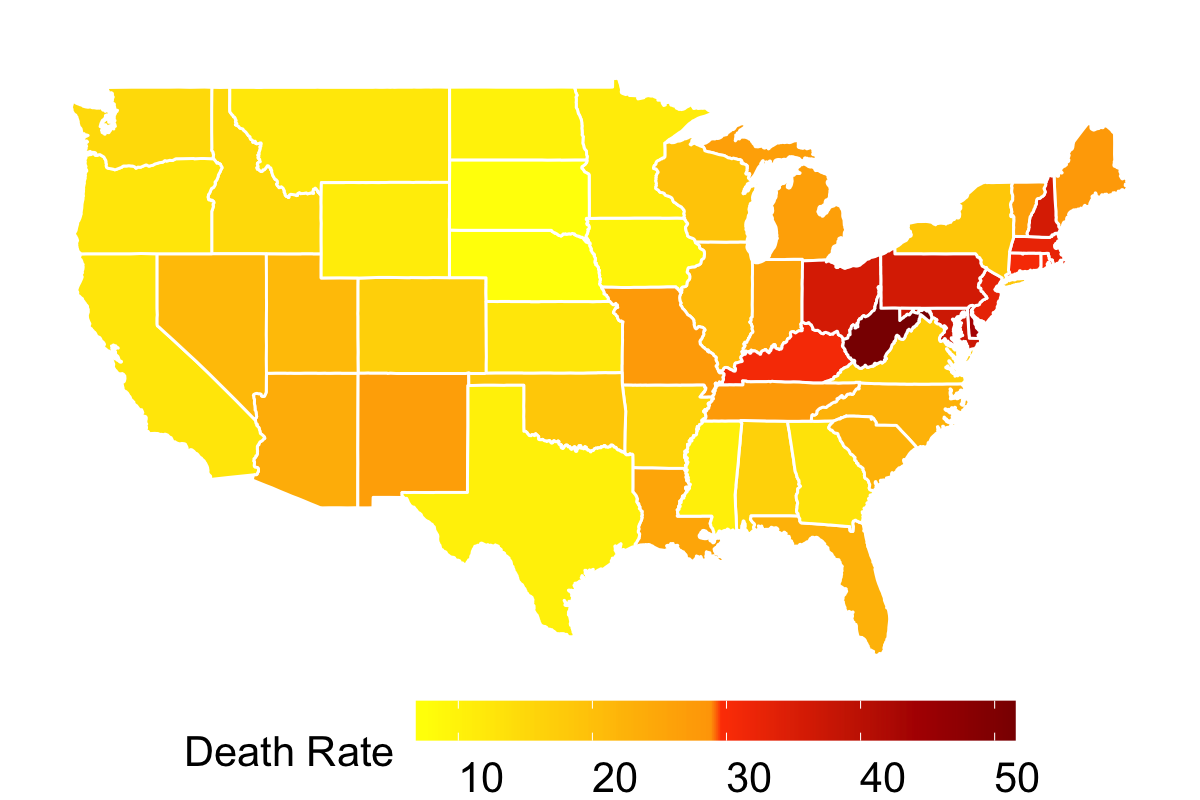}
\includegraphics[width=0.32\linewidth]{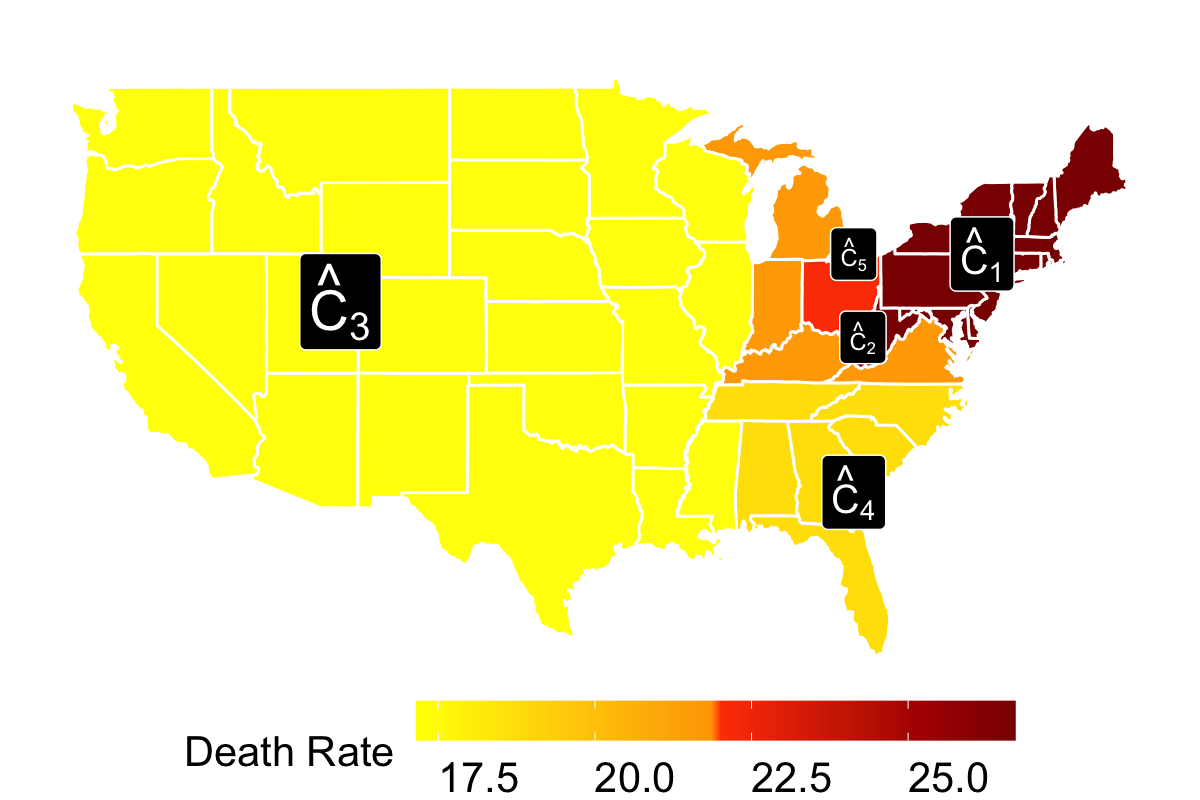}}
\raisebox{-.35\height}{\resizebox{0.28\linewidth}{!}{
\begin{tabular}{cccccc}
  \toprule
  $H_0$  & $p_{\text{Naive}} $ & $\pB$  & $\pHyun$  \\
  \toprule
$\bar{\beta}_1 = \bar{\beta}_2$ & 0.78  & 0.78  & 0.45 \\
\rowstyle{\boldmath\textbf} 
$\bar{\beta}_1 = \bar{\beta}_3$ & \boldmath{$<0.001$} & \boldmath{$<0.001$}& \boldmath{$0.890$}  \\
\rowstyle{\boldmath\textbf}
$\bar{\beta}_1 = \bar{\beta}_4$ & \textbf{0.003} & \textbf{0.024}& \textbf{0.820} \\
$\bar{\beta}_1 = \bar{\beta}_5$ & 0.12 & 0.49& 0.68   \\
$\bar{\beta}_2 = \bar{\beta}_3$ & 0.007 & 0.64& 0.58  \\
$\bar{\beta}_2 = \bar{\beta}_4$ & 0.10 & 0.70 & 0.65   \\
$\bar{\beta}_2 = \bar{\beta}_5$& 0.29 & 0.78 & 0.52  \\
$\bar{\beta}_3 = \bar{\beta}_4$& 0.03 & 0.21 & 0.36  \\
\rowstyle{\boldmath\textbf}
$\bar{\beta}_3 = \bar{\beta}_5$ & \textbf{0.003} & \textbf{0.039} & \textbf{0.650}  \\
$\bar{\beta}_4 = \bar{\beta}_5$ & 0.36  & 0.24 & 0.60 \\
 \bottomrule
\end{tabular}}}

\hspace{10mm} (d)  \\ 
\centering
\includegraphics[width=\linewidth]{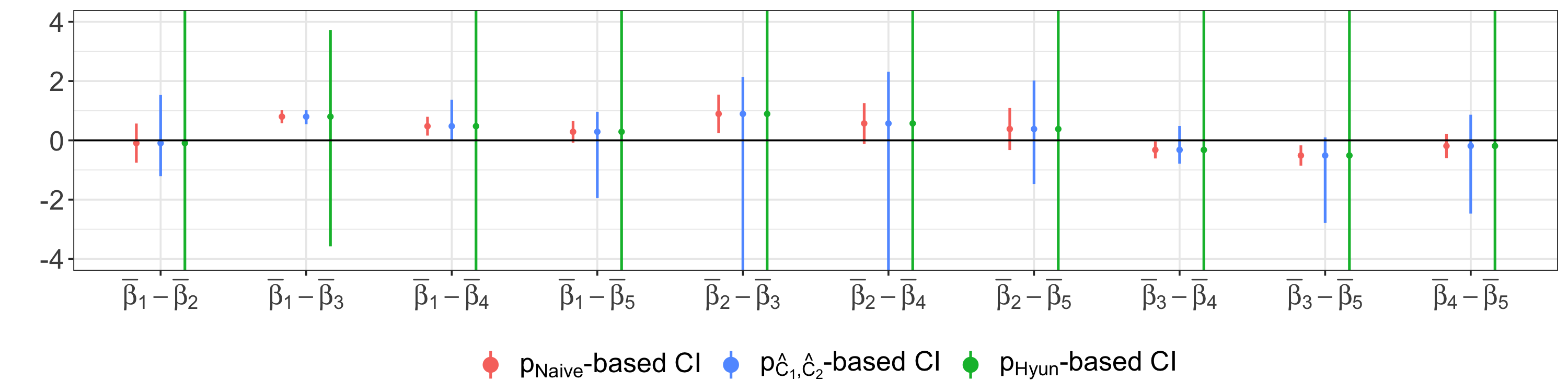}

\caption{\textit{(a):} The observed drug overdose death rates (deaths per 100,000 persons) for the 48 contiguous U.S. states in the year 2018. \textit{(b):} Applying the graph fused lasso to the drug overdose data results in five estimated connected components. \textit{(c):} For each pair of estimated connected components, we  computed $p_{\text{Naive}}$ in \eqref{eq:naive_p}, $\pHyun$ in \eqref{eq:hyun_pval}, and $\pB$ in \eqref{eq:pB}. For brevity, we use the notation $\bar{\beta}_l = \sum_{j\in \hat{C}_l}\beta_j/|\hat{C}_l|$.  \textit{(d):} For each pair of estimated connected components, we constructed confidence intervals for the difference in means, corresponding to $p_{\text{Naive}}$, $\pHyun$, and $\pB$.}
\label{fig:real_data_drug}
\end{figure}

Figure~\ref{fig:real_data_drug}(a) displays the drug overdose death rate in a color map. We solved \eqref{eq:graph_fused_lasso} with $K=30$ steps in the dual path algorithm, which resulted in five connected components (see Appendix~\ref{appendix:real_data_different_K} for results with other choices of $K$); the results are displayed in Figure~\ref{fig:real_data_drug}(b). We have estimated a constant drug overdose death rate in five geographical regions, which we refer to as the Northeast ($\hat{C}_1$), Ohio ($\hat{C}_2$), the West and Mountain region ($\hat{C}_3$), the Southeast ($\hat{C}_4$), and the Midwest ($\hat{C}_5$). Among these regions, the Northeast and Midwest have the highest estimated drug overdose rates.

We assess the equality of the means of each pair of connected components using $p_{\text{Naive}}$ in \eqref{eq:naive_p}, $\pHyun$ in \eqref{eq:hyun_pval}, and $\pB$ in \eqref{eq:pB}. The results are  in Figure~\ref{fig:real_data_drug}(c). The subset of pairs for which $\pB$ is below $0.05$ and $\pHyun$ is not is displayed in bold. For instance, the Northeast ($\hat{C}_1$) and the Southeast ($\hat{C}_4$) have a statistically significant difference in mean drug overdose death rates using the test based on $\pB$, but not using the test based on $\pHyun$, at level $\alpha=0.05$. Confidence intervals corresponding to these $p$-values are displayed in Figure~\ref{fig:real_data_drug}(d). Intervals based on $\pHyun$ are much wider than those based on $\pB$ across all ten pairs of connected components. In addition, the confidence intervals based on $\pB$ are not much wider than those based on $p_{\text{Naive}}$, even though the latter do not have correct coverage for the true parameter $\nu^\top\beta$. 

\subsection{Teenage birth rates in the contiguous U.S. in 2018}
\label{sec:teen_birth}

\begin{figure}[!htbp]
\hspace{15mm} (a) \hspace{45mm} (b) \hspace{45mm} (c) \\
\centering
\raisebox{-.5\height}{\includegraphics[width=0.32\linewidth]{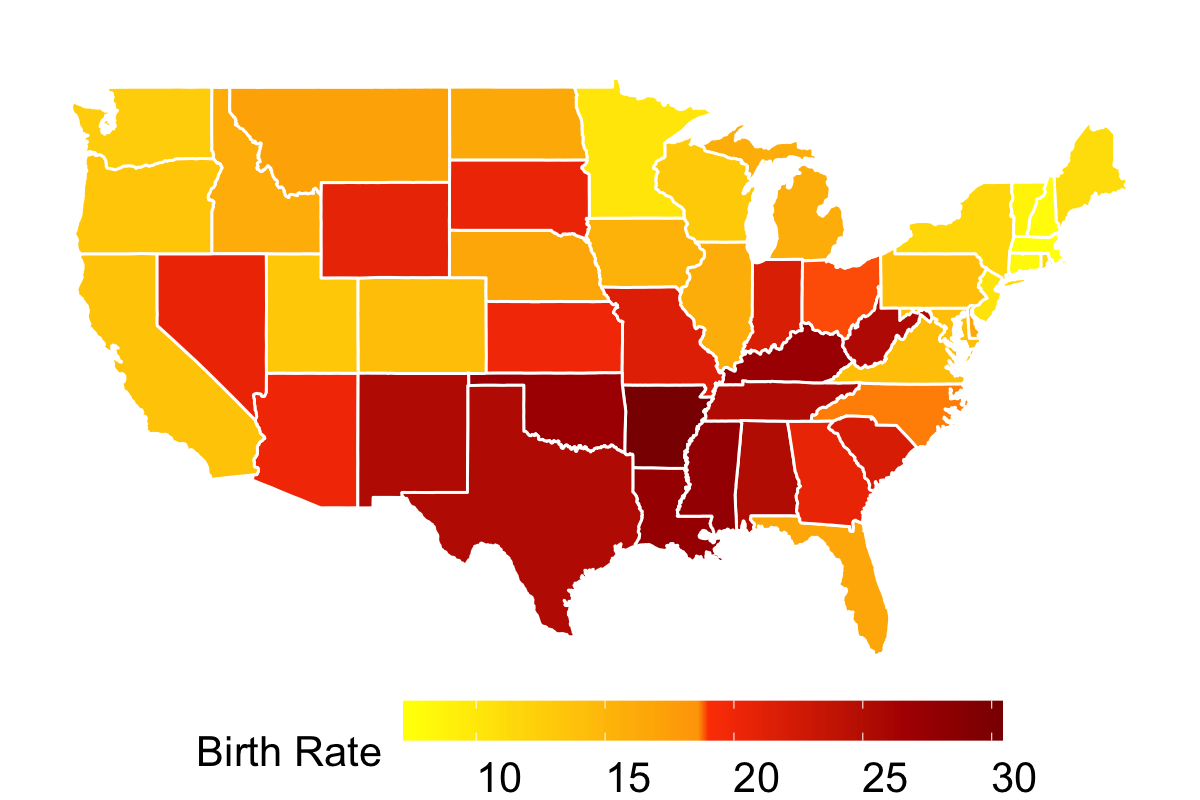}
\includegraphics[width=0.32\linewidth]{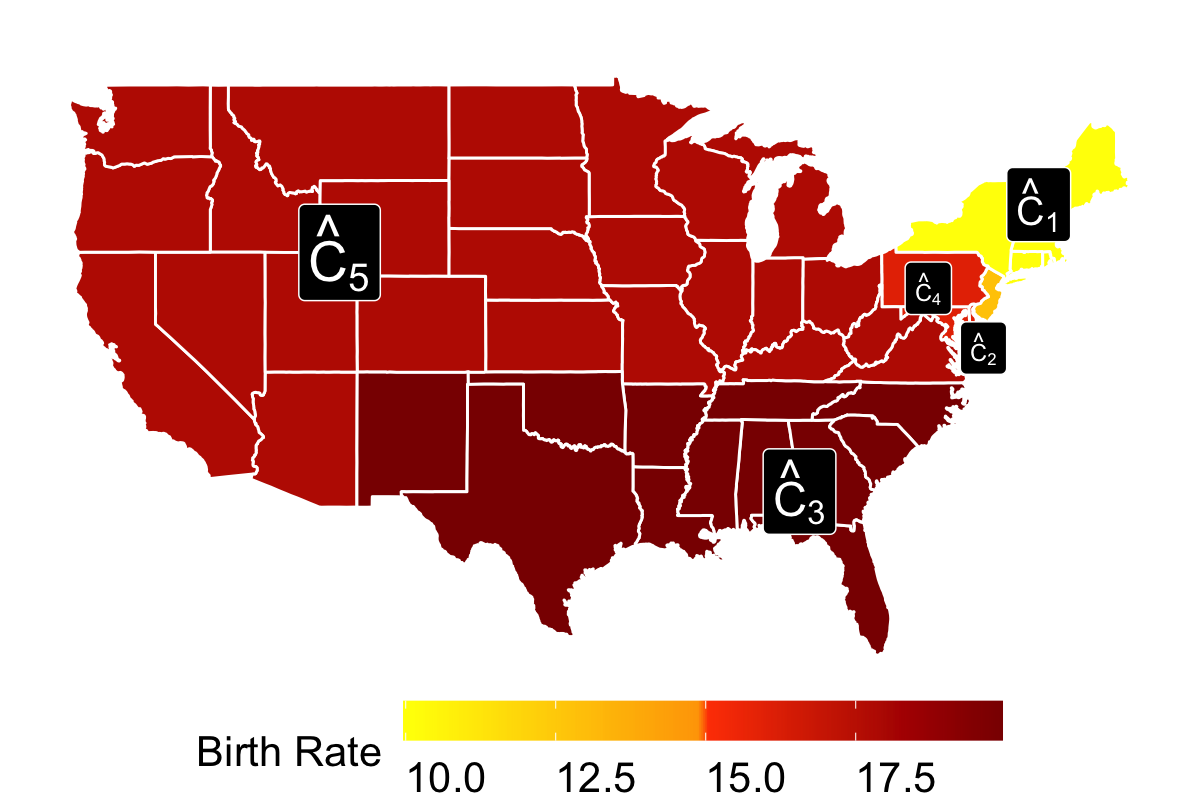}}
\raisebox{-.35\height}{\resizebox{0.28\linewidth}{!}{%
\begin{tabular}{cccccc}
  \toprule
  $H_0$  & $p_{\text{Naive}} $ & $\pB$ & $p_{\text{Hyun}}$   \\
  \toprule
$\bar{\beta}_1 = \bar{\beta}_2$ & 0.67  & 0.99 & 0.74 \\
\rowstyle{\boldmath\textbf} 
$\bar{\beta}_1 = \bar{\beta}_3$ & \boldmath{$<0.001$}& \boldmath{$<0.001$} & \boldmath{$0.57$}  \\
\rowstyle{\boldmath\textbf}
$\bar{\beta}_1 = \bar{\beta}_4$ & \boldmath{$0.001$}  & \textbf{0.005} & \textbf{0.26} \\
\rowstyle{\boldmath\textbf} 
$\bar{\beta}_1 = \bar{\beta}_5$ &  \boldmath{$<0.001$}& \boldmath{$<0.001$} & \boldmath{$0.30$}  \\
\rowstyle{\boldmath\textbf} 
$\bar{\beta}_2 = \bar{\beta}_3$ & \boldmath{$<0.001$} & \boldmath{$0.02$}  & \boldmath{$0.260$}  \\
$\bar{\beta}_2 = \bar{\beta}_4$ & 0.13 & 0.24 & 0.26   \\
$\bar{\beta}_2 = \bar{\beta}_5$& 0.02 & 0.10 & 0.24  \\
\rowstyle{\boldmath\textbf}
$\bar{\beta}_3 = \bar{\beta}_4$& \boldmath{$<0.001$}  & \textbf{0.04} & \textbf{0.71} \\
\rowstyle{\boldmath\textbf}
$\bar{\beta}_3 = \bar{\beta}_5$  & \boldmath{$<0.001$} & \textbf{0.002} & \textbf{0.61}  \\
$\bar{\beta}_4 = \bar{\beta}_5$ & 0.34 & 0.38 & 0.71  \\
  \bottomrule
\end{tabular}}}

\hspace{10mm} (d)  \\ 
\centering
\includegraphics[width=\linewidth]{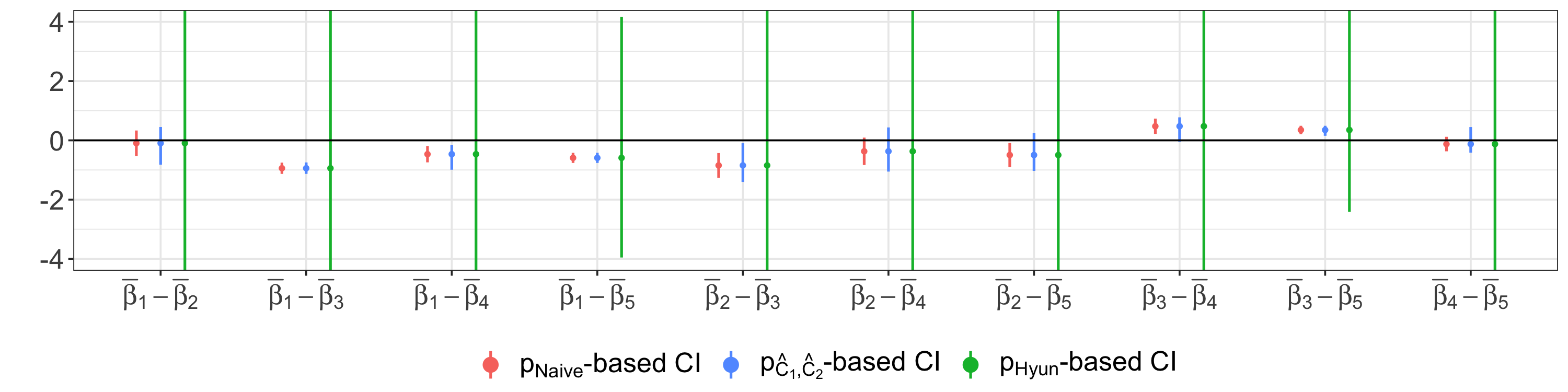}

\caption{\textit{(a):} The observed teenage birth rates (births per 1,000 females aged 15--19) for the 48 contiguous U.S. states in 2018. \textit{(b):} The graph fused lasso solution with $K=30$ results in five connected components, displayed in distinct colors. \textit{(c):} For each pair of estimated connected components, we computed $p_{\text{Naive}}$ in \eqref{eq:naive_p}, $\pHyun$ in \eqref{eq:hyun_pval}, and $\pB$ in \eqref{eq:pB}. Pairs for which the test based on $\pB$ results in a rejection at $\alpha=0.05$, but not for the test based on $\pHyun$, are in bold. \textit{(d):} Confidence intervals for the differences in means for each pair of connected components.}
\label{fig:real_data_teen_birth}
\end{figure}

Figure~\ref{fig:real_data_teen_birth}(a) displays the teenage birth rate in each of the 48 states. We solved the graph fused lasso with $K=30$ steps of the dual path algorithm, which results in five estimated connected components displayed in Figure~\ref{fig:real_data_teen_birth}(b); Appendix~\ref{appendix:real_data_different_K} contains additional  results for $K=20$. For each pair of estimated connected components, we computed the $p$-values $p_{\text{Naive}}$, $\pHyun$, and $\pB$, along with the corresponding confidence intervals for the difference in means. The results are displayed in Figures~\ref{fig:real_data_teen_birth}(c) and (d). 

As in Section~\ref{sec:drug_overdose}, at level $\alpha=0.05$, the test based on $\pB$ makes more rejections than that based on $\pHyun$. Additionally, the confidence intervals based on $\pB$ are much narrower than those based on $\pHyun$; in some cases, the former are of comparable length to those based on $p_{\text{Naive}}$.

%% file: sections-revision-v1/discussion.tex

We have proposed a new procedure for testing for the difference in the means of two connected components resulting from the graph fused lasso. Our approach conditions on less information than existing approaches, leading to substantially higher power while still controlling the selective Type I error.

Methods developed in this paper are implemented in the R package \texttt{GFLassoInference}. Instructions on how to download and use this package can be found at \url{https://yiqunchen.github.io/GFLassoInference}. Code and files to reproduce the results in the paper can be found at \url{https://github.com/yiqunchen/GFLassoInference-experiment}. 


\subsection{Incorporating the selection of the tuning parameter}

Throughout this paper, we have chosen $K$, the number of steps in the dual path algorithm for \eqref{eq:graph_fused_lasso}, without making use of the data. However, in practice, the tuning parameter $K$ is often selected based on the data. For instance, we could choose the value of $K$ that minimizes the modified Bayesian information criterion~\citep{Hyun2018-ta,Zhang2007-rr}. We can extend our idea in Section~\ref{section:method} to obtain a $p$-value similar to \eqref{eq:pB} that also conditions on the value of adaptively-chosen $K$ (or $L$, the number of connected components in $\hat{\beta}$).

\subsection{Extension to other generalized lasso problems}

Ideas in this paper apply beyond the setting of the piecewise constant model in \eqref{eq:model} and the graph fused lasso estimator in \eqref{eq:graph_fused_lasso}. For instance, we can consider extending our proposal to the trend filtering problem, which postulates that the underlying signal is ordered and piecewise polynomial~\citep{Tibshirani2014-gg,Kim2009-ty}. Because trend filtering is a special case of \eqref{eq:genlasso} and can be solved using the dual path algorithm, an extension of the approach in Section~\ref{section:method} can be applied. 

In addition, we can extend our proposal from an identity matrix in \eqref{eq:genlasso} to any design matrix $X \in \mathbb{R}^{n\times q}$ with full column rank, i.e., $\hat{\beta} = \text{argmin}_{\beta \in \mathbb{R}^n} \left\{ \frac{1}{2} \Vert y- X\beta \Vert_2^2+\lambda \Vert D \beta \Vert_1 \right\}$. \citet{Hyun2018-ta} showed that a $p$-value similar to \eqref{eq:hyun_pval} can be used in this case to test the hypothesis \eqref{eq:null}. Therefore, we can directly apply the computational insights in Section~\ref{section:method} to obtain a more powerful test.

We leave the details of outlined extensions, as well as comparisons to recent selective inference tools for trend filtering (e.g.,~\citet{Mehrizi2021-jv,Leiner2021-es}), to future work.  

\subsection{Relaxing assumptions in \eqref{eq:model}}

While the idea of conditioning on less information to improve the power of a selective inference procedure applies regardless of the distributions of the observations, the assumptions in model \eqref{eq:model} are critical to the proof of Proposition~\ref{prop:pval}, and therefore, the efficient computation of $\pB$. A line of recent work in selective inference has focused on relaxing these assumptions in high-dimensional linear modeling~\citep{Tibshirani2018-rr,Tian2018-js,Charkhi2018-lz}, and may be applicable to the generalized lasso. Alternatively, we can extend \eqref{eq:model} to other exponential family distributions by leveraging the recent developments in generalized data carving~\citep{Rasines2021-jc,Leiner2021-es,Schultheiss2021-zb}.

\section{Acknowledgments}
We thank the authors of \citet{Le_Duy2021-iy} for providing us with their software implementation. This work was partially supported by National Institutes of Health grants [R01EB026908, R01DA047869] and a Simons Investigator Award to D.W.

%% file: sections-revision-v1/appendix_a1.tex

\subsection{Dual path algorithm for \eqref{eq:genlasso} with $X=I$ \citep{Tibshirani2011-fq}}
\label{appendix:dual_path}

{\small
\begin{algorithm}[!htp]
\setstretch{0.2}
\SetKwInOut{Input}{Input}
\SetKwInOut{Output}{Output}
 \Input{Input data $y\in\mathbb{R}^n$, Penalty matrix $D\in\mathbb{R}^{m\times n}$, Number of steps $K$}
 \Output{Solution $\hat\beta$}
 Initialize $k=0$, $\lambda_0 = \infty$, Boundary set $B_0 = \emptyset$, sign vector for the boundary set $s_{B_0}= \emptyset$ \; 
 
 \begin{enumerate}
  \item
  Compute $\hat{u} = \qty(DD^\top)^\dagger y$.
  \item
  Compute  $\lambda_1 = \max_{i\in[m]} |\hat{u}_i|, i^* = \argmax_{i\in [m]} |\hat{u}_i|$. 
  \item
  Update $B_{k+1}\leftarrow B_{k}\cup i^* , s_{B_{k+1}}\leftarrow s_{B_{k}}\cup \text{sign}(\hat{u}_{i^*})$
  \item
  Record the solution $\forall \lambda \in [\lambda_{1},\infty]$:
\begin{align*}
\hat{u}(\lambda)  = \hat{u},  \quad \hat\beta(\lambda) = y-D^\top \hat{u}(\lambda).
 \end{align*} 
  \item
  $k \leftarrow k+1$
 \end{enumerate}

 \While{$\lambda_k>0$ and $k\leq K$}{
 \begin{enumerate}
     \item
  Compute
  \begin{align*} 
   &a = (D_{-B_k}D^\top_{-B_k})^\dagger D_{-B_k} y, \\
  &b = (D_{-B_k}D^\top_{-B_k})^\dagger D_{-B_k} D_{B_k}^\top s_{B_k}.
  \end{align*} 
 \item
  Compute hitting times 
 $t_i^{\text{(hit)}} = \max\qty{\frac{a_i}{b_i+ 1},\frac{a_i}{b_i- 1}}, \forall i \in [m]\setminus B_k$.
 \item
  Set next hitting time $h_{k+1} = \max_{i \notin B_k} t_i^{\text{(hit)}}$, with the hitting coordinate $i^* = \argmax_{i \notin B_k} t_i^{\text{(hit)}}$.
 \item 
  Compute  
   \begin{align*} 
  &c = \text{diag}(s_{B_k})D_{B_k}\qty(y-D^\top_{-B_k}a), \\
  &d = \text{diag}(s_{B_k})D_{B_k}\qty(D^\top_{B_k}s_{B_k}-D^\top_{-B_k}b).
  \end{align*} 

 \item 
 Compute leaving times
 $t_i^{\text{(leave)}} = \frac{c_i}{d_i}\cdot 1_{\{c_i<0\}} \cdot 1_{\{d_i<0\}}, \forall i \in B_k$.
 \item
 Set next leaving time $l_{k+1} = \max_{i\in B_k} t_i^{\text{(leave)}}$, leaving coordinate $i^{\diamond}  = \argmax_{i\in B_k} t_i^{\text{(leave)}}$.
 \item
  Set $\lambda_{k+1} = \max\{h_{k+1},l_{k+1}\}$. 
   \item
   \eIf{$h_{k+1}\geq l_{k+1}$}{ \hspace{10mm} $B_{k+1}\leftarrow B_{k}\cup i^* $ \\
    \hspace{10mm}  $s_{B_{k+1}}\leftarrow s_{B_{k}}\cup \text{sign}(t^{(\text{hit})}_{i^*})$
    }{
     \hspace{10mm}  $B_{k+1}\leftarrow B_{k}\setminus i^\diamond$ \\
    \hspace{10mm}  $s_{B_{k+1}}\leftarrow s_{B_{k}}\setminus \text{sign}(t^{(\text{leave})}_{i^\diamond})$    
   }
 \item
 Record the solution: $\forall \lambda \in [\lambda_{k+1},\lambda_k]$:
\begin{align*}
\hat{u}(\lambda)  = a - \lambda b, \quad \hat\beta(\lambda) = y-D^\top \hat{u}(\lambda).
 \end{align*} 
 \item
 $k\leftarrow k+1$
 \end{enumerate}
 }
 \caption{Dual path algorithm for the generalized lasso problem \eqref{eq:genlasso} when $X=I$.}
    \label{alg:dual_path}
\end{algorithm}
}

Algorithm~\ref{alg:dual_path} details the dual path algorithm  for computing the generalized lasso problem \eqref{eq:genlasso} with an identity design matrix~\citep{Tibshirani2011-fq}. For a matrix $A \in \mathbb{R}^{m\times m}$, we use $A^\dagger$ to denote its Moore-Penrose inverse; for a vector $a \in \mathbb{R}^n$, we use $\text{diag}(a)$ to denote the $n\times n$ diagonal matrix with elements of $a$ on the diagonal. 

\newpage
\subsection{Proof of Proposition~\ref{prop:piecewise_beta}}
\label{sec:appendix:proof_of_piecewise_lemma}

The proof of Proposition~\ref{prop:piecewise_beta} is similar to the arguments in Section 6.2 of \citet{Tibshirani2011-fq}. 

We first prove the ``if'' direction:
\begin{align}
\label{eq:appendix:if_prop_beta}
j,j' \in C_l \text{ for some } l \in [L] \implies \hat{\beta}_j = \hat{\beta}_{j'}. 
\end{align}
First recall the result in \eqref{eq:primaldual2}, which states that $\hat{\beta}  = P_{\text{Null}(D_{-B_k})} \qty(y-\lambda D^{\top}_{B_k}s_{B_k})$. It follows that $\hat\beta_j =  \qty(\qty[P_{\text{Null}(D_{-B_k})}]_j)^\top \qty(y-\lambda D^{\top}_{B_k}s_{B_k})$ and $\hat\beta_{j'} =  \qty(\qty[P_{\text{Null}(D_{-B_k})}]_{j'})^\top \qty(y-\lambda D^{\top}_{B_k}s_{B_k})$, where $\qty[P_{\text{Null}(D_{-B_k})}]_{j}$ denotes the $j$th row of the matrix $P_{\text{Null}(D_{-B_k})}$, written as a column vector. Therefore, to prove \eqref{eq:appendix:if_prop_beta}, it suffices to prove that 
\begin{align}
\label{eq:appendix:if_prop_beta_null}
j,j' \in C_l \text{ for some } l \in [L] \implies \qty[P_{\text{Null}(D_{-B_k})}]_{j} = \qty[P_{\text{Null}(D_{-B_k})}]_{j'}.
\end{align}
To prove \eqref{eq:appendix:if_prop_beta_null}, we first compute $P_{\text{Null}(D_{-B_k})}$. The null space $\text{Null}(D_{-B_k})$ has dimension $L$ and admits the basis $\{\textbf{1}_{C_1},\ldots, \textbf{1}_{C_L}\}$~\citep{Tibshirani2011-fq}, where the $j$th element of $\textbf{1}_{C_l}$ equals $1_{(\text{Node } j \in C_l)}, \quad j = 1,\ldots,n.$ Therefore, denoting $|C_{l}|$ the cardinality of the set $C_{l}$, we have 
$$P_{\text{Null}(D_{-B_k})} = \sum_{l=1}^L \frac{\textbf{1}_{C_{l}}\textbf{1}_{C_{l}}^\top}{|C_{l}|}.$$ It follows from algebra that
\begin{align}
\label{eq:appendix:pnull_hat_expression}
\qty[P_{\text{Null}(D_{-B_k})}]_{j} = \sum_{l=1}^L  \frac{[\textbf{1}_{C_l}]_{j}}{|C_l|}\textbf{1}_{C_l}, \quad \qty[P_{\text{Null}(D_{-B_k})}]_{j'} = \sum_{l=1}^L  \frac{[\textbf{1}_{C_l}]_{j'}}{|C_l|}\textbf{1}_{C_l}.
\end{align} 
Now, according to the definition of $\textbf{1}_{C_l}$ and the assumption that $j$ and $j'$ are in the same connected component, we have $[\textbf{1}_{C_l}]_{j} = [\textbf{1}_{C_l}]_{j'},\forall l \in [L],$ which implies that $\qty[P_{\text{Null}(D_{-B_k})}]_{j} = \qty[P_{\text{Null}(D_{-B_k})}]_{j'}$ and therefore completes the proof for \eqref{eq:appendix:if_prop_beta_null}.

Next, we will prove the ``only if'' direction: that with probability one, 
\begin{align}
\label{eq:appendix:only_if_prop_beta}
\hat{\beta}_j = \hat{\beta}_{j'} \implies j,j' \in C_l \text{ for some }l \in [L].
\end{align} 
 Combining the results in \eqref{eq:primaldual2} and \eqref{eq:appendix:pnull_hat_expression}, we have that
\begin{align}
\label{eq:appendix:beta_hat_expression}
\hat\beta_j = \sum_{l=1}^L  \frac{[\textbf{1}_{C_l}]_{j}}{|C_l|}\qty(\textbf{1}_{C_l})^\top \qty(y-\lambda D^{\top}_{B_k}s_{B_k}), \quad \hat\beta_{j'} = \sum_{l=1}^L  \frac{[\textbf{1}_{C_l}]_{j'}}{|C_l|}(\textbf{1}_{C_l})^\top \qty(y-\lambda D^{\top}_{B_k}s_{B_k}).
\end{align} 
We proceed to prove \eqref{eq:appendix:only_if_prop_beta} by contradiction. Suppose without loss of generality that $j\in C_1,j'\in C_2$ and $C_1\cap C_2 = \emptyset,$ we will show that, with probability one, $\hat\beta_j \neq \hat\beta_{j'}$. 
Combining our assumption and \eqref{eq:appendix:beta_hat_expression}, we have that
\begin{align}
\label{eq:appendix_contradiction}
\hat\beta_j-\hat\beta_{j'} &= \frac{1}{|C_1|}\qty(\textbf{1}_{C_1})^\top \qty(y-\lambda D^{\top}_{B_k}s_{B_k})-\frac{1}{|C_2|}\qty(\textbf{1}_{C_2})^\top \qty(y-\lambda D^{\top}_{B_k}s_{B_k}) \nonumber \\
&= \qty(\frac{\textbf{1}_{C_1} }{|C_1|}-\frac{\textbf{1}_{C_2} }{|C_2|})^\top \qty(y-\lambda D^{\top}_{B_k}s_{B_k}). 
\end{align}
By our assumption, $\frac{\textbf{1}_{C_1} }{|C_1|}-\frac{\textbf{1}_{C_2}}{|C_2|}$ is a non-zero vector. In addition, it follows from algebra that entries of the vector $\lambda D_{B_k}^\top s_{B_k}$ can only take values that are integer multiples of $\lambda$. Under model \eqref{eq:model}, $y\sim \mathcal{N}(0,\sigma^2I_n)$, which implies that the inner product in \eqref{eq:appendix_contradiction} will be non-zero with probability one. Therefore, we have proven that with probability one, \eqref{eq:appendix:only_if_prop_beta} holds.

%% file: sections-revision-v1/appendix_a1_b.tex

\subsection{Algorithm for computing $\SB$ in \eqref{eq:SB}}
\label{appendix:S_B_algo}

We now present an algorithm to compute $\SB$ \eqref{eq:SB}. 
\begin{algorithm}[!htpb]
\setstretch{0.1}
\SetKwInOut{Input}{Input}
    \SetKwInOut{Output}{Output}
 \Input{Data $y$, $\hat{C}_1$, $\hat{C}_2$, number of steps $K$ for Algorithm~\ref{alg:dual_path}, $\eta=10^{-4}$}
 
 \Output{$\SB$ in \eqref{eq:SB}}
  \begin{enumerate}
    \item
     $\mathcal{J} \leftarrow \{0\}$, $i\leftarrow 1$, $j\leftarrow 0$. 
 \item 
   $[a_0,a_1]=\left\{ \phi\in\mathbb{R}:  \bigcap_{k=1}^K \left\{ M_k(y'(\phi)) = M_k(y)\right\} \right\}$, where $M_k$ was defined in \eqref{eq:M}. 
  \item
   \While{$a_{i}<\infty$}{
     \begin{enumerate}
     \item
   Compute   $[\tilde{a}_i,a_{i+1}] = \left\{ \phi\in\mathbb{R}: \bigcap_{k=1}^K \left\{ M_k(y'(\phi)) = M_k(y'(a_i + \eta))\right\} \right\}$. 
     \item
     \While{$\tilde{a}_i\neq a_{i}$}{
     \begin{enumerate}
      \item
     $\eta \leftarrow \frac{1}{2}\cdot \eta ,$
     \item
      $[\tilde{a}_i,a_{i+1}]= \left\{ \phi\in\mathbb{R}:  \bigcap_{k=1}^K \left\{ M_k(y'(\phi)) = M_k(y'(a_i + \eta))\right\} \right\}$.
    \end{enumerate}
     }
    \item 
    \If{$\hat{C}_1,\hat{C}_2 \in \mathcal{CC}_{K}(y'(a_i+\eta))$}{\hspace{20mm} 
      $\mathcal{J}\leftarrow \mathcal{J}\cup \{i\}$.
    }
    \item
    $i \leftarrow i + 1 $.
     \end{enumerate}
     }
  
    \item 
   \While{$a_{j}>-\infty$}{
     \begin{enumerate}
     \item
      $[a_{j-1},\tilde{a}_{j}] = \left\{ \phi\in\mathbb{R}:  \bigcap_{k=1}^K \left\{ M_k(y'(\phi)) = M_k(y'(a_j-\eta))\right\} \right\}$. \\
     \item
     \While{$\tilde{a}_j\neq a_{j}$}{
     \begin{enumerate}
      \item
     $\eta \leftarrow \frac{1}{2}\cdot \eta ,$\\
     \item
     $[a_{j-1},\tilde{a}_{j}]= \left\{ \phi\in\mathbb{R}:  \bigcap_{k=1}^K \left\{ M_k(y'(\phi)) = M_k(y'(a_j-\eta))\right\} \right\}$.
   \end{enumerate}
     }
    \item 
    \If{$\hat{C}_1, \hat{C}_2 \in \mathcal{CC}_{K}(y'(a_j-\eta))$}{
\hspace{20mm} 
$\mathcal{J}\leftarrow \mathcal{J}\cup \{j-1\}$.
    }
    \item
    $j \leftarrow j - 1 .$ \\
     \end{enumerate}
     }
 \end{enumerate}
 $\SB \leftarrow \bigcup_{i \in \mathcal{J}} \qty[a_i, a_{i+1}]$. 
 \caption{Characterizing the set $\SB$ in \eqref{eq:SB}}
 \label{alg:SB} 
\end{algorithm}

In Algorithm~\ref{alg:SB}, we initialize with $\eta = 10^{-4}$, and apply Corollary~\ref{cor:hyun_single_union} to obtain the set {\footnotesize$[\tilde{a}_1,a_2] = \left\{ \phi\in\mathbb{R}: \bigcap_{k=1}^K \left\{ M_k\qty(y'\qty(\phi)) = M_k\qty(y'\qty(a_1 + \eta))\right\} \right\}$} (see step 4(a) in Algorithm~\ref{alg:SB} with $i=1$). If the left endpoint $\tilde{a}_1>a_1+\epsilon$ (where $\epsilon$ is a small constant set to $1.5\times 10^{-8}$ by default), we repeat with a smaller value of $\eta$ (see step 4(b)i of Algorithm~\ref{alg:SB}).

\begin{figure}[!ht]
\centering
\begin{subfigure}{0.45\textwidth}
  \caption{}
  \includegraphics[width=\linewidth]{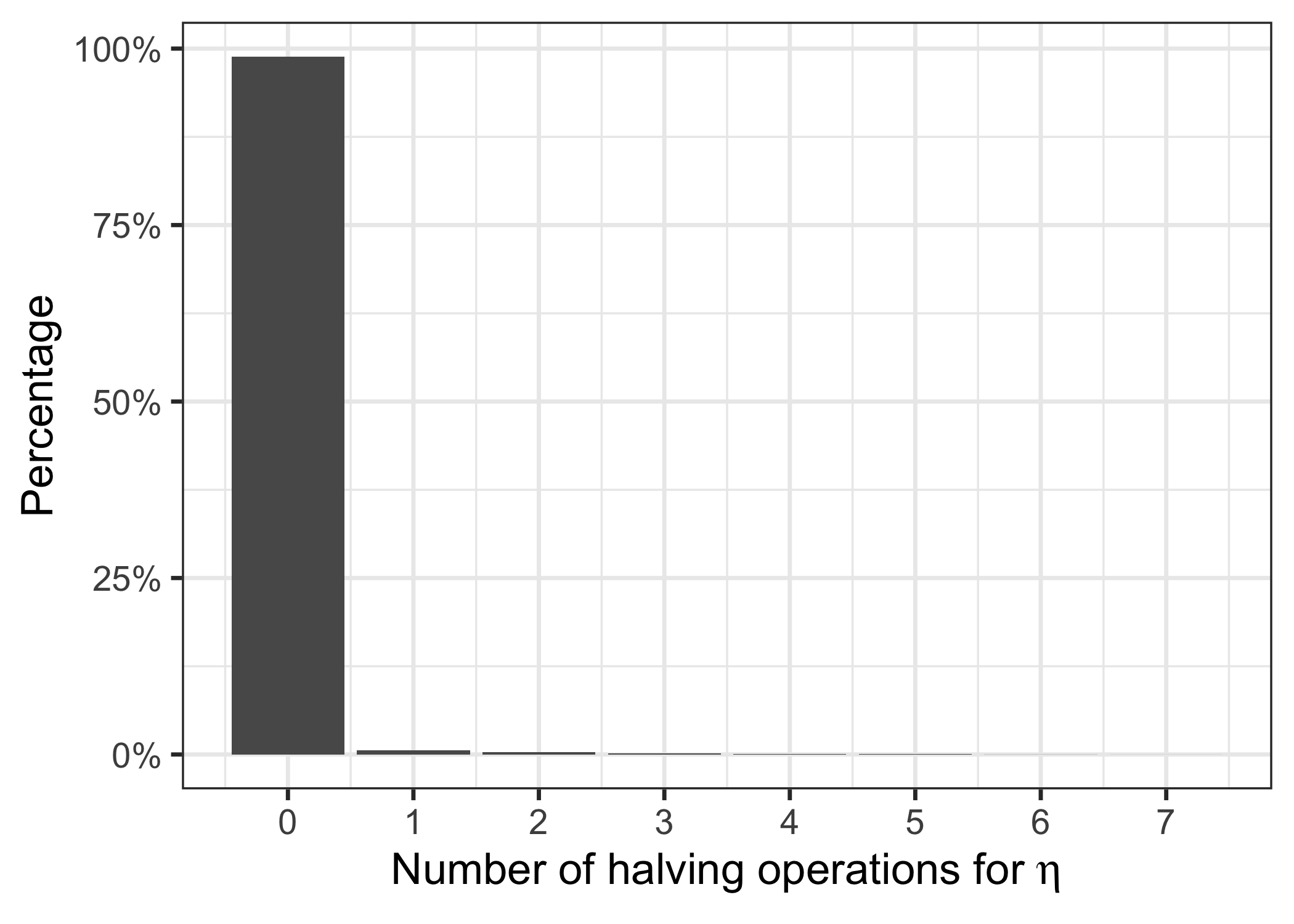}
\end{subfigure}
\begin{subfigure}{0.45\textwidth}
  \caption{}
  \includegraphics[width=\linewidth]{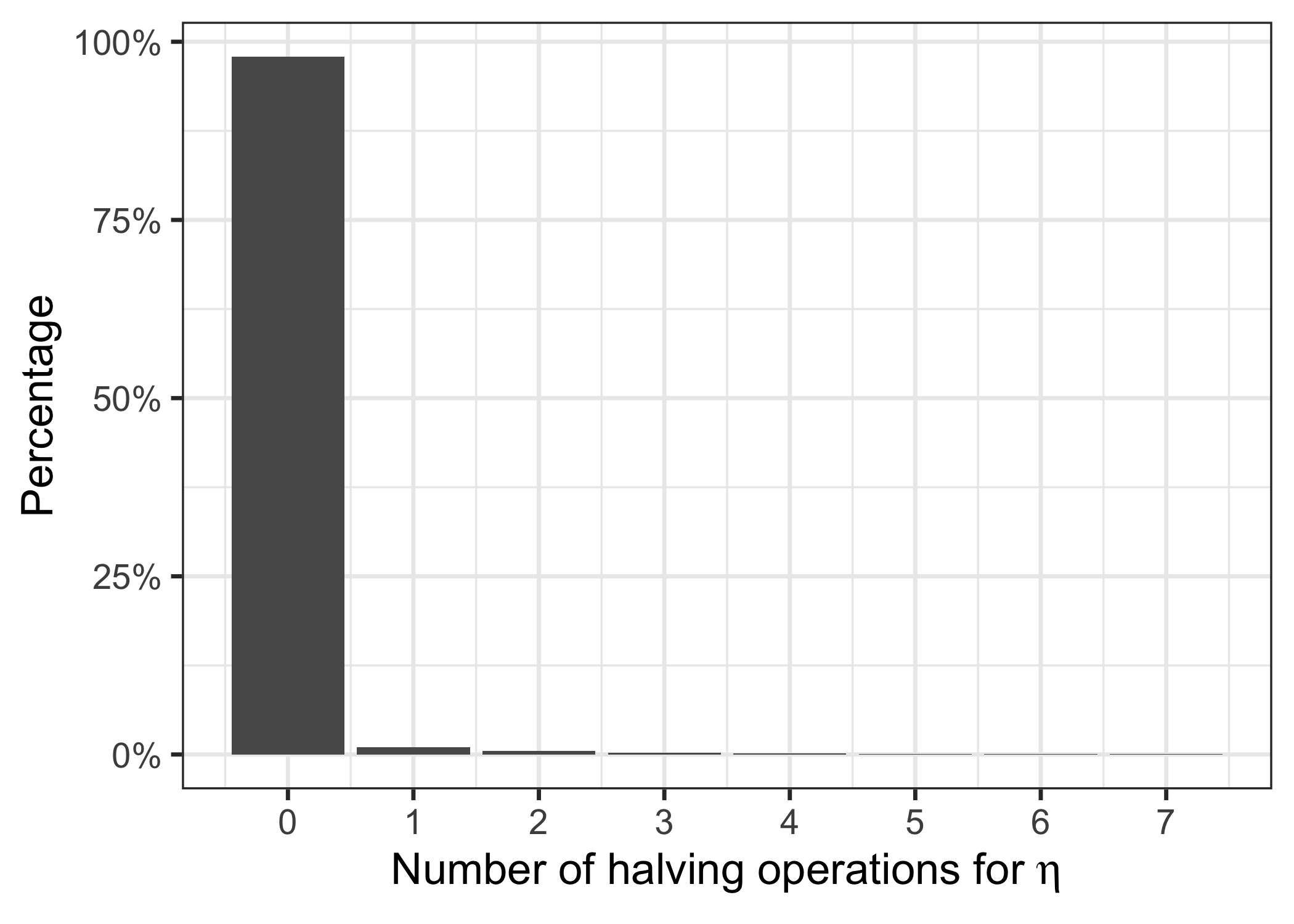}
\end{subfigure}

\caption{ \textit{(a): } Number of halving operations for $\eta$ (defined in Section~\ref{section:method_pb_algo} and Algorithm~\ref{alg:SB} of Appendix~\ref{appendix:S_B_algo}) in the one-dimensional fused lasso simulations described in Section~\ref{section:sim_one_d_type_1} with $\sigma=1$. \textit{(b): } Same as (a), but for the two-dimensional fused lasso simulations described in Section~\ref{section:sim_two_d_type_1} with $\sigma=1$.}
\label{fig:half_eta}
\end{figure}

In a simulation study, we investigate how often, using the initialization $\eta=10^{-4}$, we need to perform the halving operation in step 4(b)i of Algorithm~\ref{alg:SB}. Results are aggregated in Figure~\ref{fig:half_eta}. We almost never have to halve the initial value $10^{-4}$, and the number of halving operations never exceeds seven.

%% file: sections-revision-v1/appendix_a2.tex

\subsection{Proof of Proposition~\ref{prop:pval}}
\label{appendix:prop_pval}


We first prove the statement \eqref{eq:single_param_null}. The following equalities hold for an arbitrary vector $\nu \in \mathbb{R}^n$:
\begin{equation}
\begin{aligned}
&\mathbb{P}\left(|\nu^{\top} Y| \geq |\nu^{\top} y| \;\middle\vert\; \hat{C}_1(y), \hat{C}_2(y) \in \mathcal{CC}_K(Y), \Pi_{\nu}^{\perp} Y= \Pi_{\nu}^{\perp} y \right) \\
 &\overset{a.}{=} \mathbb{P}\left(|\nu^{\top} Y| \geq |\nu^{\top} y| \;\middle\vert\; \hat{C}_1(y), \hat{C}_2(y) \in \mathcal{CC}_K(\Pi_{\nu}^{\perp}y +\Pi_{\nu} Y), \Pi_{\nu}^{\perp} Y= \Pi_{\nu}^{\perp} y  \right) \\
 &\overset{b.}{=}  \mathbb{P} \left( |\nu^\top Y| \geq  |\nu^{\top}y|  \;\middle\vert\; \hat{C}_1(y), \hat{C}_2(y)  \in \mathcal{CC}_K\left(y'\qty(\nu^\top Y)\right) ,\Pi_{\nu}^{\perp} Y= \Pi_{\nu}^{\perp} y  \right) \\
 &\overset{c.}{=}  \mathbb{P} \left( |\nu^\top Y| \geq  |\nu^{\top}y|  \;\middle\vert\; \hat{C}_1(y), \hat{C}_2(y)  \in \mathcal{CC}_K\left(y'\qty(\nu^\top Y)\right) \right).
\end{aligned}
\label{eq:appendix_block_phi}
\end{equation}
Here, $a.$ follows from the fact that $Y = \Pi_{\nu}^{\perp}Y+\Pi_{\nu}Y$, and the fact that we have conditioned on the event $\Pi_{\nu}^{\perp}Y = \Pi_{\nu}^{\perp}y$. To prove $b.$, we first note that $I = \Pi_{\nu} + \Pi_{\nu}^{\perp}$ and $\Pi_{\nu} Y=\frac{\nu\nu^{\top}}{||\nu||_2^2}Y$, which implies $$\Pi_{\nu}^{\perp}y +\Pi_{\nu} Y = y- \Pi_{\nu}y+\Pi_{\nu} Y = y-\frac{\nu^{\top}y}{||\nu||_2^2}\nu+\frac{\nu^{\top}Y}{||\nu||_2^2}\nu  = y'(\nu^\top Y),$$ where the notation $y'(\cdot)$ is defined in \eqref{eq:phi}. Finally, $c.$ follows from the fact that $Y \sim \mathcal{N}(\beta, \sigma^2 I)$ implies independence of $\nu^{\top}Y$ and $\Pi_\nu^\perp Y$.

 Now under $H_0$ in \eqref{eq:null},  we have that
 \begin{align*}
\pB &\overset{a.}{=} \mathbb{P}_{H_0}\left(|\nu^{\top} Y| \geq |\nu^{\top} y| \;\middle\vert\; \hat{C}_1(y), \hat{C}_2(y) \in \mathcal{CC}_K(Y), \Pi_{\nu}^{\perp} Y= \Pi_{\nu}^{\perp} y \right) \\
&\overset{b.}{=}   \mathbb{P}_{H_0} \left( |\nu^\top Y| \geq  |\nu^{\top}y|  \;\middle\vert\; \hat{C}_1(y), \hat{C}_2(y)  \in \mathcal{CC}_K\left(y'\qty(\nu^\top Y)\right) \right) \\
&\overset{c.}{=} \mathbb{P} \left( |\phi| \geq  |\nu^{\top}y|  \;\middle\vert\; \hat{C}_1(y), \hat{C}_2(y)  \in \mathcal{CC}_K\left(y'\qty(\phi)\right) \right). 
\end{align*}
Here, step $a.$ is the definition of $\pB$ in \eqref{eq:pB}, and step $b.$ follows from applying the result in \eqref{eq:appendix_block_phi}. In $c.$, we used the fact that under $H_0$, $\nu^\top Y \sim \mathcal{N}\qty(0, \sigma^2 ||\nu||_2^2) \overset{d}{=} \phi$, which completes the proof.

Next, we will prove that the test that rejects $H_0:\nu^\top \beta = 0$ when $\pB\leq \alpha$ controls the selective Type I error as in \eqref{eq:selective_type_1}. First of all, a test for the null hypothesis in \eqref{eq:null} controls the selective Type I error in \eqref{eq:selective_type_1} if
\begin{align}
\label{eq:appendix:selective_type_1}
\mathbb{P}_{H_0}\qty(\text{reject $H_0$ at level $\alpha$} \;\middle\vert\; \hat{C}_1(y), \hat{C}_2(y)  \in \mathcal{CC}_K\left(Y\right) )\leq \alpha,\; \forall \alpha \in (0,1).
\end{align}
In what follows, we will write $\pB$ as $\pB\qty(\nu^\top y)$ to emphasize that $\pB$ is a function of the observed difference in means $\nu^\top y$. Using the result in \eqref{eq:single_param_null}, we have that
\begin{align}
\label{eq:appendix:pB_CDF}
\pB\qty(\nu^\top y) = 1-\tilde{F}_{0,\sigma^2||\nu||_2^2}^{\SB}\qty(|\nu^\top y|),
\end{align}
where $\tilde{F}_{0,\sigma^2||\nu||_2^2}^{\SB}(\cdot)$ is the cumulative distribution function of \emph{the magnitude} of a $\mathcal{N}(0,\sigma^2||\nu||_2^2)$ random variable, truncated to the set $\SB$. 

Therefore, we have that, $\forall \alpha \in (0,1)$,
\begin{equation}
\label{eq:appendix:pB_property}
\begin{aligned}
\mathbb{P}_{H_0}&\qty(\pB(\nu^\top Y)\leq \alpha \;\middle\vert\;  \hat{C}_1(y),  \hat{C}_2(y) \in \CC_K(Y), \Pi_{\nu}^{\perp} Y= \Pi_{\nu}^{\perp} y ) \\
&\overset{a.}{=} \mathbb{P}\qty(1-\tilde{F}_{0,\sigma^2||\nu||_2^2}^{\SB}\qty(|\nu^\top Y|)\leq \alpha \;\middle\vert\;  \hat{C}_1(y),  \hat{C}_2(y) \in \CC_K(Y), \Pi_{\nu}^{\perp} Y= \Pi_{\nu}^{\perp} y ) \\
&\overset{b.}{=} \mathbb{P}\qty(\tilde{F}_{0,\sigma^2||\nu||_2^2}^{\SB}\qty(Z)\geq 1-\alpha) \\
&\overset{c.}{\leq}\alpha.
\end{aligned}
\end{equation}
Step $a.$ follows from \eqref{eq:appendix:pB_CDF} and step $b.$ follows from \eqref{eq:single_param_null} and letting $Z$ denote the magnitude of a $\mathcal{N}(0,\sigma^2||\nu||_2^2)$ random variable, truncated to the set $\SB$. Finally, in $c.$, we use the probability integral transform, which states
that for a continuous random variable $X$, $F_X(X)$ is distributed as a Uniform(0,1) distribution.

Now for the test that rejects $H_0$ if $\pB\leq \alpha$, we have that 
\begin{equation}
\label{eq:appendix:pb_defn}
\begin{aligned}
\mathbb{P}_{H_0}&\qty(\pB \leq \alpha \;\middle\vert\;  \hat{C}_1(y),  \hat{C}_2(y) \in \CC_K(Y))  \\
&=\mathbb{E}_{H_0}\qty[1\{\pB \leq \alpha\} \;\middle\vert\;  \hat{C}_1(y),  \hat{C}_2(y) \in \CC_K(Y)]\\
&\overset{a.}{=}\mathbb{E}_{H_0}\qty[\mathbb{E}_{H_0}\qty[1\{\pB \leq \alpha\} \;\middle\vert\;  \hat{C}_1(y),  \hat{C}_2(y) \in \CC_K(Y), \Pi_{\nu}^{\perp} Y= \Pi_{\nu}^{\perp} y] \;\middle\vert\;\hat{C}_1(y),  \hat{C}_2(y) \in \CC_K(Y)] \\
&\overset{b.}{\leq} \mathbb{E}_{H_0}\qty[ \alpha \;\middle\vert\;\hat{C}_1(y),  \hat{C}_2(y) \in \CC_K(Y)] \\
&= \alpha.
\end{aligned}
\end{equation}

In the proof above, step $a.$ follows from the tower property of conditional expectation. To prove $b.$, we apply the results from \eqref{eq:appendix:pB_property}.

By inspection of \eqref{eq:appendix:selective_type_1} and \eqref{eq:appendix:pb_defn}, we conclude that the test based on $\pB$ controls the selective Type I error, which completes the proof.

\newpage
\subsection{Proof of Corollary~\ref{cor:hyun_single_union} and Proposition~\ref{prop:finite_union}}
\label{appendix:prop_finite_union}

We first prove Corollary~\ref{cor:hyun_single_union}: that is,  $\left\{ \phi\in\mathbb{R}:  \bigcap_{k=1}^K \left\{ M_k(y'(\phi)) = M_k(y)\right\} \right\}$ is an interval, where $M_k$ is defined in \eqref{eq:M}.

\begin{proof}
{\small
\begin{align*}
 \qty{\phi \in \mathbb{R}:  \bigcap_{k=1}^K \qty{M_{k}(y'(\phi)) = M_{k}(y)} }  &\overset{a.}{=} \qty{\phi \in \mathbb{R}: Ay'(\phi) \leq 0 } \\  
  &\overset{b.}{=} \qty{\phi \in \mathbb{R}: A\qty(y-\frac{\nu^\top y}{||\nu||_2^2}\nu+\frac{\phi}{||\nu||_2^2}\nu) \leq 0 } \\
  &\overset{c.}{=}   \qty{\phi \in \mathbb{R}: \phi \cdot \qty(A\nu) \leq (\nu^\top y) A\nu - ||\nu||^2_2 Ay  } \\
  &\overset{d.}{=}
   \qty[\max_{i: (A\nu)_i <0 } \frac{ \qty(\nu^\top y) (A\nu)_i - ||\nu||^2_2 (Ay)_i }{(A\nu)_i} ,
 \min_{i: (A\nu)_i >0  } \frac{ \qty(\nu^\top y) (A\nu)_i - ||\nu||^2_2 (Ay)_i }{(A\nu)_i} ].
  \end{align*}
  }
 Here, $a.$ follows from Proposition~\ref{prop:generalized_hyun}, which states that the set $\qty{Y \in \mathbb{R}^n: \bigcap_{k=1}^K M_{k}(Y) = M_{k}(y)} = \qty{Y \in \mathbb{R}^n :AY\leq 0 }$ for some matrix $A$, where $\leq$ is interpreted as the component-wise inequality. Next, $b.$ follows from the definition of $y'(\phi)$ in \eqref{eq:phi}. Finally, $c.$ and $d.$ follow from solving the linear inequality in $\phi$. Note that in $d.$, we implicitly assumed that $\exists i$ (or $i'$) such that $(A\nu)_i < 0$ (or $(A\nu)_{i'} > 0$); if that doesn't hold, the corresponding expression in $d.$ is replaced by $-\infty$ (or $+\infty$).

\end{proof}

We proceed to prove Proposition~\ref{prop:finite_union}. 
\begin{align*}
 \SB &\overset{a.}{=} \qty{\phi\in\mathbb{R}:  \hat{C}_{1},\hat{C}_{2}\in \CC_K\qty(y'\qty(\phi))} \\
 &\overset{b.}{=} \bigcup_{(m_1,\ldots m_K)\in \mathcal{M}_K} \qty{\phi\in\mathbb{R}:  \hat{C}_{1},\hat{C}_{2}\in \CC_K\qty(y'\qty(\phi)), \bigcap_{k=1}^K \qty{{M}_k\qty(y'\qty(\phi)) = m_k } } \\
 &\overset{c.}{=}  \bigcup_{(\tilde{m}_1,\ldots \tilde{m}_K)\in \mathcal{I}} \qty{\phi\in\mathbb{R}:  \hat{C}_{1},\hat{C}_{2}\in \CC_K\qty(y'\qty(\phi)),  \bigcap_{k=1}^K \qty{{M}_k\qty(y'\qty(\phi)) = \tilde{m}_k} } \\
  &\overset{d.}{=} \bigcup_{(\tilde{m}_1,\ldots \tilde{m}_K)\in \mathcal{I}} \qty{\phi\in\mathbb{R}: \bigcap_{k=1}^K \qty{{M}_k\qty(y'\qty(\phi)) = \tilde{m}_k} } \\
  &\overset{e.}{=} \bigcup_{i\in \mathcal{J}} [a_i,a_{i+1}],
\end{align*}
where in $b.$, $\mathcal{M}_K$ is the set of all possible outputs of the first $K$ steps of the dual path algorithm. In the proof above, $a.$ is the definition of $\SB$, and $b.$ follows from the definition of $\mathcal{M}_K$. Furthermore, steps $c.$ and $d.$ follows from the definition of the index set $\mathcal{I}$ (see \eqref{eq:I_s_b_set}). Finally, to prove $e.$, we apply Corollary~\ref{cor:hyun_single_union}, which implies that for each $\tilde{m}_k$, $\qty{\phi\in\mathbb{R}: \bigcap_{k=1}^K \qty{{M}_k\qty(y'\qty(\phi)) = \tilde{m}_k} }$ is an interval. This concludes the proof of Proposition~\ref{prop:finite_union}.

\subsection{Proof and an empirical analysis of Proposition~\ref{prop:conservative_p}}
\label{appendix:prop_conservative_p}

The proof of Proposition~\ref{prop:conservative_p} is similar to the proof of Proposition 4 in \citetappendix{jewell2019testing}. First, note that by the definition of $\SBconserv$ in Proposition~\ref{prop:conservative_p}, $\mathbb{P}\qty(\phi \in \SB)>0\implies \mathbb{P}\qty(\phi \in \SBconserv)>0$. Next, we have that
\begin{align*}
\mathbb{P}\left( |\phi|\geq |\nu^\top y| \;\middle\vert\; \phi \in \SBconserv \right) &\overset{a.}{=} \frac{\mathbb{P}\left( |\phi|\geq |\nu^\top y| , \phi \in \SBconserv \right)}{\mathbb{P}\qty(\phi \in \SBconserv) } \\
&\overset{b.}{=}  \frac{\mathbb{P}\left( |\phi|\geq |\nu^\top y| , \phi \in \SB \right)+\mathbb{P}\left( |\phi|\geq |\nu^\top y| , \phi \in \SBconserv\setminus\SB \right)}{\mathbb{P}\qty(\phi \in \SB)+\mathbb{P}\qty(\phi \in \SBconserv\setminus\SB)} \\
&\overset{c.}{=}  \frac{\mathbb{P}\left( |\phi|\geq |\nu^\top y| , \phi \in \SB \right)+\mathbb{P}\left( \phi \in \SBconserv\setminus\SB \right)}{\mathbb{P}\qty(\phi \in \SB)+\mathbb{P}\qty(\phi \in \SBconserv\setminus\SB)} \\
&\overset{}{\geq}  \frac{\mathbb{P}\left( |\phi|\geq |\nu^\top y| , \phi \in \SB \right)}{\mathbb{P}\qty(\phi \in \SB)} \\
&\overset{}{=}  \mathbb{P}\left( |\phi|\geq |\nu^\top y| \;\middle\vert\; \phi \in \SB \right).
\end{align*}
Here, $a.$ follows from Bayes' rule and the fact that $\mathbb{P}\qty(\phi \in \SBconserv)>0$, and $b.$ is a direct consequence of the law of total probability. Step $c.$ follows from the definition of $\SBconserv$, which implies that $\qty{\phi:|\phi|\geq |\nu^\top y|}\cap \qty{\phi:\phi \in \SBconserv\setminus\SB} = \qty{\phi:\phi \in \SBconserv\setminus\SB}$. 

\begin{figure}[!htbp]
\centering
\begin{subfigure}{0.4\textwidth}
  \caption{}
  \includegraphics[width=\linewidth]{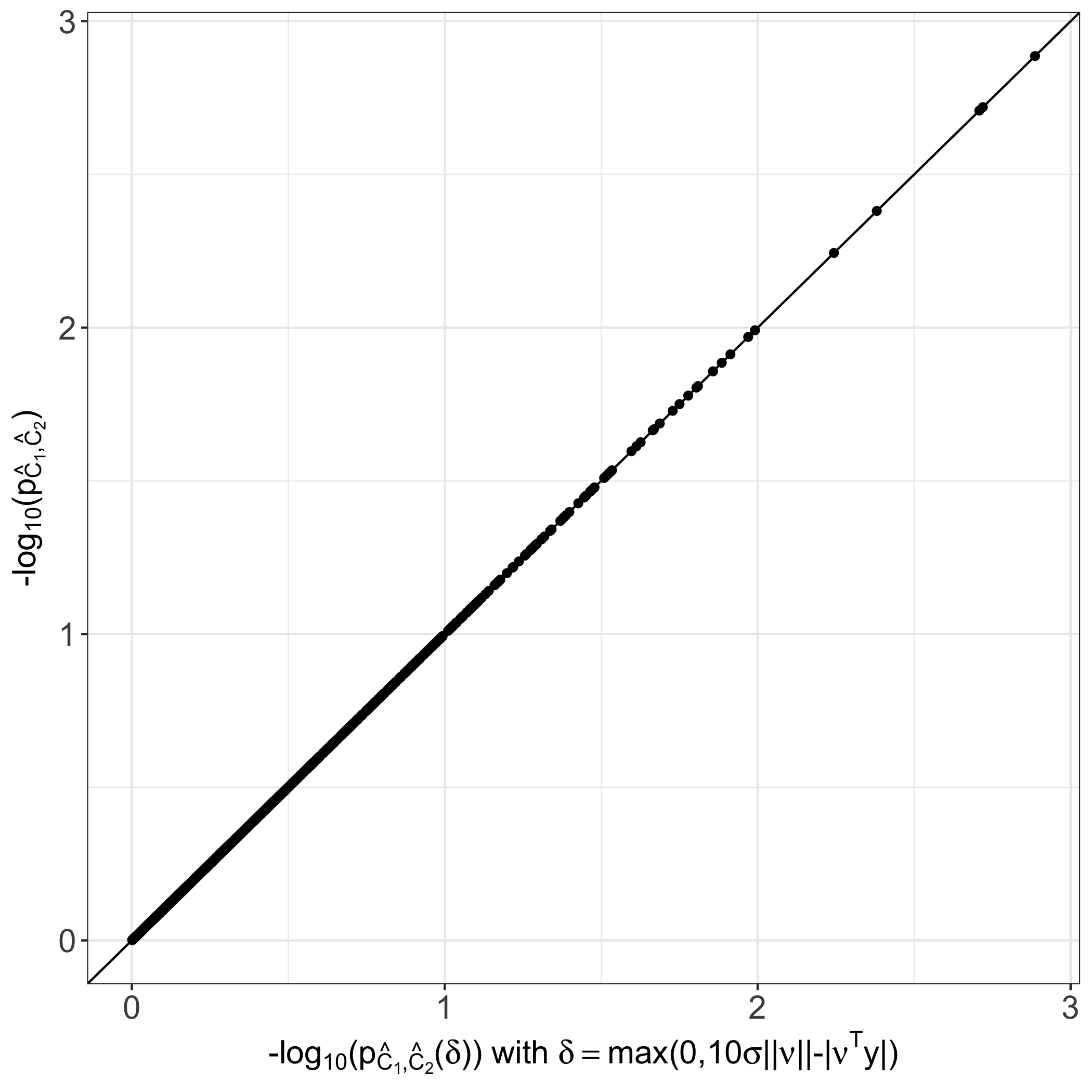}
\end{subfigure}
\begin{subfigure}{0.4\textwidth}
  \caption{}
  \includegraphics[width=\linewidth]{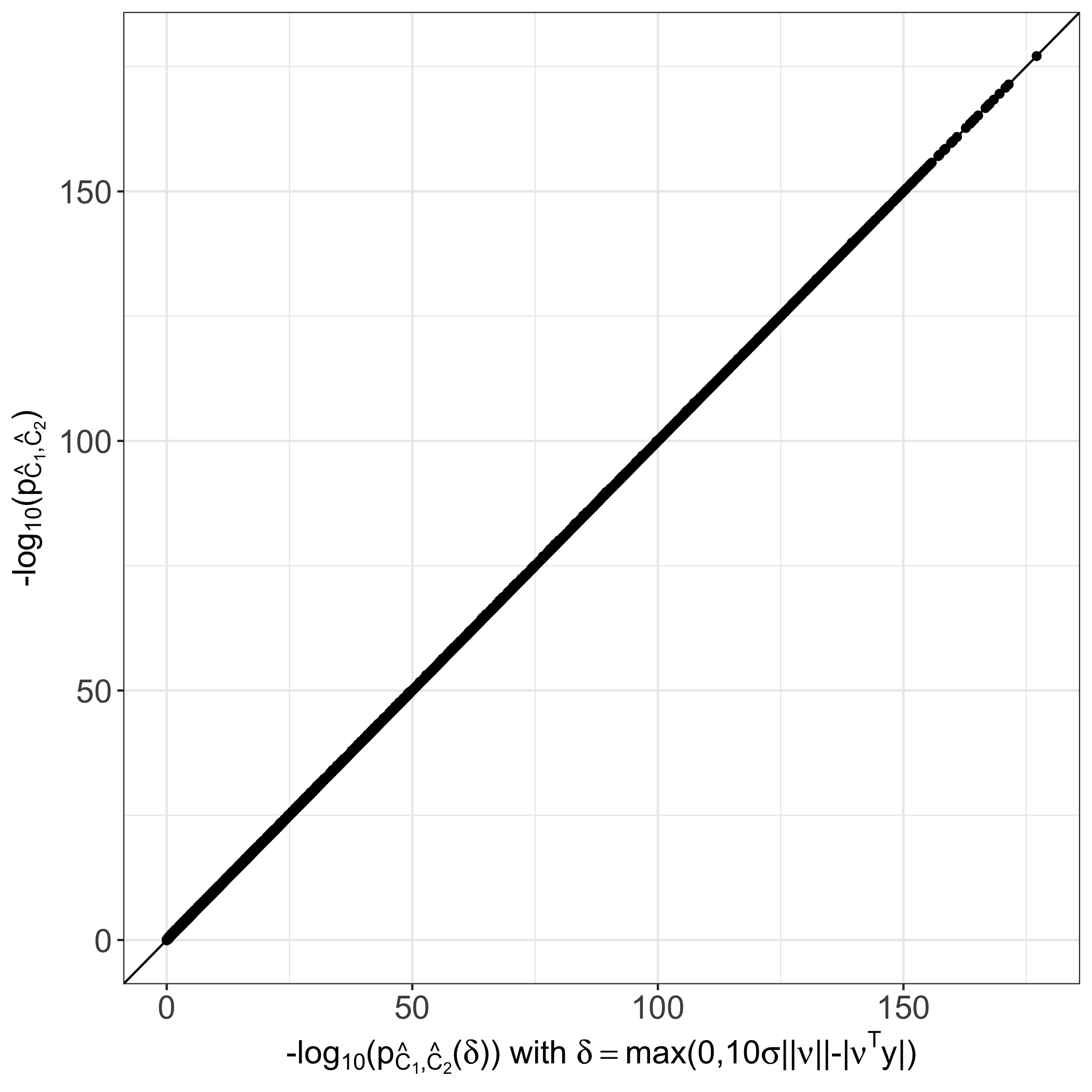}
\end{subfigure}
\caption{ \textit{(a): } For the one-dimensional fused lasso simulations described in Section~\ref{section:sim_one_d_type_1}, the $p$-values $\pB$ and $\pB(\delta)$  with $\delta = \max\{0,10\sigma\Vert \nu \Vert_2 - |\nu^\top y| \}$ on the $-\log_{10}$ scale. \textit{(b): } Same as (a), but for the simulations described in Section~\ref{section:sim_one_d_power}.}
\label{fig:early_stop_pval}
\end{figure}

Next, we investigate the approximation error of using $\SBconserv$ to compute the $p$-value instead of $\SB$. In what follows, we denote $\mathbb{P}\left( |\phi|\geq |\nu^\top y| \;\middle\vert\; \phi \in \SBconserv \right)$ as $\pB(\delta)$ for brevity. In prior work, several choices of $\delta$ has been proposed (e.g., $\max\qty{0, 10\sigma\Vert \nu \Vert_2 -|\nu^\top y|}$~\citep{jewell2019testing} and $\max\qty{0, 20 \sigma -|\nu^\top y|}$~\citep{Liu2018-zx}). In this section, we computed $\pB(\delta)$ with $\delta = \max\{0,10\sigma\Vert \nu \vert_2 - |\nu^\top y| \}$ for the experiments in the one-dimensional fused lasso case (Section~\ref{section:sim_one_d} of the main text). Results are aggregated and displayed in Figure~\ref{fig:early_stop_pval}. Panel (a) displays the $p$-values $\pB$ versus $\pB(\delta)$  with $\delta = \max\{0,10\sigma\Vert \nu \Vert_2 - |\nu^\top y| \}$ on the $-\log_{10}$ scale under the global null. We see that the two set of $p$-values are nearly identical; the same holds for the datasets simulated with true changepoints (see Figure~\ref{fig:early_stop_pval}(b)). Regarding computational efficiency, this choice of $\delta$ reduces the overall running time of Algorithm 2 by 15--20\%, depending on the specific simulation parameters.
In conclusion, we suggest using $\delta = \max\{0,10\sigma\Vert \nu \Vert_2 - |\nu^\top y| \}$ to balance the  inferential accuracy and computational efficiency. 

\newpage
\subsection{Proof of Proposition~\ref{prop:ci}}
\label{appendix:prop_ci}

The proof of Proposition~\ref{prop:ci} is similar to the proof of Theorem 6.1 of \citetappendix{Lee2016-te} and the proof of Corollary 3.1 of \citetappendix{Chen2020-rh}.

Recalling the definition that $F_{\mu,\sigma^2||\nu||_2^2}^{\SB}(t)$ is the cumulative distribution function of a $\mathcal{N}(0,\sigma^2||\nu||_2^2)$ random variable, truncated to the set $\SB$ defined in \eqref{eq:SB}, we have that $F^{\SB}_{\mu, \sigma^2||\nu||_2^2}(t)$ is a monotonically decreasing function of $\mu$ for each $t\in \SB$ (see, e.g., Lemma A.2. of \citetappendix{Kivaranovic2020-ug}). 
Since $\frac{\alpha}{2}<1-\frac{\alpha}{2}$, it follows that $\theta_l(t)$ and $\theta_u(t)$ defined in \eqref{eq:LCB_UCB} are unique, and that $\theta_l(t)< \theta_u(t)$.

In addition, monotonicity implies that for all $t \in \SB$,  (i) $\nu^{\top}\beta > \theta_l(t)$ if and only if $F^{\SB}_{\nu^{\top}\beta, \sigma^2||\nu||_2^2}(t) < 1-\alpha/2$; and (ii) $\nu^{\top}\beta < \theta_u(t)$ if and only if $F^{\SB}_{\nu^{\top}\beta, \sigma^2||\nu||_2^2}(t) > \alpha/2$. In other words, we have that
\begin{align}
\label{eq:ci_cdf_equiv}
\qty{\nu^{\top}\beta: \nu^{\top}\beta\in \left[\theta_l(t),\theta_u(t)\right] } = \left\{\nu^{\top}\beta: \frac{\alpha}{2} \leq F^{\SB}_{\nu^{\top}\beta, \sigma^2||\nu||_2^2}(t) \leq 1- \frac{\alpha}{2}   \right\},\; \forall t \in \SB.
\end{align}

Now, under \eqref{eq:model}, $Y\sim \mathcal{N}(\beta,\sigma^2 I_n)$, which implies that
\begin{align*}
&\mathbb{P} \qty( \nu^{\top}\beta \in \left[\theta_l(\nu^{\top} Y),\theta_u(\nu^{\top} Y)\right]  \;\middle\vert\;  \hat{C}_1,\hat{C}_2 \in \mathcal{CC}_K(Y), \Pi_{\nu}^{\perp} Y = \Pi_{\nu}^{\perp} y) \\
&\overset{a.}{=} \mathbb{P} \qty( \frac{\alpha}{2} \leq F^{\SB}_{\nu^{\top}\beta, \sigma^2\Vert\nu\Vert_2^2}(\nu^{\top}Y) \leq 1- \frac{\alpha}{2}   \;\middle\vert\; \hat{C}_1,\hat{C}_2 \in \mathcal{CC}_K(Y), \Pi_{\nu}^{\perp} Y = \Pi_{\nu}^{\perp} y ) \\
&\overset{b.}{=} \mathbb{P}\qty(F^{\SB}_{\nu^{\top}\beta, \sigma^2\Vert\nu\Vert_2^2}\qty(Z)\in \qty[\frac{\alpha}{2},1-\frac{\alpha}{2}]) \\
&\overset{c.}{=} 1-\alpha.
\end{align*}
In step $a.$, we use the fact that \eqref{eq:ci_cdf_equiv} holds for all  $t \in \SB$; therefore it holds for $\nu^\top Y$ conditioning on the event $\qty{\hat{C}_1,\hat{C}_2 \in \mathcal{CC}_K(Y), \Pi_{\nu}^{\perp} Y = \Pi_{\nu}^{\perp} y}$ as well. Next, step $b.$ follows from \eqref{eq:single_param_null} in the proof of Proposition~\ref{prop:pval}, and letting $Z$ denote a $\mathcal{N}(\nu^\top\beta,\sigma^2||\nu||_2^2)$ random variable, truncated to the set $\SB$. The last step follows from the probability integral transform.

%% file: sections-revision-v1/appendix_a3.tex

\subsection{Modification of Algorithm~\ref{alg:SB} to compute $\pD$}
\label{section:appendix:pC_PD}


For the graph fused lasso with an arbitrary graph, it may be the case that for two integers $K,K'$, $|\mathcal{CC}_{K}(Y)|=|\mathcal{CC}_{K'}(Y)|$ and $\mathcal{CC}_{K}(Y) \neq \mathcal{CC}_{K'}(Y)$~\citepappendix{Tibshirani2011-fq}. In other words, $\mathcal{CC}(Y)$ in \eqref{eq:pD} need not to be unique. Therefore, we cannot directly apply the recipes in Section~\ref{section:method} to characterize $\pD$ in \eqref{eq:pD}. 

In what follows, we propose to characterize a variant of $\pD$ that makes use of the smallest value of $K$ to yield exactly $L$ connected components:
\begin{equation}
\label{eq:pDvar}
\pDvar \equiv \mathbb{P}_{H_0}\qty(|\nu^\top Y |\geq |\nu^\top y | \;\middle\vert\; \hat{C}_1(y), \hat{C}_2(y) \in \mathcal{CC}_{\argmin_{k}\{k:|\CC_k(Y)|=L\}}(Y) ,  \Pi_{\nu}^{\perp} Y= \Pi_{\nu}^{\perp} y ).
\end{equation}
We remark that the definitions of $\pDvar$ in \eqref{eq:pDvar} and $\pD$ in \eqref{eq:pD} coincide in the special case of the one-dimensional fused lasso, when the number of estimated connected components $L$ is uniquely determined by the number of steps in the dual path algorithm $K$, and, as a result, $\mathcal{CC}(Y)$ is uniquely defined. 

Using a similar argument to Proposition~\ref{prop:pval}, we have that for $\phi\sim \mathcal{N}\qty(0,\sigma^2||\nu||_2^2)$, 
\begin{equation}
\pDvar = \mathbb{P}\qty(|\phi|\geq |\nu^\top y | \;\middle\vert\; \hat{C}_1(y), \hat{C}_2(y) \in \mathcal{CC}_{\argmin_{k}\{k:|\CC_k(y'(\phi))|=L\}}(y'(\phi)) ).
\end{equation}
 Therefore, the key to computing $\pD$ in \eqref{eq:pD} is to characterize the set 
\begin{equation}
\label{eq:SD}
 \SD \equiv \qty{\phi: \hat{{C}}_1(y), \hat{{C}}_2(y) \in \mathcal{CC}_{\argmin_{k}\{k:|\CC_k(y'(\phi))|=L\}}(y'(\phi)) }.
\end{equation}
The algorithm for characterizing the set $\SD$ in \eqref{eq:SD} requires only minor modifications to Algorithm~\ref{alg:SB}. Details are provided in Algorithm~\ref{alg:SD}.

\begin{algorithm}[!htpb]
\setstretch{0.1}
\SetKwInOut{Input}{Input}
    \SetKwInOut{Output}{Output}
 \Input{Data $y$, $\hat{C}_1$, $\hat{C}_2$, number of connected components $L$, $\eta=10^{-4}$}

 \Output{$\SD$ in \eqref{eq:SD}}

  \begin{enumerate}
    \item
    $\mathcal{J}\leftarrow  \{0\}$, $i\leftarrow 1$, $j\leftarrow 0$. 
	\item 
   $[a_0,a_1]=\left\{ \phi\in\mathbb{R}:  \bigcap_{k=1}^{K^*} \left\{ M_k(y'(\phi)) = M_k(y)\right\} \right\}$, where $K^*=\argmin_{k} \qty{|\CC_k(y)| = L}.$
  \item
   \While{$a_{i}<\infty$}{
     \begin{enumerate}
     \item
   Compute   $[\tilde{a}_i,a_{i+1}] = \left\{ \phi\in\mathbb{R}: \bigcap_{k=1}^{K^*} \left\{ M_k(y'(\phi)) = M_k(y'(a_i + \eta))\right\} \right\}$, where $K^*=\argmin_{k} \qty{|\CC_k(y'(a_i + \eta))| = L}.$ 

     \item
     \While{$\tilde{a}_i\neq a_{i}$}{
     \begin{enumerate}
      \item
     $\eta \leftarrow \frac{1}{2}\cdot \eta, $
     \item
      $[\tilde{a}_i,a_{i+1}]= \left\{ \phi\in\mathbb{R}:  \bigcap_{k=1}^{K^*} \left\{ M_k(y'(\phi)) = M_k(y'(a_i + \eta))\right\} \right\}$, where $K^*=\argmin_{k} \qty{|\CC_k(y'(a_i + \eta))| = L}.$
    \end{enumerate}
     }
    \item 
    \If{$\hat{C}_1,\hat{C}_2 \in \mathcal{CC}_{K^*}(y'(a_i+\eta))$}{\hspace{20mm} $\mathcal{J} \leftarrow \mathcal{J} \cup \{i\}.$}
    \item
    $i \leftarrow i + 1 .$
     \end{enumerate}
     }
    \item 
   \While{$a_{j}>-\infty$}{
     \begin{enumerate}
     \item
      $[a_{j-1},\tilde{a}_{j}] = \left\{ \phi\in\mathbb{R}:  \bigcap_{k=1}^{K^*} \left\{ M_k(y'(\phi)) = M_k(y'(a_j-\eta))\right\} \right\}$, where $K^*=\argmin_{k} \qty{|\CC_k(y'(a_j-\eta))| = L}.$ \\
     \item
     \While{$\tilde{a}_j\neq a_{j}$}{
     \begin{enumerate}
      \item
     $\eta \leftarrow \frac{1}{2}\cdot \eta, $\\
     \item
     $[a_{j-1},\tilde{a}_{j}]= \left\{ \phi\in\mathbb{R}:  \bigcap_{k=1}^{K^*} \left\{ M_k(y'(\phi)) = M_k(y'(a_j-\eta))\right\} \right\}$, where $K^*=\argmin_{k} \qty{|\CC_k(y'(a_j-\eta))| = L}.$
   \end{enumerate}
     }
    \item 
    \If{$\hat{C}_1, \hat{C}_2 \in \mathcal{CC}_{K^*}(y'(a_j-\eta))$}{
\hspace{20mm} $\mathcal{J} \leftarrow \mathcal{J} \cup \{j-1\}.$\\
    }
    \item
    $j \leftarrow j - 1.$ \\
     \end{enumerate}
     }
 \end{enumerate}
 $\SD \leftarrow \bigcup_{i\in \mathcal{J}}[a_i, a_{i+1}]$
 \caption{Characterizing the set $\SD$ in \eqref{eq:SD}}
 \label{alg:SD}
\end{algorithm}

\begin{figure}[!htbp]
\centering
\begin{subfigure}{0.4\textwidth}
  \caption{}
  \includegraphics[width=\linewidth]{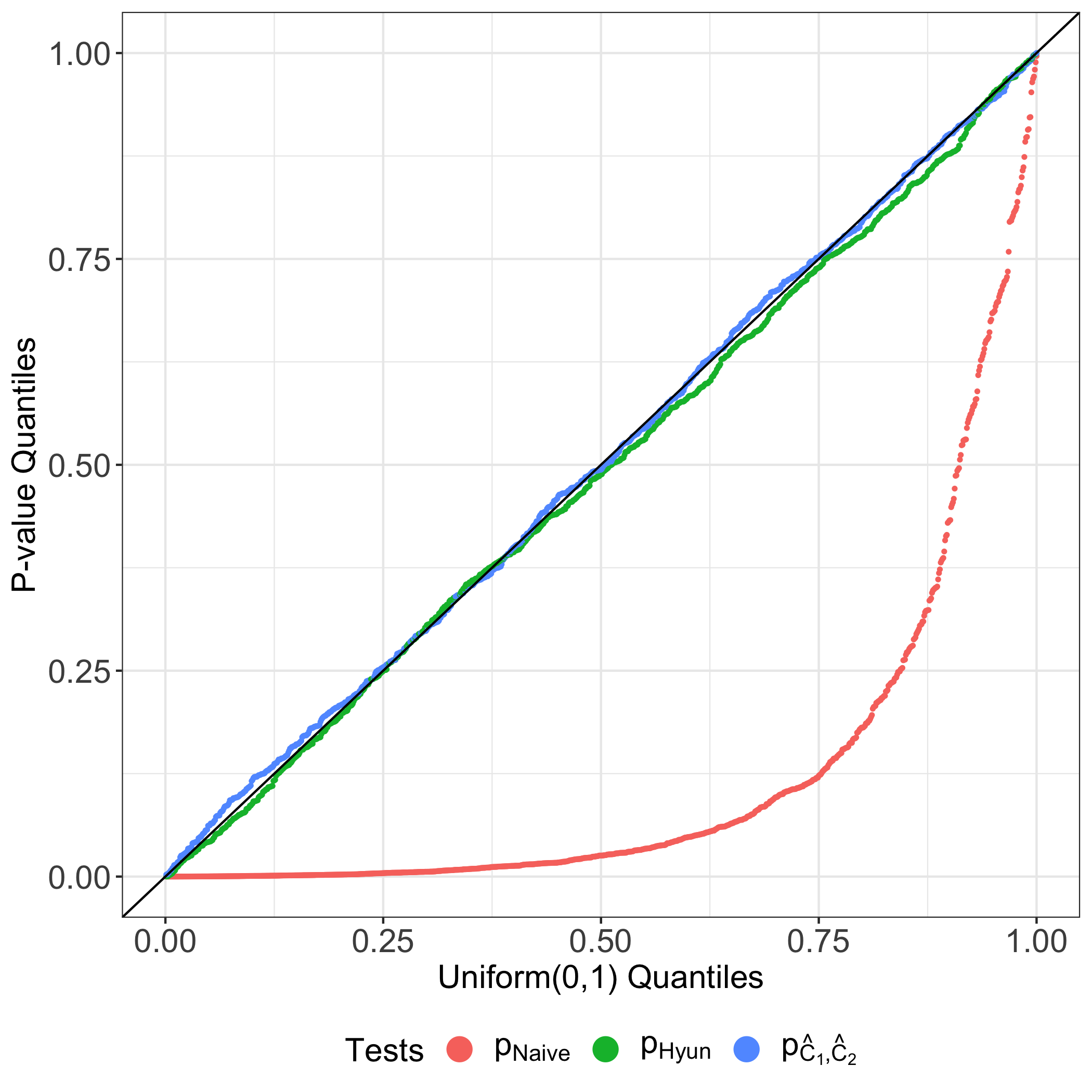}
\end{subfigure}
\begin{subfigure}{0.4\textwidth}
  \caption{}
  \includegraphics[width=\linewidth]{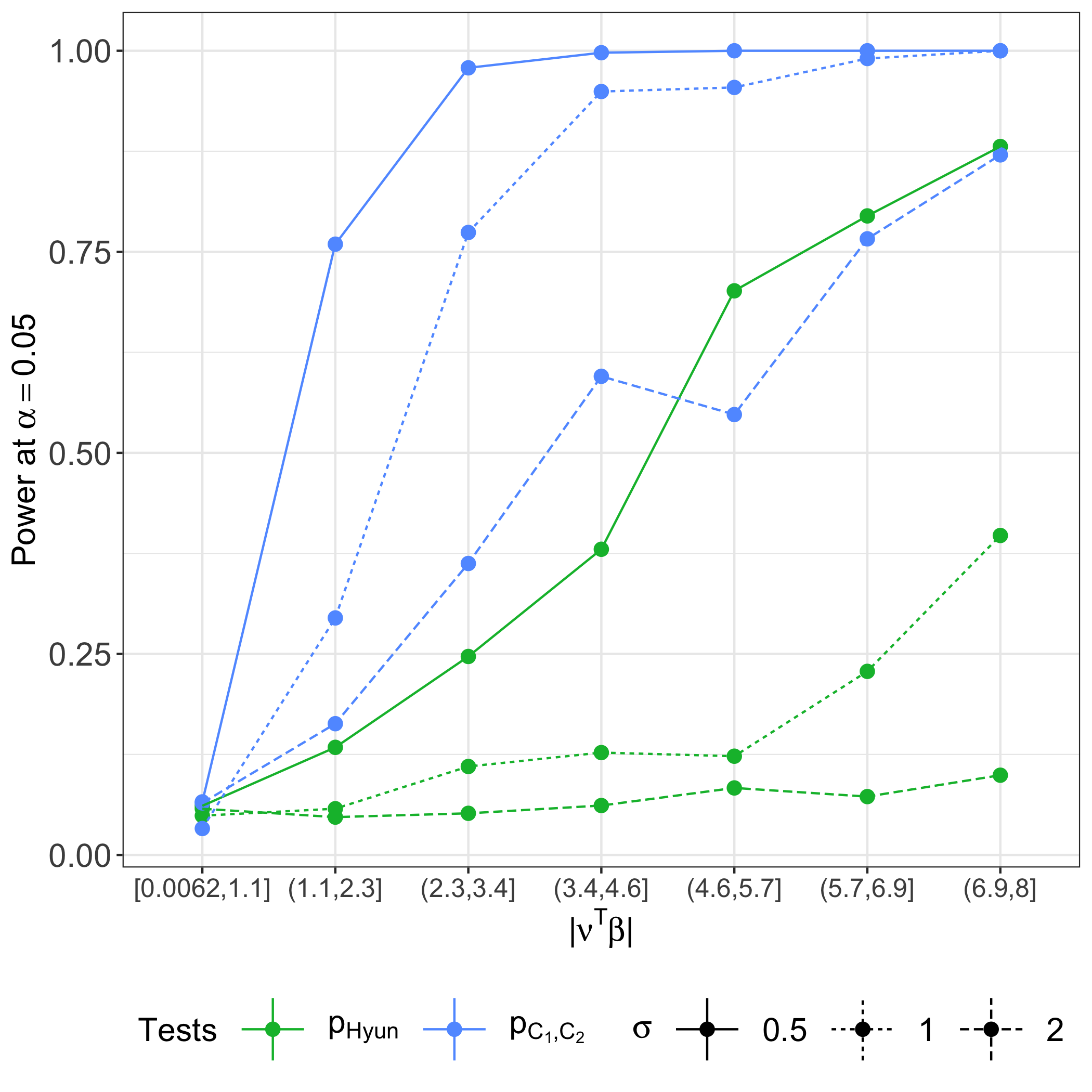}
\end{subfigure}
\begin{subfigure}{0.4\textwidth}
  \caption{}
  \includegraphics[width=\linewidth]{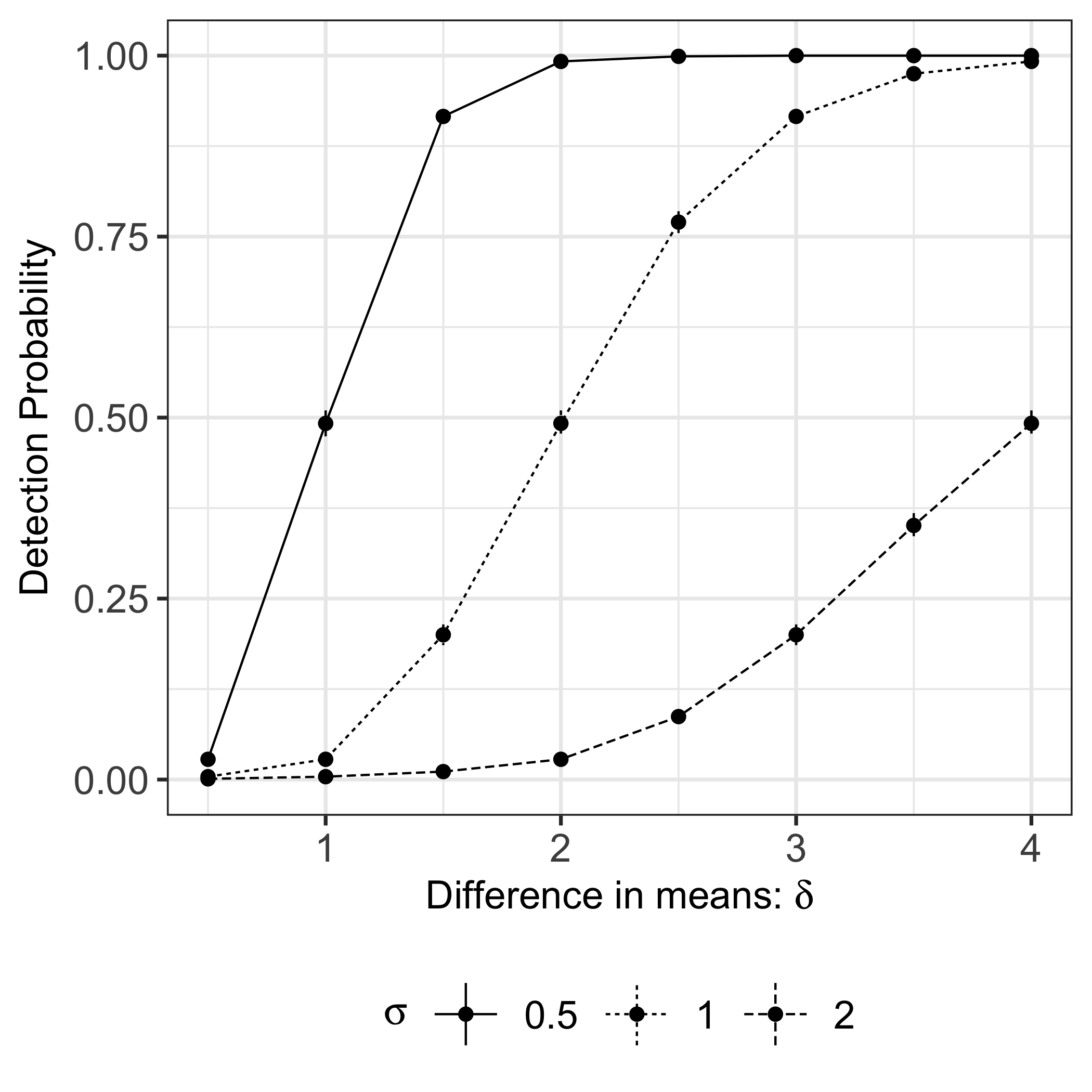}
\end{subfigure}
\begin{subfigure}{0.4\textwidth}
  \caption{}
  \includegraphics[width=\linewidth]{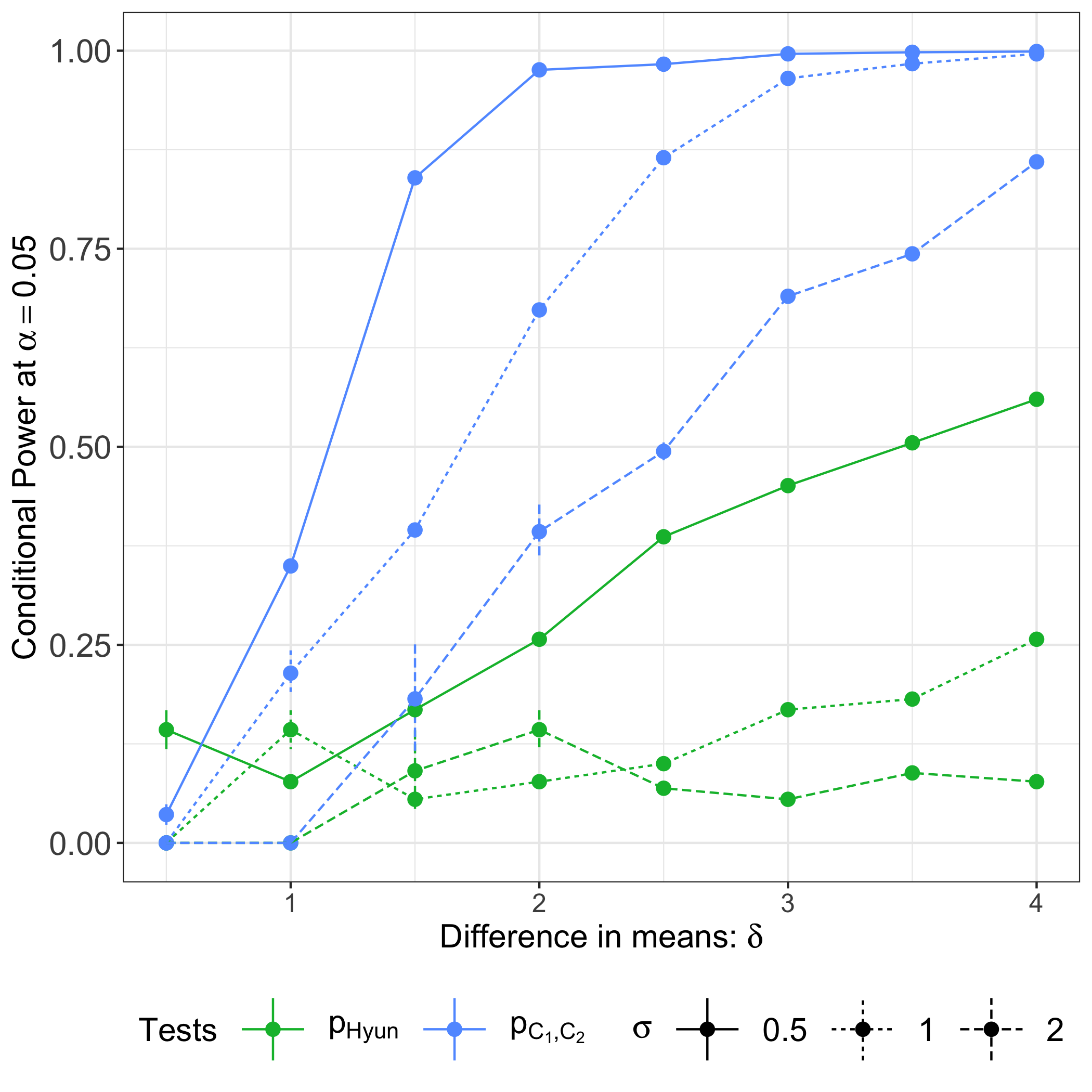}
\end{subfigure}
\caption{ \textit{(a): } For the two-dimensional fused lasso model in \eqref{sim:2d_eq}, when $\delta=0$, tests based on $\pHyun$ and $\pD$ control the selective Type I error. By contrast, the naive $p$-value leads to an inflated selective Type I error. \textit{(b): }  Under the simulation setup in Section~\ref{section:sim_two_d_power}, the power of tests based on $\pHyun$ and $\pD$ increases as a function of $|\nu^\top\beta|$. For a given bin of $|\nu^\top\beta|$, the test based on $\pD$ has substantially higher power than that based on $\pHyun$.  \textit{(c): } Under the same setup as (b), the detection probability defined in \eqref{eq:detect_p} increases as a function of the difference in means between two piecewise constant segments ($\delta$ in \eqref{sim:2d_eq}). Moreover, a larger value of noise variance $\sigma^2$ leads to a smaller detection probability.  \textit{(d): } Under the same setup as (b), the test based on $\pD$ has substantially higher conditional power (defined in \eqref{eq:conditional_power}) than that based on $\pHyun$. For both tests, power increases as a function of $\delta$ (defined in \eqref{eq:detect_p}).}
\label{fig:fixed_L_results}
\end{figure}

Figure~\ref{fig:fixed_L_results} displays the results for the test based on $\pD$ with $L=3$ for the two-dimensional fused lasso simulations described in Section~\ref{section:sim_two_d}. Panel (a) demonstrates that the test based on $\pD$ controls the selective Type I error. In panel (b) of Figure~\ref{fig:fixed_L_results}, we see that the test based on $\pD$ has substantially higher power than the test based on $\pHyun$. Figures~\ref{fig:fixed_L_results}(c) and (d) investigate the detection probability (defined in \eqref{eq:detect_p}) and conditional power (defined in \eqref{eq:conditional_power}) of $\pD$. 

%% file: sections-revision-v1/appendix_a4.tex

\subsection{Additional power comparisons}
\label{appendix:conditional_power}

In Section~\ref{section:sim}, we considered the power of the tests based on $\pB$ and $\pHyun$ as a function of the binned effect size $|\nu^\top\beta|$. Here, we conduct three additional analyses on the same simulated datasets from Section~\ref{section:sim}. 

In the first analysis, we separately consider two quantities considered in prior work~\citepappendix{Gao2020-yt,jewell2019testing,Hyun2018-gx}: (i) the \emph{detection probability} (i.e., the probability that $\hat{C}_1$ and  $\hat{C}_2$ are true piecewise constant regions in the original signal $\beta$) of the graph fused lasso estimator in Eq. (3) of the main manuscript, and (ii) the \emph{conditional power} of the tests based on $\pB$ and $\pHyun$ (i.e., the probability of rejecting $H_0:\nu^\top \beta = 0$, \emph{given that} $\hat{C}_1$ and $\hat{C}_2$ are true piecewise constant regions). 

Given $M$ simulated datasets, we can estimate the conditional power as 
\begin{align}
\label{eq:conditional_power}
\text{Conditional power} = \frac{\sum_{m=1}^M 1\qty{\exists j,j' \text{ s.t. } \hat{{C}}^{(m)}_1={{C}}_j,\hat{{C}}^{(m)}_2 ={{C}}_{j'}, p^{(m)}\leq \alpha}}{\sum_{m=1}^M 1\qty{\exists j,j' \text{ s.t. } \hat{{C}}^{(m)}_1={{C}}_j,\hat{{C}}^{(m)}_2 ={{C}}_{j'} }},
\end{align}
where $p^{(m)}$ and $\hat{{C}}^{(m)}_1,\hat{{C}}^{(m)}_2$ correspond to the $p$-values and estimated connected components under consideration for the $m$th simulated dataset. Here, ${{C}}_j$ is the $j$th true piecewise constant segment in $\beta$. Because the quantity in \eqref{eq:conditional_power} conditions on the event that  $\hat{{C}}^{(m)}_1$ and $\hat{{C}}^{(m)}_2$ correspond to true piecewise constant segments, we also estimate how often this occurs:
\begin{align}
\label{eq:detect_p}
\text{Detection probability} = \frac{\sum_{m=1}^M 1\qty{\exists j,j' \text{ s.t. } \hat{{C}}^{(m)}_1={{C}}_j,\hat{{C}}^{(m)}_2 ={{C}}_{j'}}}{M}.
\end{align}

\begin{figure}[!htbp]
\centering
\begin{subfigure}{0.4\textwidth}
  \caption{}
  \includegraphics[width=\linewidth]{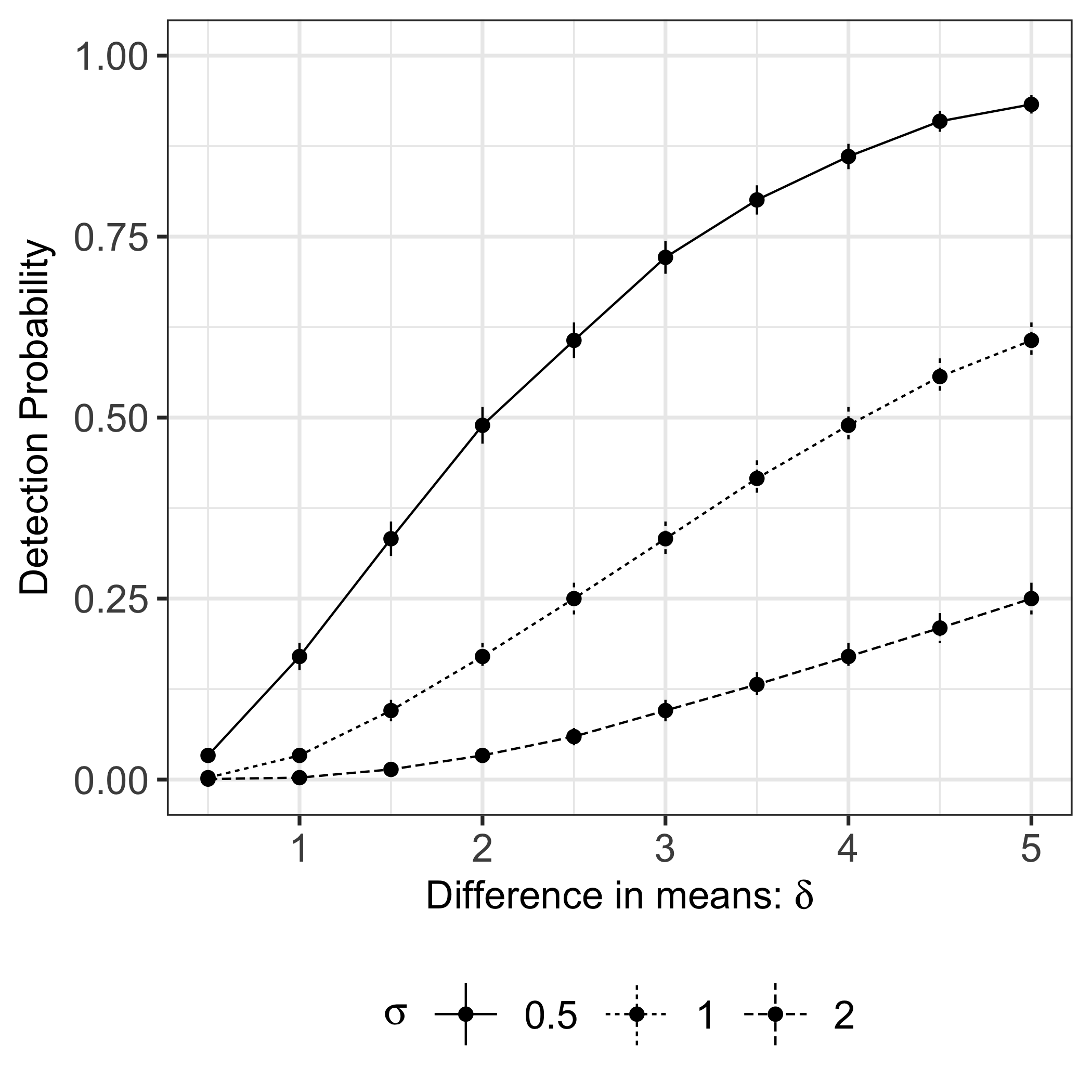}
\end{subfigure}
\begin{subfigure}{0.4\textwidth}
  \caption{}
  \includegraphics[width=\linewidth]{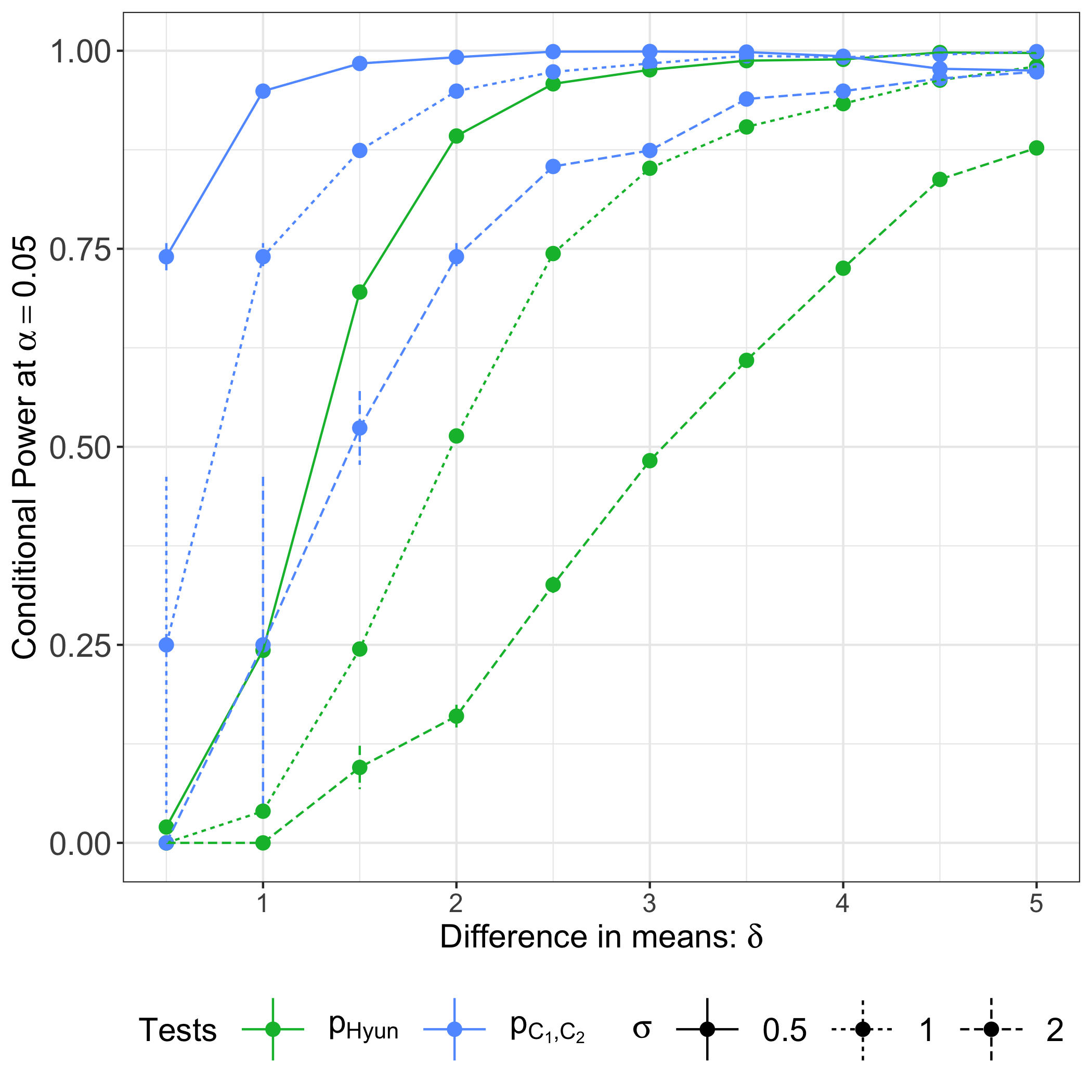}
\end{subfigure}
\begin{subfigure}{0.4\textwidth}
  \caption{}
  \includegraphics[width=\linewidth]{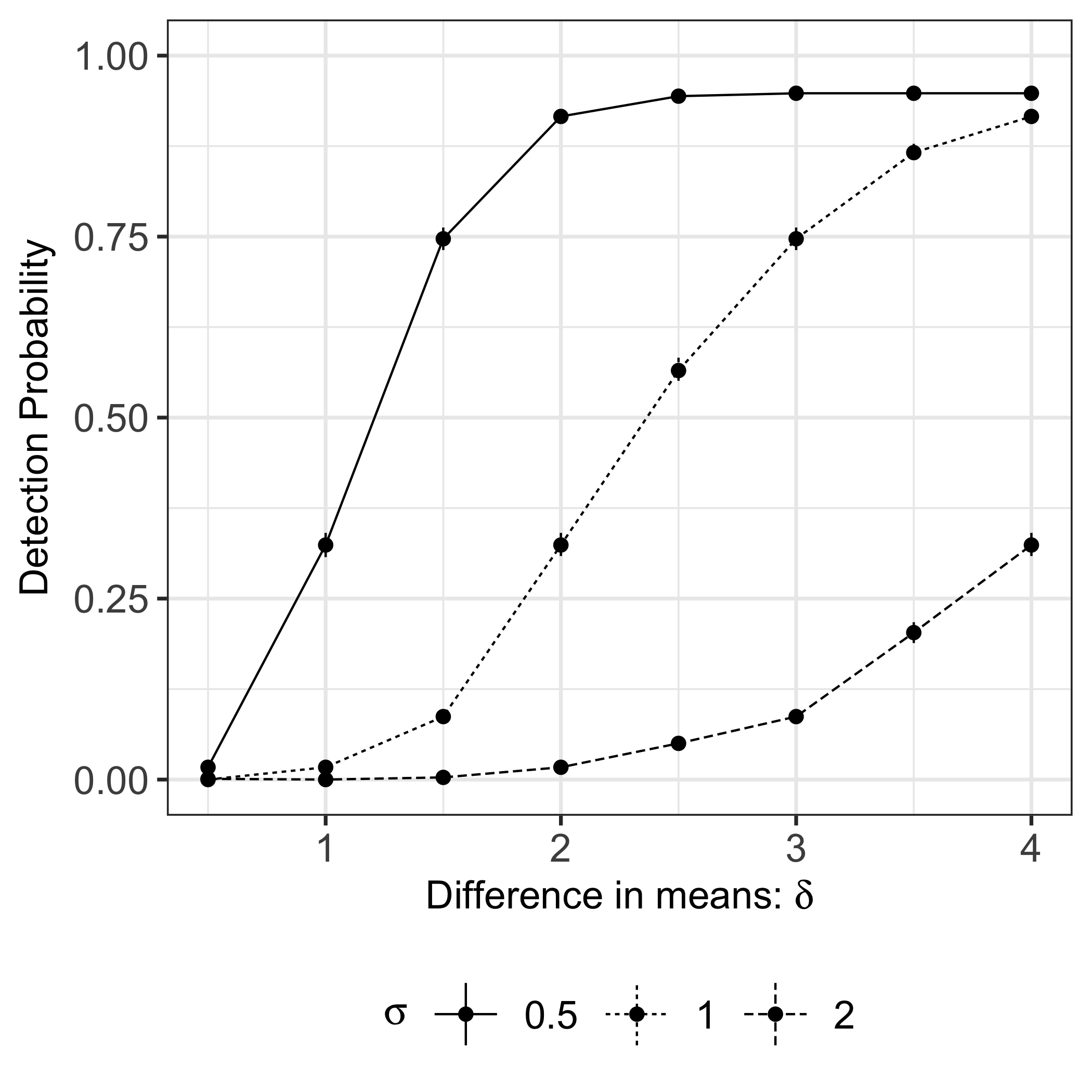}
\end{subfigure}
\begin{subfigure}{0.4\textwidth}
  \caption{}
  \includegraphics[width=\linewidth]{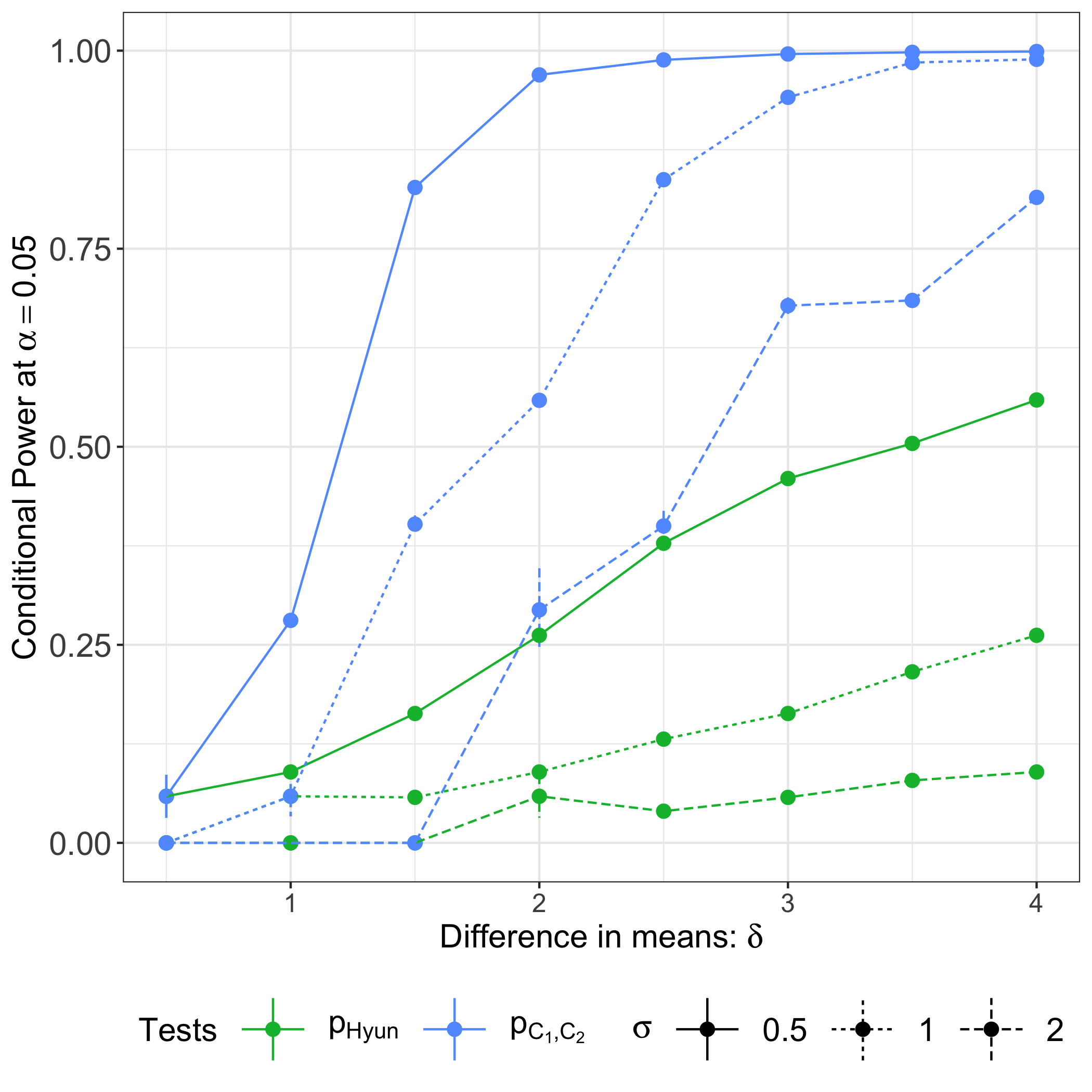}
\end{subfigure}
\caption{ \textit{(a): } For the one-dimensional fused lasso simulations described in Section~\ref{section:sim_one_d_power}, the detection probability of the tests based on both $\pHyun$ and $\pB$ increases as a function of the difference in means between two piecewise constant segments, $\delta$, across all values of $\sigma$. \textit{(b): } For the one-dimensional fused lasso simulations described in Section~\ref{section:sim_one_d_power}, the conditional power of the tests based on both $\pHyun$ and $\pB$ increases as a function of $|\nu^\top\beta|$, across all values of $\sigma$. For a given bin of $|\nu^\top\beta|$ and a given value of $\sigma$, the test based on $\pB$ has higher conditional power than the test based on $\pHyun$. \textit{(c): } Same as (a), but for the two-dimensional fused lasso simulations described in Section~\ref{section:sim_two_d_power} \textit{(d): } Same as (b), but for the two-dimensional fused lasso simulations described in Section~\ref{section:sim_two_d_power}.}
\label{fig:detect_p_cond_power}
\end{figure}

We evaluated detection probability and conditional power on data generated from the one-dimensional and two-dimensional fused lasso models, with the same simulation setup as in Sections~\ref{section:sim_one_d} and ~\ref{section:sim_two_d}, respectively. Results aggregated over 1,500 simulations are displayed in Figure~\ref{fig:detect_p_cond_power}. Panels (a) and (b) display the detection probability and conditional power (with $\alpha=0.05$) for the one-dimensional fused lasso, respectively. Both quantities
increase as the difference in means between the two piecewise constant segments ($\delta$ in \eqref{sim:1d_eq}) increases. By contrast, both quantities decrease as the variance $\sigma^2$ increases. In addition, for a given value of $\sigma$ and $\delta$, the conditional power of the test based on $\pB$ is higher than that based on $\pHyun$. We observe similar trends in the two-dimensional fused lasso case; see Figure~\ref{fig:detect_p_cond_power}(c)--(d).

\begin{figure}[hb!]
\centering
\includegraphics[width=0.45\linewidth]{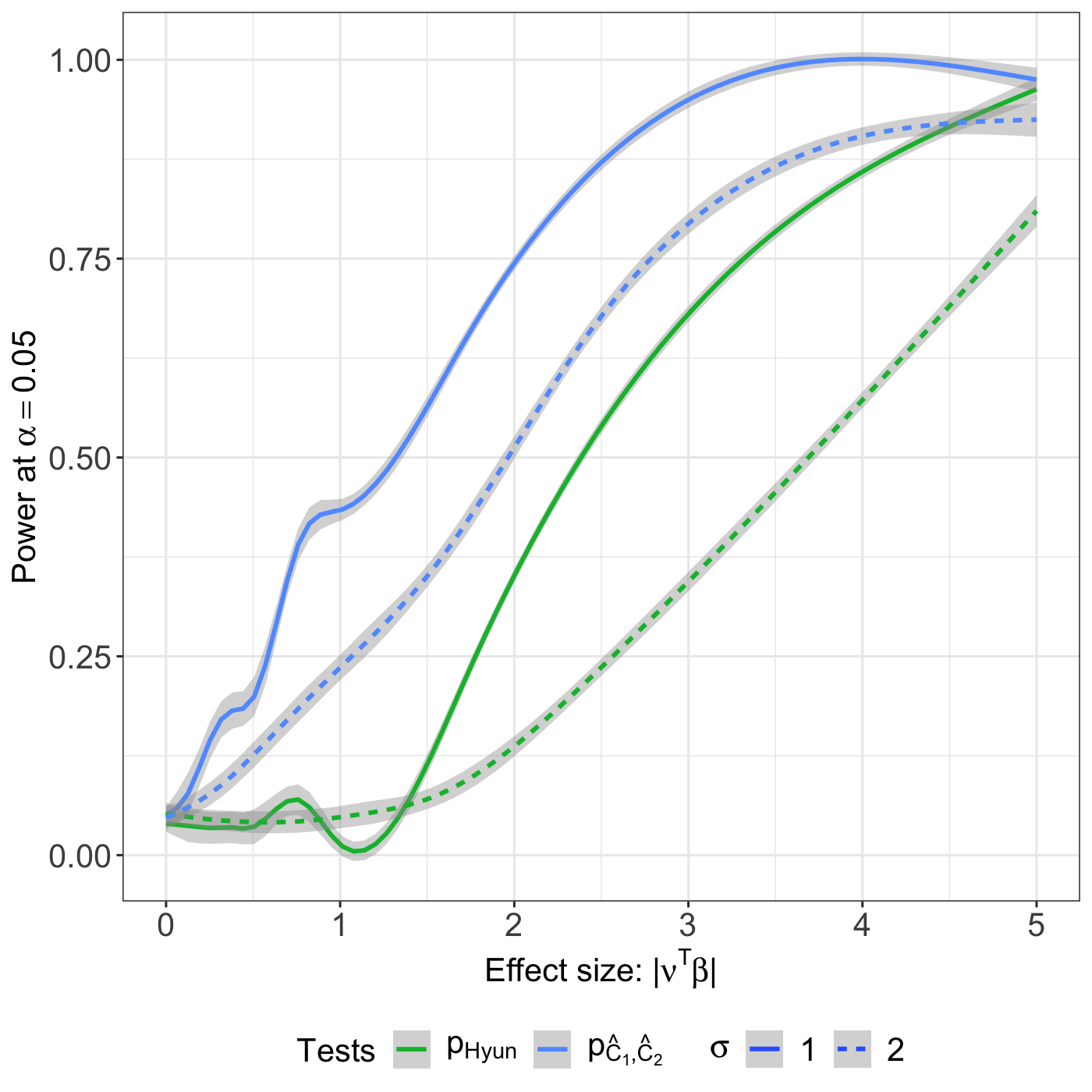}
\caption{Additional analysis of the data in Section~\ref{section:sim_one_d_power}. We used a generalized additive model to obtain the power of the tests based on $\pHyun$ in \eqref{eq:hyun_pval} and $\pB$ in \eqref{eq:pB} as a smooth function of $|\nu^\top\beta|$. }
\label{fig:one_d_cond_power}
\end{figure}

In the second analysis, instead of binning $|\nu^\top\beta|$, we fit a regression spline using the \texttt{gam} function in the \texttt{R} package \texttt{mgcv} \citepappendix{wood_2017} to obtain a smooth estimate for the one-dimensional fused lasso simulations in Section~\ref{section:sim_one_d_power}. The results are in Figure~\ref{fig:one_d_cond_power}. As in Figure~\ref{fig:one_d_sim}(c), the power of the tests that reject $H_0$ if $\pB$ or $\pHyun$ is below $\alpha=0.05$ increases as $|\nu^\top\beta|$ increases. For a given value of $|\nu^\top\beta|$, the test based on $\pB$ has higher power than that based on $\pHyun$.

Finally, in the third analysis, we assess the sensitivity of our conclusions to the choice of $K$ in the dual path algorithm, using the one-dimensional fused lasso model in \eqref{sim:1d_eq}. Recall that in Section~\ref{section:sim_one_d}, we choose $K=2$ so the number of estimated connected components resulting from the one-dimensional fused lasso equals the true number of connected components in \eqref{sim:1d_eq}. 

\begin{figure}[!htbp]
\centering
\begin{subfigure}{0.4\textwidth}
  \caption{}
  \includegraphics[width=\linewidth]{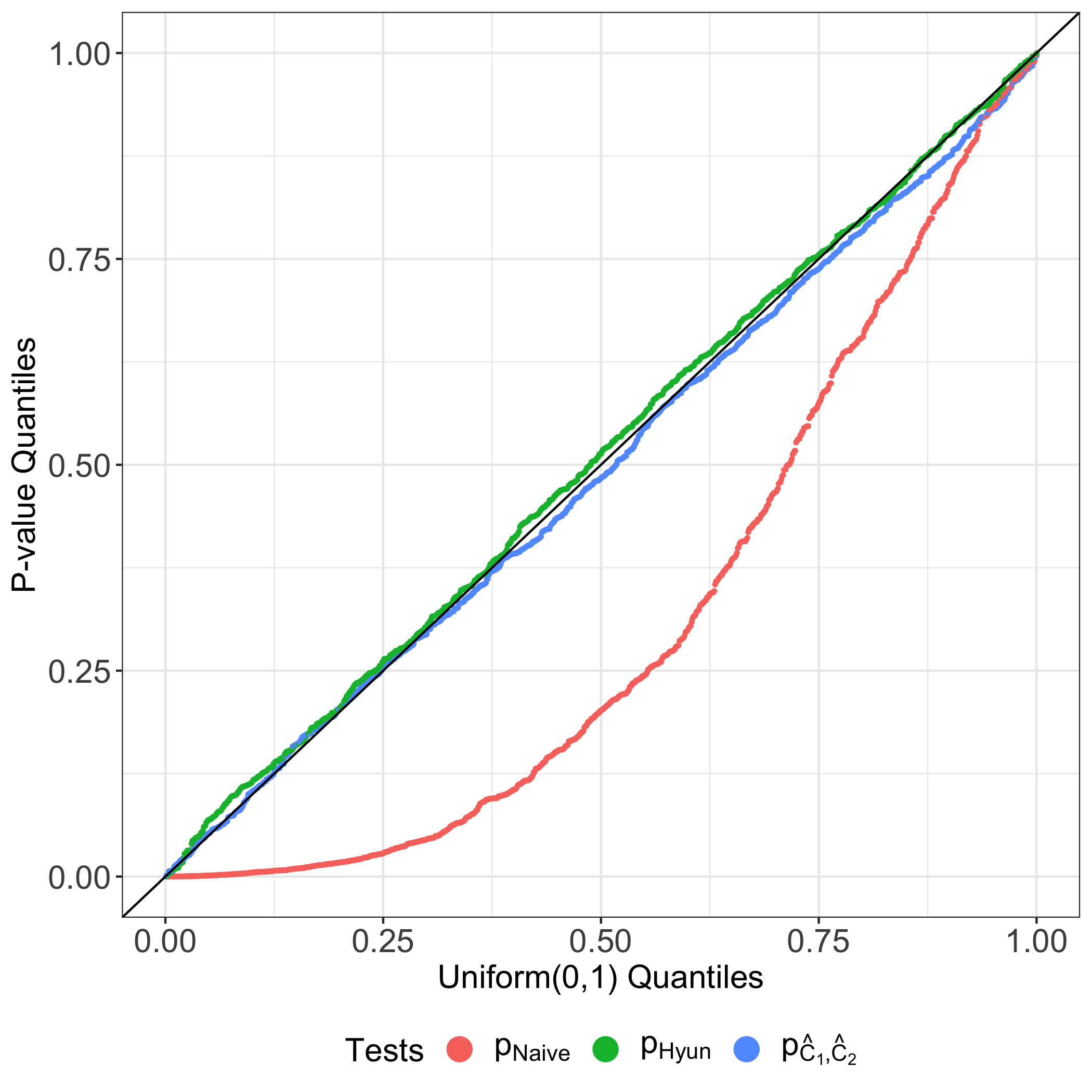}
\end{subfigure}
\begin{subfigure}{0.4\textwidth}
  \caption{}
  \includegraphics[width=\linewidth]{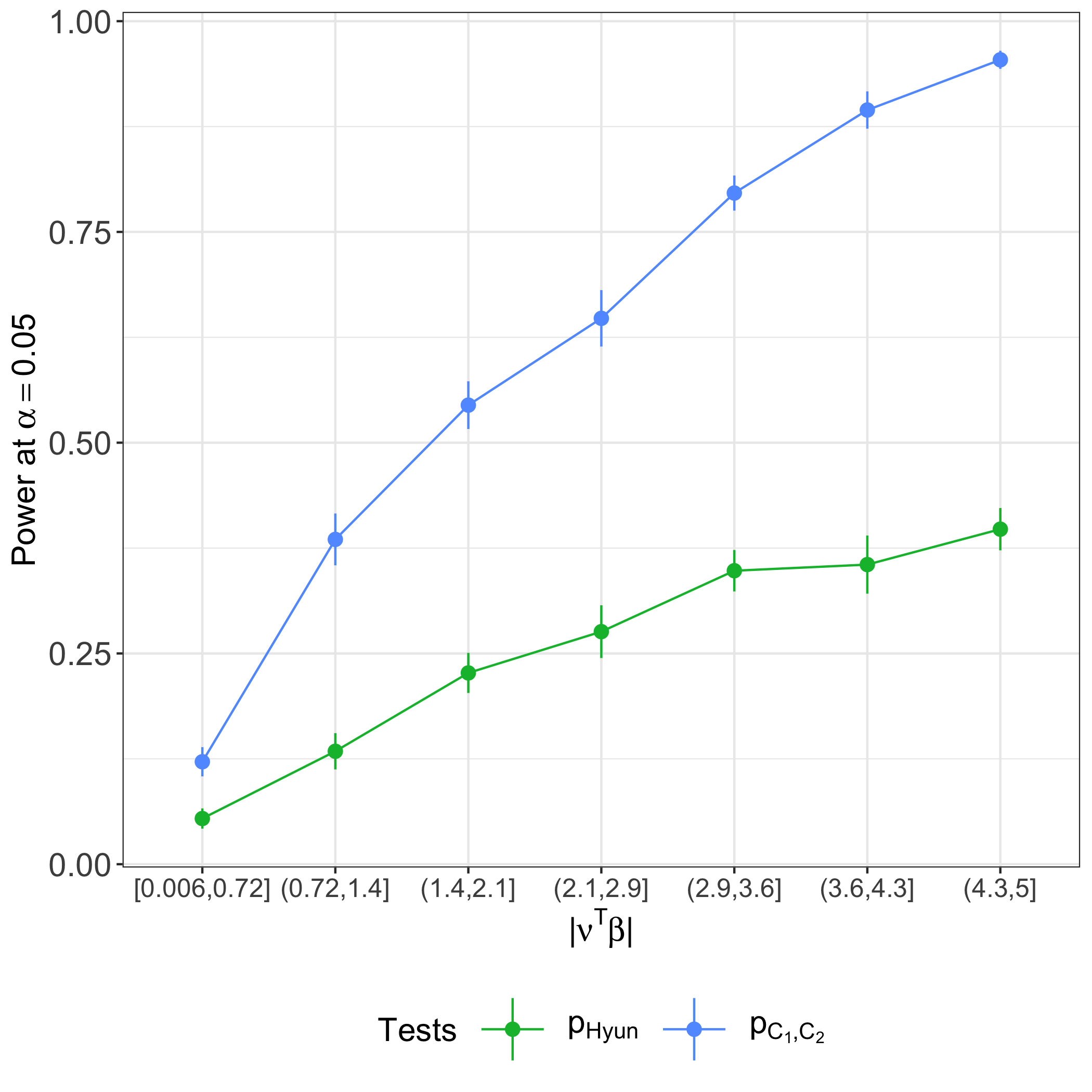}
\end{subfigure}
\caption{ \textit{(a): } For the one-dimensional fused lasso simulations  in Section~\ref{section:sim_one_d_type_1}, when $\delta=0$ and the graph fused lasso is solved using the dual path algorithm with $K=4$, tests based on $\pHyun$ and $\pB$ control the selective Type I error. By contrast, the naive $p$-value leads to an inflated selective Type I error. \textit{(b): } For the one-dimensional fused lasso simulations in Section~\ref{section:sim_one_d_power} with $\sigma=1$,  the power of tests based on $\pHyun$ and $\pB$ increases as a function of $|\nu^\top\beta|$. For a given bin of $|\nu^\top\beta|$, the test based on $\pB$ has substantially higher power than that based on $\pHyun$. }
\label{fig:type_1_power_different_K}
\end{figure}

Here, we repeat the experiments in Section~\ref{section:sim_one_d} with $K=4$, which yields five estimated connected components. Results are displayed in Figure~\ref{fig:type_1_power_different_K}. Panel (a) displays the observed $p$-value quantiles versus Uniform(0,1) quantiles, aggregated over 1,000 simulated datasets. As in the case of $K=2$, only tests based on $\pHyun$ or $\pB$ control the selective Type I error. In Figure~\ref{fig:type_1_power_different_K}(b), we see that the power of the tests based on $\pHyun$ or $\pB$ increases as $|\nu^\top \beta|$ increases. For a given value of $|\nu^\top \beta|$, the test based on  $\pB$ has substantially higher power than the test based on $\pHyun$. In other words, the substantial increase in power, as well as the selective Type I error control, of the test based on $\pB$, \emph{does not} depend on correctly specifying $K$.

\newpage
\subsection{Estimation of the error variance $\sigma^2$ in \eqref{eq:model}}
\label{appendix:estimated_sigma}

Throughout this section, we have assumed that $\sigma^2$ in \eqref{eq:model} is known. In practice, we can plug in an estimate $\hat\sigma$ when computing the $p$-values $\pB$ and $\pHyun$. That is, we use 
\begin{align}
\pB(\hat\sigma) = \mathbb{P}\qty(|\phi(\hat\sigma)|\geq |\nu^\top y| \;\middle\vert\; \hat{C}_1(y),\hat{C}_2(y) \in \CC_K(y'(\phi))),
\end{align}
where $\phi(\hat\sigma)\sim \hat\sigma\cdot\mathcal{N}(0,||\nu||_2^2)$. 

In this section, we investigate the empirical selective Type I error control and power of the following estimators of $\sigma^2$:
\begin{itemize} 
\item
$\hat\sigma^2_{\text{Residual}} = \frac{1}{n-L}\sum_{l=1}^{L}\sum_{j\in \hat{C}_l}\qty(y_j-\qty(\sum_{j'\in\hat{C}_l}y_{j'})/|\hat{C}_l|)^2$, where $\hat{C}_1,\ldots, \hat{C}_L$ are the $L$ estimated connected components of the graph fused lasso solution $\hat\beta$;

\item
$\hat\sigma^2_{\text{Sample}} = \frac{1}{n-1}\sum_{i=1}^n\qty(y_j-\bar{y})^2$, where $\bar{y}$ is the mean of the data $y$; and

\item
$\hat\sigma^2_{\text{MAD}} = \frac{\text{median}_{i=2,\ldots,n}\qty(|z_i-\tilde{z}|)}{\sqrt{2}\Phi^{-1}(3/4)}$, where $\tilde{z}=\text{median}_{i=2,\ldots,n}(z_i)$, and $z_i = y_{i}-y_{i-1}$.
\end{itemize}

\begin{figure}[!htbp]
\centering
\begin{subfigure}{0.31\textwidth}
  \caption{}
  \includegraphics[width=\linewidth]{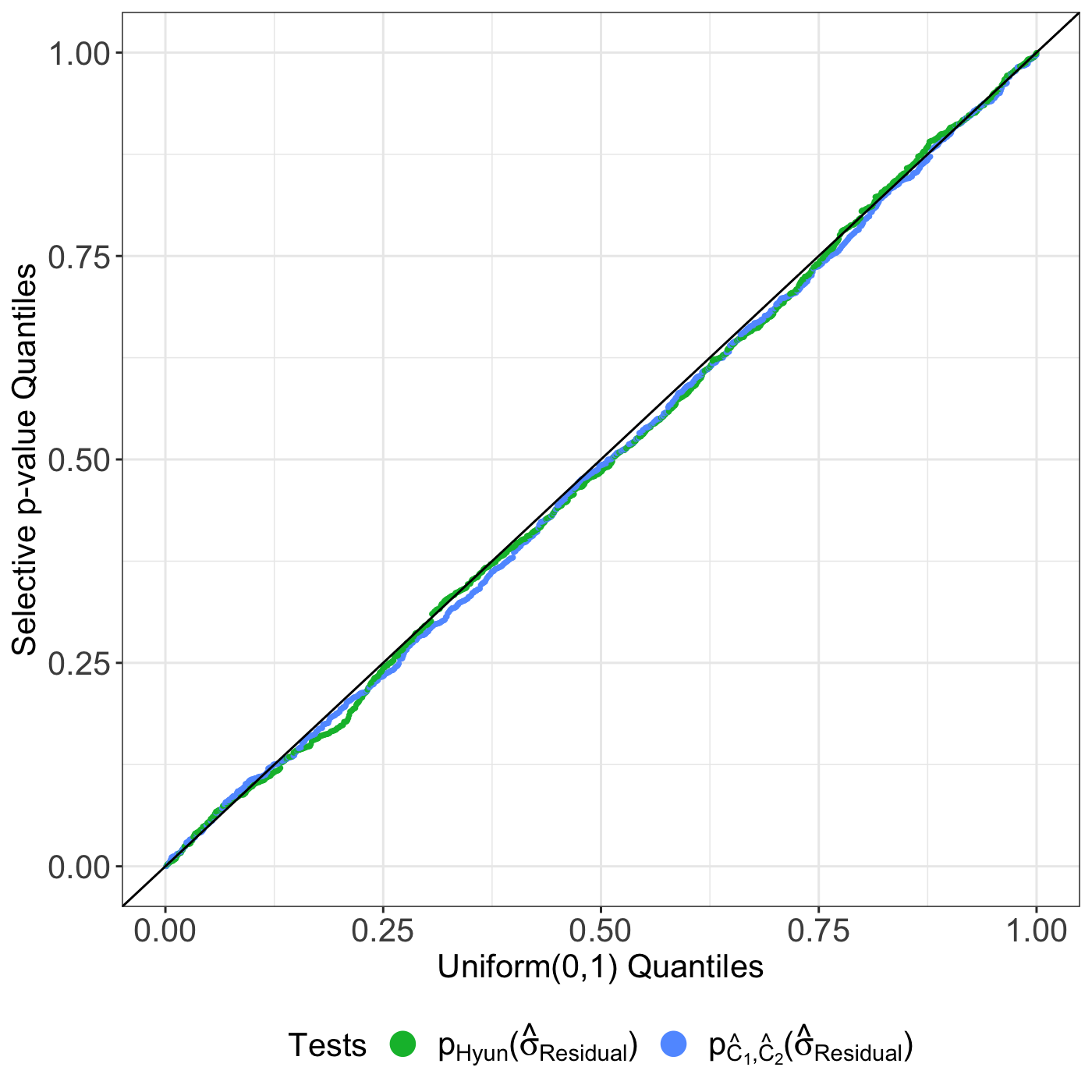}
\end{subfigure}
\begin{subfigure}{0.31\textwidth}
  \caption{}
  \includegraphics[width=\linewidth]{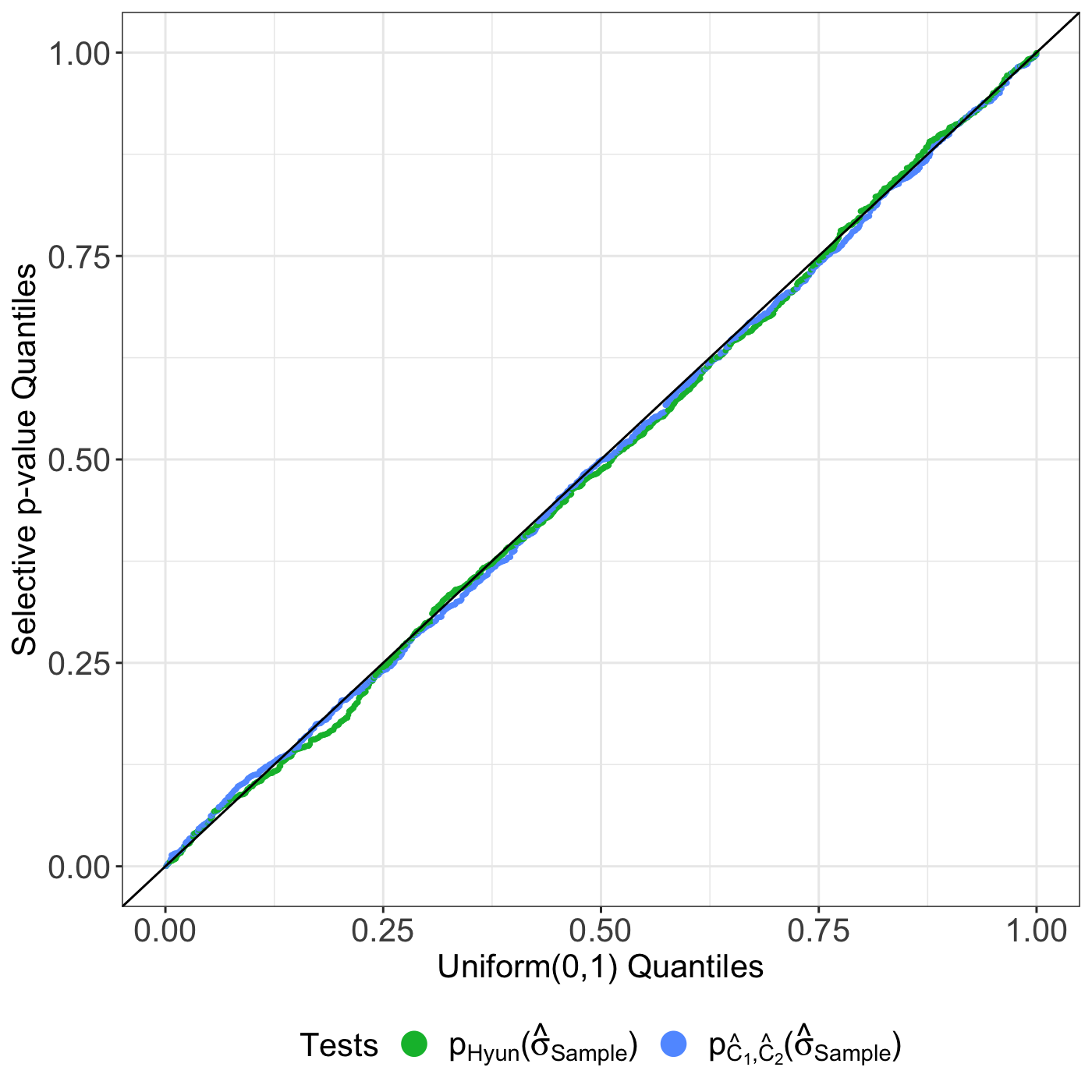}
\end{subfigure}
\begin{subfigure}{0.31\textwidth}
  \caption{}
  \includegraphics[width=\linewidth]{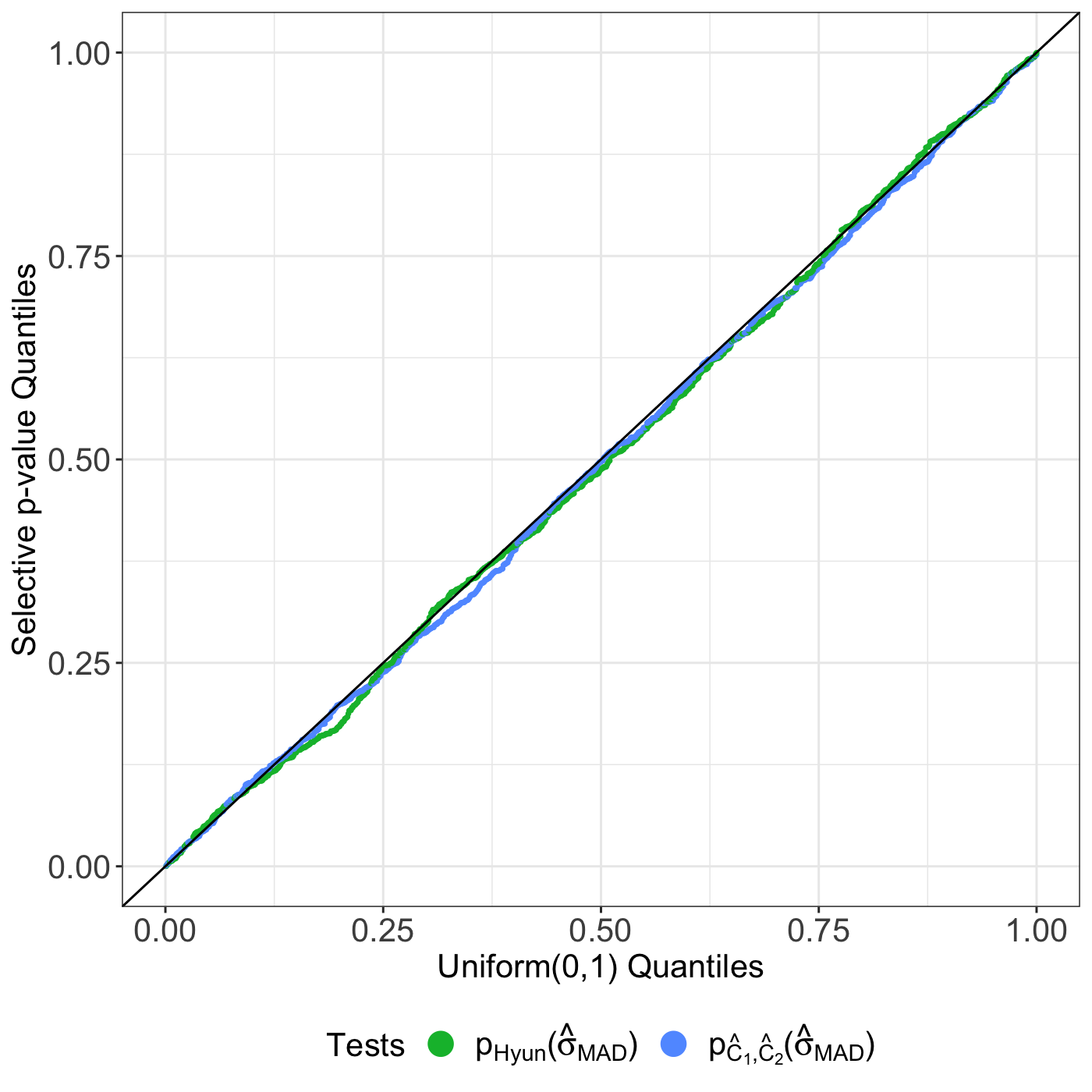}
\end{subfigure}
\caption{ \textit{(a): } Quantile-quantile plot for $p$-values $\pHyun$ and $\pB$ computed using the estimated variances $\hat\sigma^2_{\text{Residual}}$  for the one-dimensional fused lasso simulations described in Section~\ref{section:sim_one_d_type_1} of the manuscript. Results are based on 1,000 simulated datasets under the global null. \textit{(b): } Same as (a), but for the variance estimator $\hat\sigma^2_{\text{Sample}}$. \textit{(c): } Same as (a), but for the variance estimator $\hat\sigma^2_{\text{MAD}}$.}
\label{fig:type_1_sigma_hat}
\end{figure}

Figure~\ref{fig:type_1_sigma_hat} displays the quantiles of the $p$-values $\pHyun$ and $\pB$ computed using the estimated variances with the same simulation setup as in Section~\ref{section:sim_one_d_type_1}. 
All three estimators ($\hat\sigma^2_{\text{Residual}}$ with $L=3$, $\hat\sigma^2_{\text{Sample}}$, and $\hat\sigma^2_{\text{MAD}}$) lead to selective Type I error control under the global null.

\begin{figure}[!htbp]
\centering
   \includegraphics[width=0.9\linewidth]{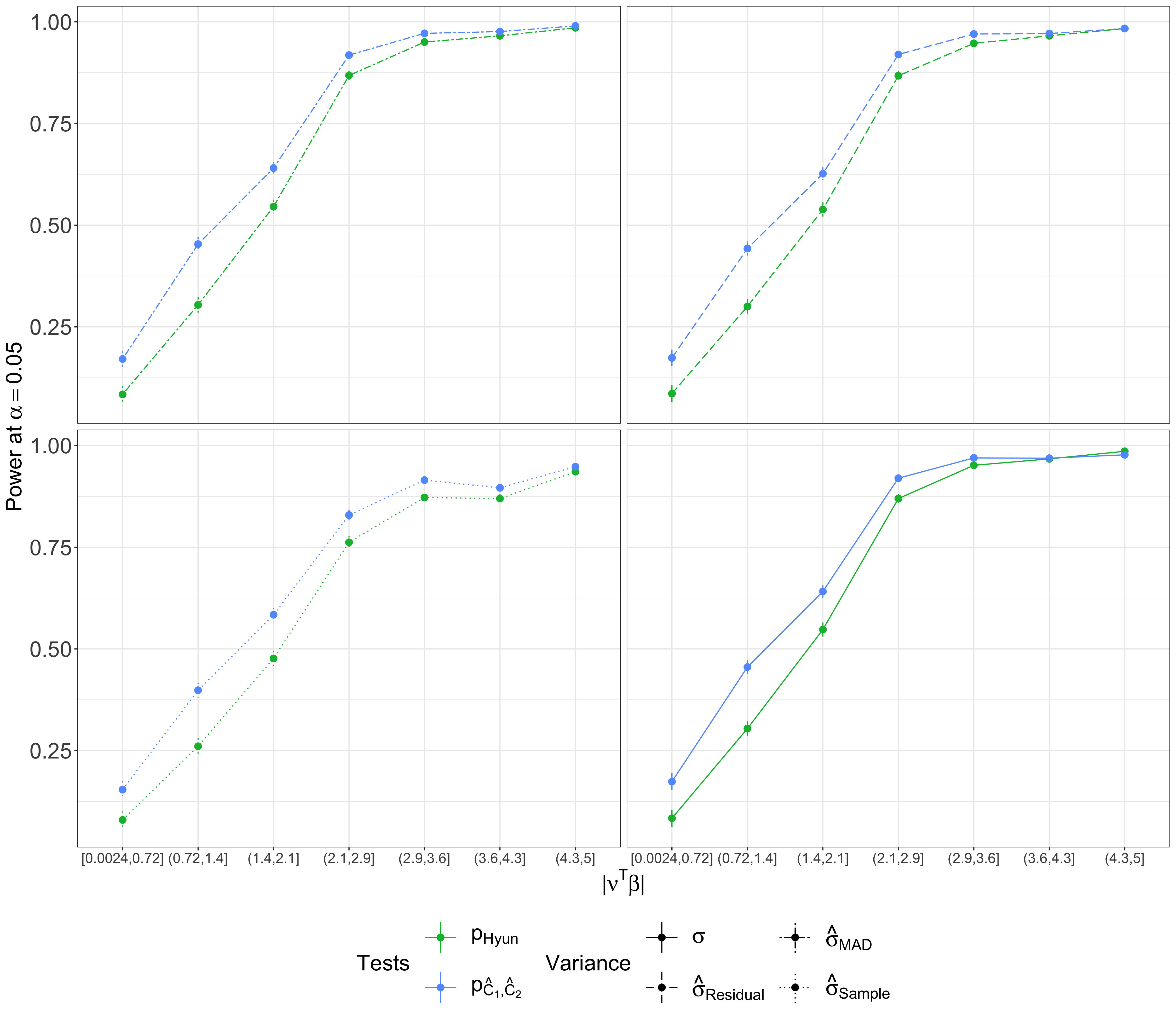}
\caption{The power of the tests based on both $\pHyun$ and $\pB$ computed using the true variance and three different variance estimators at level $\alpha= 0.05$, for the one-dimensional fused lasso simulations described in Section~\ref{section:sim_one_d_power} of the manuscript. Tests based on the true variance have the highest power, followed closely by the ones based on $\hat\sigma^2_{\text{MAD}}$ and $\hat\sigma^2_{\text{Residual}}$. Tests based on $\hat\sigma^2_{\text{Sample}}$ have the lowest power.}
\label{fig:power_sigma_hat}
\end{figure}

In addition, we compared the power of the tests based on estimated variance with that obtained using the true variance. Results from a simulation study with the same setup as in Section~\ref{section:sim_one_d_power} are aggregated in Figure~\ref{fig:power_sigma_hat}. We see that the tests based on $\hat\sigma^2_{\text{Residual}}$ with $L=3$ or $\hat\sigma^2_{\text{MAD}}$ result in nearly identical power to the test based on the true variance $\sigma^2$. By contrast, using $\hat\sigma^2_{\text{Sample}}$ leads to a less powerful test, especially for larger values of $|\nu^\top \beta|$. 
Moreover, the test based on $\pB$ is more powerful than the counterpart based on $\pHyun$, \emph{regardless of the chosen variance estimator}. 

Our empirical results agree with observations made in related problems for selective inference: (i) when the global null does not hold, $\hat\sigma^2_{\text{Sample}}$ is a conservative estimator of the true variance $\sigma^2$~\citepappendix{Tibshirani2018-rr,Hyun2018-gx,Gao2020-yt,Rugamer2020-fz}; (ii) $\hat\sigma^2_{\text{Residual}}$ has good empirical performance when $L$ is correctly specified; and (iii) in the case of the one-dimensional fused lasso, $\hat\sigma^2_{\text{MAD}}$ is an asymptotically consistent estimator under appropriate assumptions~\citepappendix{Kovacs2020-qb,jewell2019testing}.

%% file: sections-revision-v1/appendix_a5.tex

\subsection{Timing complexity for Algorithm~\ref{alg:SB}}
\label{appendix:timing_complexity}

\begin{figure}[!htbp]
\centering
\begin{subfigure}{0.45\textwidth}
\caption{}
  \includegraphics[width=\linewidth]{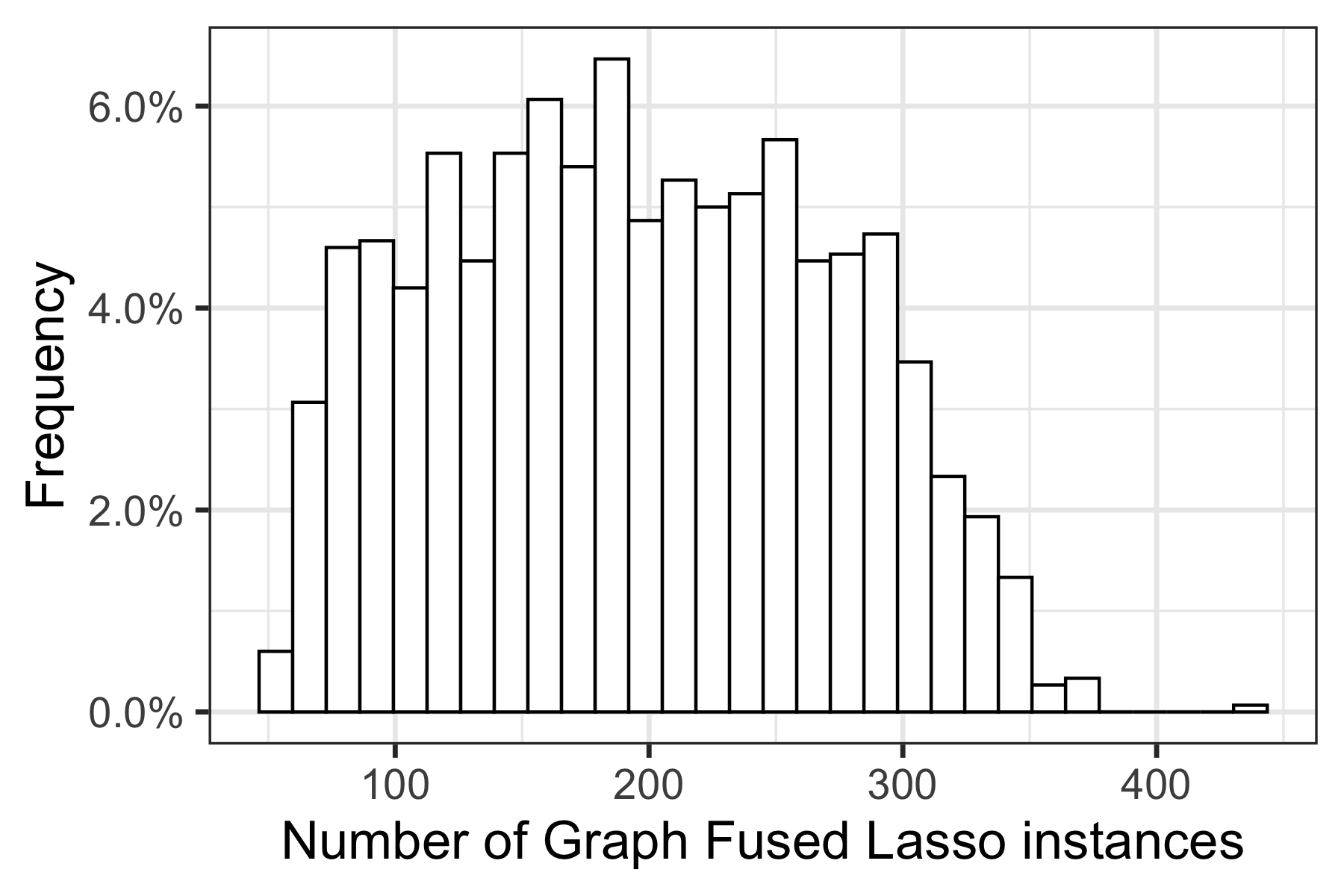}
 \end{subfigure}
\begin{subfigure}{0.45\textwidth}
\caption{}
  \includegraphics[width=\linewidth]{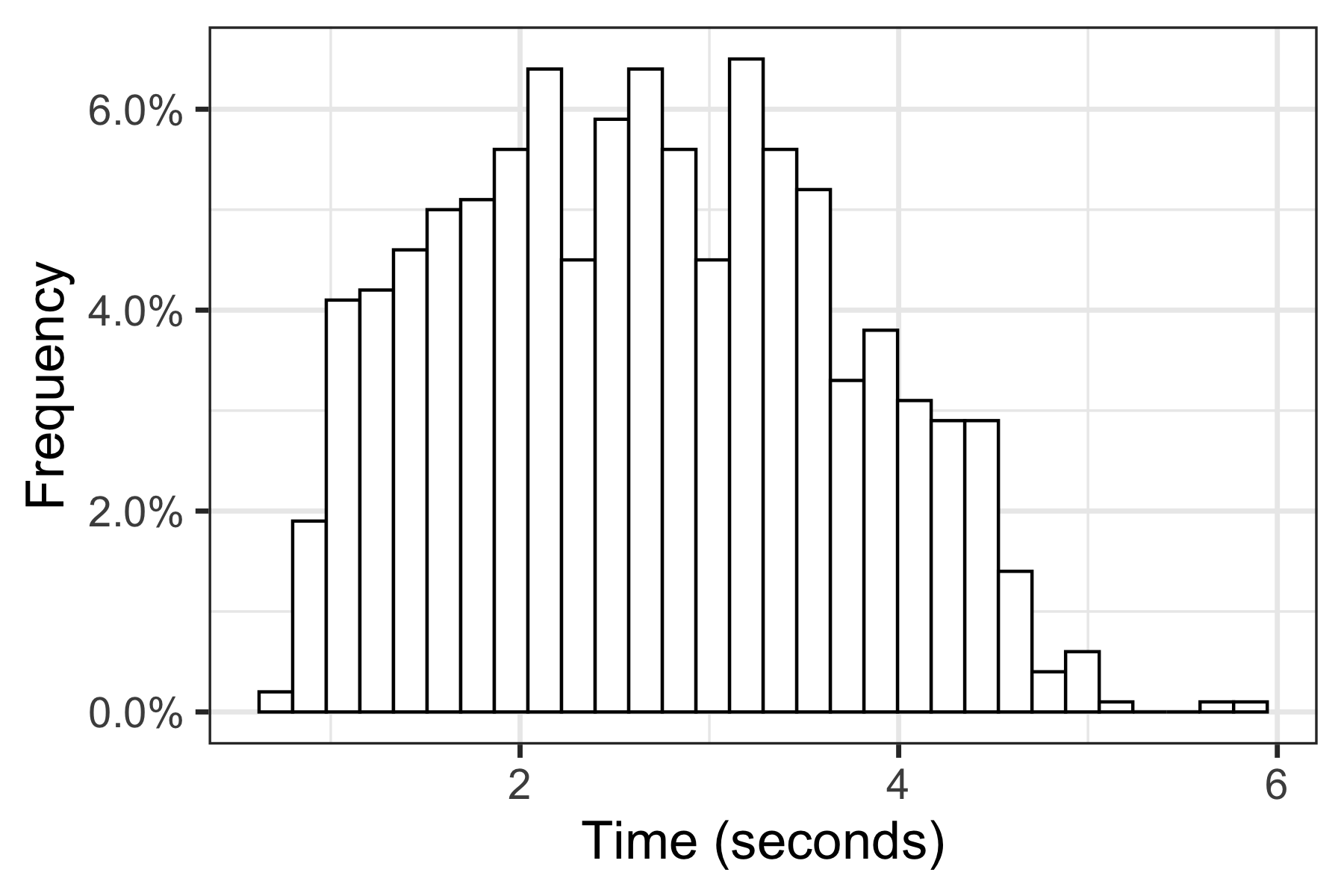}
 \end{subfigure}
\caption{\textit{(a): } Empirical distribution of $|\tilde{\mathcal{I}}|$  over 1,000 replicate datasets. Each dataset is simulated according to the one-dimensional fused lasso model described in Section~\ref{section:sim_one_d_type_1}. We solved the graph fused lasso problem with $K=2$ steps in the dual path algorithm. \textit{(b): } Same as (a), but for the running time of Algorithm~\ref{alg:SB}.}
\label{fig:algo_2_timing}
\end{figure}

In this section, we first characterize the computational complexity of Algorithm~\ref{alg:SB} using the following Proposition.
\begin{proposition}
\label{prop:comp_efficiency}
Recall that 
${M}_k(y) = \left(B_k(y), s_{B_k}(y), R_k(y), L_k(y) \right)$ is the output of the $k$th step of the dual path algorithm (see Algorithm~\ref{alg:dual_path} in Appendix~\ref{appendix:dual_path}). Define  
\begin{align}
\label{eq:I_s_b_set}
\tilde{\mathcal{I}} \equiv \qty{ (m_1,\ldots,m_K): \exists \alpha \in \mathbb{R} \text{ such that } \bigcap_{k=1}^K \qty{{M}_k\qty(y'\qty(\alpha)) = m_k}}.
\end{align}
Then, computing the $p$-value $\pB$ using Algorithm~\ref{alg:SB} in Appendix~\ref{appendix:S_B_algo} requires running the dual path algorithm $\mathcal{O}(|\tilde{\mathcal{I}}|)$ times, where $|\cdot|$ denotes the cardinality of a set.
\end{proposition}
We omit the proof of Proposition~\ref{prop:comp_efficiency}, as it directs follows from Algorithm~\ref{alg:SB} and the definition of $\tilde{\mathcal{I}}$. Proposition~\ref{prop:comp_efficiency} implies that the time complexity for Algorithm~\ref{alg:SB} is instance-dependent, and can in principle be prohibitively large. For instance, in the one-dimensional fused lasso case, $|\tilde{\mathcal{I}}|$ is upper bounded by $2^K \cdot \frac{n!}{(n-K)!}$, where $n$ is the number of observations, and $K$ is the number of steps in the dual path algorithm. \textbf{However, in practice, we are nowhere near this worst case scenario:} in the experiments in Section~\ref{section:sim}, $|\tilde{\mathcal{I}}|$ is of reasonable size. In particular, for the one-dimensional fused lasso simulations described in Section~\ref{section:sim_one_d_type_1}, the upper bound postulates that $|\tilde{\mathcal{I}}|$ can be as large as 
$158,400$ ($n=200, K=2$). However, as displayed in Figure~\ref{fig:algo_2_timing}(a), empirically, $|\tilde{\mathcal{I}}|$ falls under 500 in all instances.

In addition, because we re-implemented the polyhedron approach in \citetappendix{Hyun2018-ta} using the ideas from \citetappendix{Arnold2016-ue}, each graph fused lasso instance and its corresponding intervals of the form $[a_i,a_{i+1}]$ (see Section~\ref{section:method} for more details) can be computed efficiently. 

Figure~\ref{fig:algo_2_timing}(b) displays the running time of Algorithm 2, computed on a MacBook Pro with a 1.4 GHz Intel Core i5 processor, over 1,000 replicate datasets simulated according to the one-dimensional fused lasso model in Section~\ref{section:sim_one_d_type_1} with $n=200, \delta=0, \sigma=1$. The graph fused lasso problem is solved using the the dual path algorithm with $K=2$. The average running time for running Algorithm~\ref{alg:SB} to test each hypothesis $H_0: \nu^\top \beta = 0$ is 2.7 seconds. 

When the empirical size of $\tilde{\mathcal{I}}$ defined in \eqref{eq:I_s_b_set} is large, we could alternatively use an importance sampling approach to obtain an approximate $p$-value with ideas from, e.g., \citetappendix{Yang2016-km,Rugamer2022-yz,Rugamer2020-fz}. 

\subsection{Additional results for data applications}
\label{appendix:real_data_different_K}

\begin{figure}[!htbp]
\hspace{15mm} (a) \hspace{45mm} (b) \hspace{45mm} (c) \\
\centering
\raisebox{-.5\height}{\includegraphics[width=0.32\linewidth]{./figure/drug_overdose_obs.png}
\includegraphics[width=0.32\linewidth]{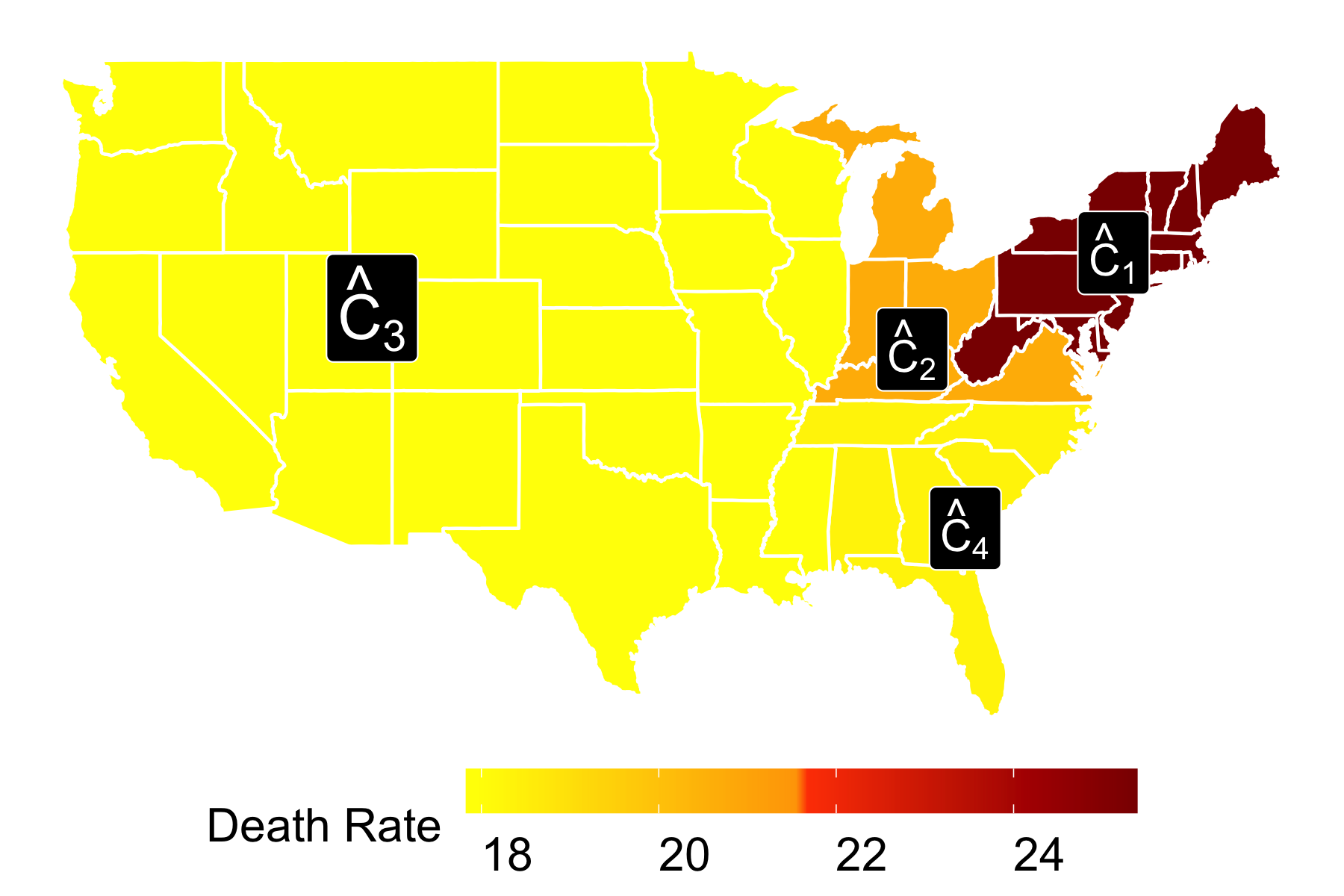}}
\raisebox{-.35\height}{\resizebox{0.28\linewidth}{!}{
\begin{tabular}{cccccc}
  \toprule
  $H_0$  & $p_{\text{Naive}} $ & $\pB$  & $\pHyun$  \\
  \toprule
$\bar{\beta}_1 = \bar{\beta}_2$ & 0.24  & 0.22  & 0.59 \\
\rowstyle{\boldmath\textbf} 
$\bar{\beta}_1 = \bar{\beta}_3$ & \boldmath{$<0.001$} & \boldmath{$<0.001$}& \boldmath{$0.90$}  \\
\rowstyle{\boldmath\textbf}
$\bar{\beta}_1 = \bar{\beta}_4$ & \textbf{0.005} & \textbf{0.02}& \textbf{0.82} \\
\rowstyle{\boldmath\textbf}
$\bar{\beta}_2 = \bar{\beta}_3$ & \boldmath{$<0.001$} & \textbf{0.04}& \textbf{0.73} \\
$\bar{\beta}_2 = \bar{\beta}_4$ & 0.20 & 0.27 & 0.69   \\
$\bar{\beta}_3 = \bar{\beta}_4$& 0.04 & 0.50 & 0.36  \\
 \bottomrule
\end{tabular}}}
\hspace{10mm} (d)  \\ 
\centering
\includegraphics[width=\linewidth]{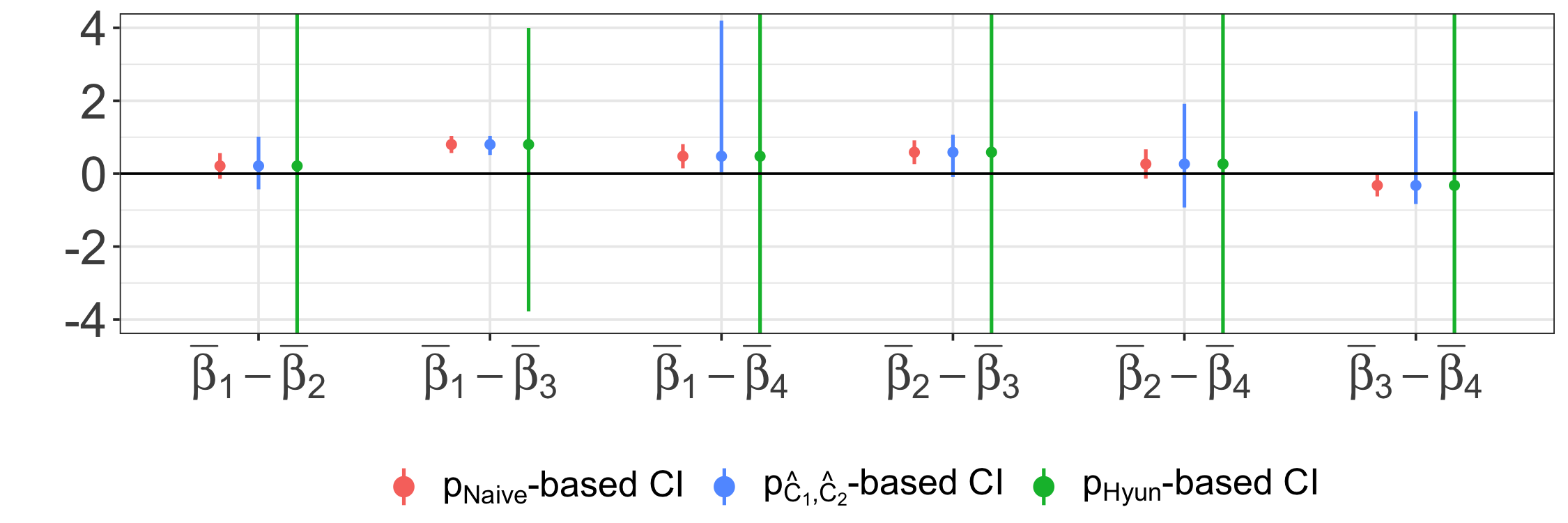}
\caption{\textit{(a):} The observed drug overdose death rates (deaths per 100,000 persons) for the 48 contiguous U.S. states in the year 2018. \textit{(b):} Applying the graph fused lasso  to the drug overdose data with $K=27$ results in four estimated connected components. \textit{(c):} For each pair of estimated connected components, we  computed $p_{\text{Naive}}$, $\pHyun$, and $\pB$. For brevity, we use the notation $\bar{\beta}_l = \sum_{j\in \hat{C}_l}\beta_j/|\hat{C}_l|$.  \textit{(d):} For each pair of estimated connected components, we constructed confidence intervals for the difference in means, corresponding to $p_{\text{Naive}}$, $\pHyun$, and $\pB$.}
\label{fig:real_data_drug_k_27}
\end{figure}

\begin{figure}[!htbp]
\hspace{15mm} (a) \hspace{45mm} (b) \hspace{45mm} (c) \\
\centering
\raisebox{-.5\height}{\includegraphics[width=0.32\linewidth]{./figure/teen_birth_obs.png}
\includegraphics[width=0.32\linewidth]{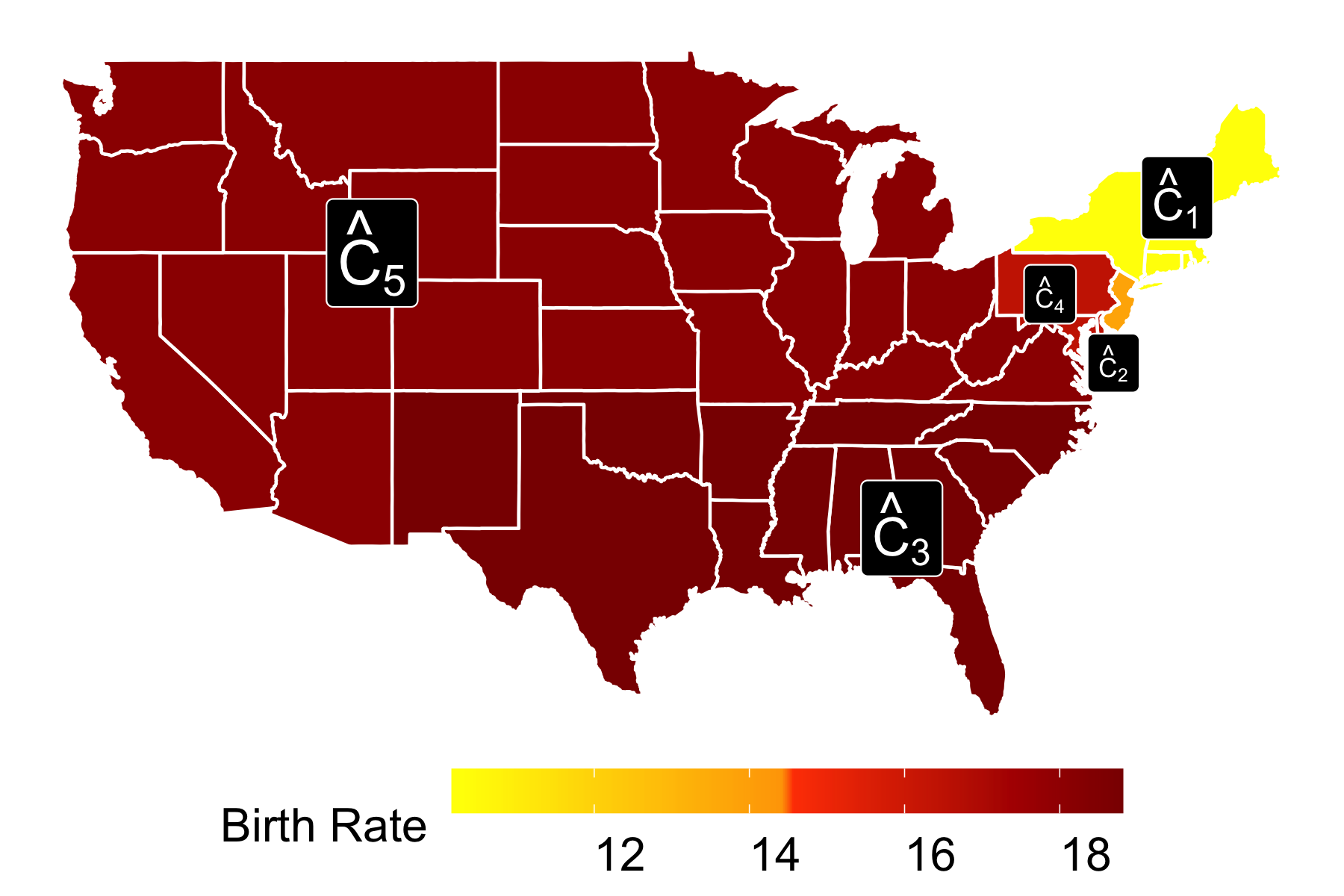}}
\raisebox{-.35\height}{\resizebox{0.28\linewidth}{!}{%
\begin{tabular}{cccccc}
  \toprule
  $H_0$  & $p_{\text{Naive}} $ & $\pB$ & $p_{\text{Hyun}}$   \\
  \toprule
$\bar{\beta}_1 = \bar{\beta}_2$ & 0.67  & 0.99 & 0.74 \\
$\bar{\beta}_1 = \bar{\beta}_3$ & $<0.001$& 0.07 & $0.57$ \\
\rowstyle{\boldmath\textbf}
$\bar{\beta}_1 = \bar{\beta}_4$ & \boldmath{$0.001$}  & \boldmath{$0.01$} & \textbf{0.26} \\
$\bar{\beta}_1 = \bar{\beta}_5$ &  {$<0.001$}& {$0.63$} & {$0.30$}  \\
$\bar{\beta}_2 = \bar{\beta}_3$ & {$<0.001$} & {$0.08$}  & {$0.27$}  \\
$\bar{\beta}_2 = \bar{\beta}_4$ & 0.13 & 0.34 & 0.26   \\
$\bar{\beta}_2 = \bar{\beta}_5$& 0.02 & 0.65 & 0.24  \\
$\bar{\beta}_3 = \bar{\beta}_4$& {$<0.001$}  & {0.12} & {0.71} \\
$\bar{\beta}_3 = \bar{\beta}_5$  & {$<0.001$} & {0.14} & {0.61}  \\
$\bar{\beta}_4 = \bar{\beta}_5$ & 0.34 & 0.78 & 0.71  \\
  \bottomrule
\end{tabular}}}
\hspace{10mm} (d)  \\ 
\centering
\includegraphics[width=\linewidth]{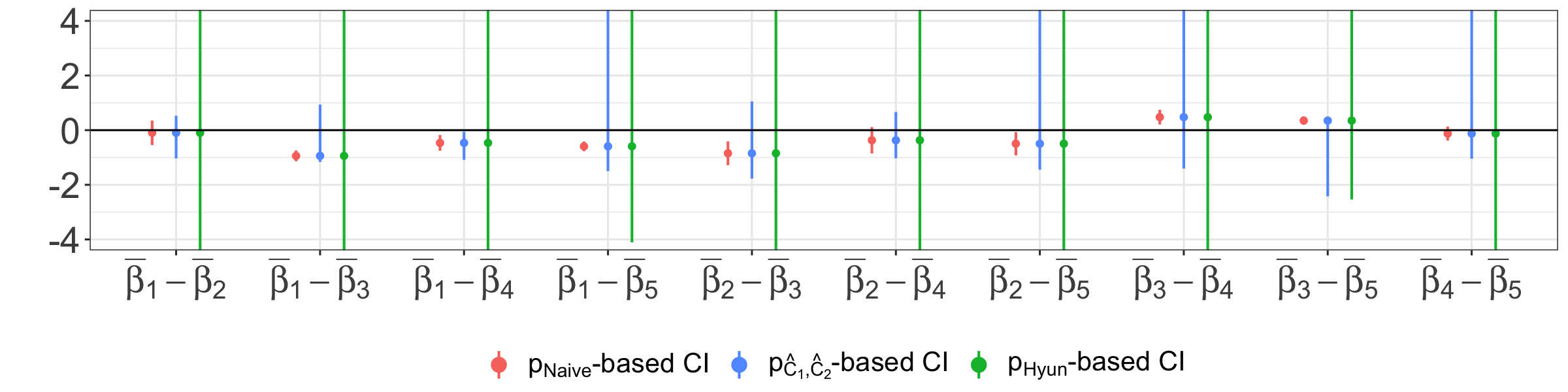}

\caption{\textit{(a):} The observed teenage birth rates (births per 1,000 females aged 15--19) for the 48 contiguous U.S. states in 2018. \textit{(b):} The graph fused lasso solution with $K=20$ results in five estimated connected components, displayed in distinct colors. \textit{(c):} For each pair of estimated connected components, we computed $p_{\text{Naive}}$, $\pHyun$, and $\pB$. Pairs for which the test based on $\pB$ results in a rejection at $\alpha=0.05$, but not for the test based on $\pHyun$, are in bold. \textit{(d):} Confidence intervals for the differences in means for each pair of connected components.}
\label{fig:real_data_teen_birth_k_20}
\end{figure}

Here, we repeat the analysis in Sections~\ref{sec:drug_overdose} and \ref{sec:teen_birth} with $K=27$ and $K=20$, respectively. Results are displayed in Figures~\ref{fig:real_data_drug_k_27} and \ref{fig:real_data_teen_birth_k_20}. Similar to the case of $K=30$, the test based on $\pB$ leads to more rejections than the test based on $\pHyun$ at $\alpha=0.05$. Furthermore, the confidence intervals based on $\pB$ are considerably shorter than those  based on $\pHyun$, and in some cases, even comparable to the naive confidence intervals that do not have proper coverage for the true parameter $\nu^\top \beta$.

%% file: sections-revision-v1/appendix_a6.tex
\subsection{A comparison of $\pB$ and $\pPP$}
\label{appendix:detailed_comp_PP}

In this section, we first briefly review the $p$-value proposal of \citetappendix{Le_Duy2021-iy} (henceforth referred to as $\pPP$), and elaborate on the conceptual differences between $\pPP$ and $\pB$ \eqref{eq:pB}. Next, we provide results from a simulation study that compares the selective Type I error \eqref{eq:selective_type_1} and the power of the tests based on $\pB$, $\pHyun$, and $\pPP$. In the simulations that follow, we will only consider the one-dimensional fused lasso problem, since the extension to a non-chain graph has not been implemented by \citetappendix{Le_Duy2021-iy} at the time of writing. We used the software for computing $\pPP$ provided by the authors at \url{https://github.com/vonguyenleduy/parametric_generalized_lasso_selective_inference}.

\subsubsection{A conceptual comparison of $\pB$ and $\pPP$}

\citetappendix{Le_Duy2021-iy} consider the $p$-value $\pPP$ defined as
\begin{equation}
\pPP \equiv \mathbb{P}_{H_0}\qty(|\nu^{\top} Y |\geq |\nu^{\top} y| \;\middle\vert\; \qty{i: \big(D\hat{\beta}(Y)\big)_i \neq 0} = \qty{i: \big(D\hat{\beta}(y)\big)_i \neq 0}, \Pi_{\nu}^{\perp} Y= \Pi_{\nu}^{\perp} y ),
\label{eq:PP_pval}    
\end{equation}  
where, with a slight abuse of notation, $\hat{\beta}(Y)$ is the solution to \eqref{eq:genlasso} with data $Y$. Using a similar argument to Proposition~\ref{prop:pval}, they showed that \eqref{eq:PP_pval} can be recast as the cumulative distribution function of a $\mathcal{N}(0,\sigma^2||\nu^2||)$ random variable, truncated to a set that can be efficiently computed. We note that in the case of the graph fused lasso \eqref{eq:graph_fused_lasso}, the set $\qty{i: \big(D\hat{\beta}(Y)\big)_i \neq 0}$ is equivalent to the set of edges used to determine the connected components of $\hat\beta$; in other words, conditioning on the event $\qty{i: \big(D\hat{\beta}(Y)\big)_i \neq 0} = \qty{i: \big(D\hat{\beta}(y)\big)_i \neq 0}$ is equivalent to conditioning on $\qty{B_K(Y)= B_K(y)}$, for an appropriate choice of $K$~\citepappendix{Tibshirani2011-fq}. 

What advantages, then, does our proposal $\pB$ in \eqref{eq:pB} offer when compared to $\pPP$? 
\begin{itemize}
\item
\emph{Smaller conditioning set:} first of all, we condition on \emph{even less} information when constructing $\pB$: $\pPP$ conditions on the set of edges that are used to determine the connected components of $\hat\beta$, and therefore implicitly, on \emph{all} of the connected components in $\hat\beta$ (see, e.g., Proposition~\ref{prop:piecewise_beta}). By contrast, in \eqref{eq:pB}, we condition only on the pair of connected components under investigation, thereby obtaining higher power.

\item
\emph{Interpretability:} the conditioning set for $\pB$ is based on the connected components of $\hat\beta$ after running the dual path algorithm for $K$ steps. By contrast, $\pPP$ conditions on the output of \eqref{eq:graph_fused_lasso} with a specific $\lambda$. We argue that, as a result, $\pB$ is more interpretable. Consider the widely-popular one-dimensional fused lasso problem and a pair of estimated piecewise constant segments $\hat{C}_1,\hat{C}_2$. $\pB$ answers the question:
\begin{quote}
\emph{Assuming that there is no difference between the \text{population} means of $\hat{C}_1$ and $\hat{C}_2$, then what's the probability of observing such a large difference in the \text{sample} means of $\hat{C}_1$ and $\hat{C}_2$, given that $\hat{C}_1$ and $\hat{C}_2$ are among the $K+1$ piecewise constant segments estimated from the data?}
\end{quote}
On the other hand, $\pPP$ answers the question:
\begin{quote}
\emph{Assuming that there is no difference between the \text{population} means of $\hat{C}_1$ and $\hat{C}_2$, then what's the probability of observing such a large difference in the \text{sample} means of $\hat{C}_1$ and $\hat{C}_2$, given that $\hat{C}_1$ and $\hat{C}_2$ are among the piecewise constant segments estimated from the data with a specific $\lambda$?}
\end{quote}

Here, $\pB$ is answering the question about $K+1$ piecewise constant segments, where $K$ is a very interpretable quantity (i.e., the number of estimated changepoints). In contrast, for $\pPP$, the meaning of $\lambda$ could vary greatly across different datasets --- the value of $\lambda$ that yields $K$ estimated changepoints on one dataset could yield far more or fewer estimated changepoints on another dataset.

\item
\emph{Numerical stability:} \citetappendix{Le_Duy2021-iy} solved the primal problem \eqref{eq:graph_fused_lasso} using an iterative solver, which in practice leads to numerical issues when identifying the set $\qty{i: \big(D\hat{\beta}(Y)\big)_i \neq 0}$ \citepappendix{Arnold2016-ue}, and consequently, in computing $\pPP$. By contrast, we chose to work with the dual problem \eqref{eq:dual}, which avoids these numerical issues and yields the \emph{exact} connected components of $\hat\beta$ when computing $\pB$. 


\end{itemize}

\subsubsection{A simulation study comparing $\pHyun$, $\pB$, and $\pPP$}
Next, we conducted a simulation study to compare the selective Type I error and power of the tests based on the following $p$-values: $\pHyun$ in \eqref{eq:hyun_pval}, $\pB$ in \eqref{eq:pB}, and $\pPP$ in \eqref{eq:PP_pval}. We tested the null hypothesis $H_0: \nu^\top \beta = 0$, where $\nu$ is defined in \eqref{eq:nu_c1_c2} for a randomly-chosen pair of \emph{adjacent} piecewise constant segments $\hat{C}_1$, $\hat{C}_2$ in the solution to the one-dimensional fused lasso problem. 

\begin{figure}[ht!]
\centering
\begin{subfigure}[b]{0.9\textwidth}
\caption{}
\includegraphics[width=\linewidth]{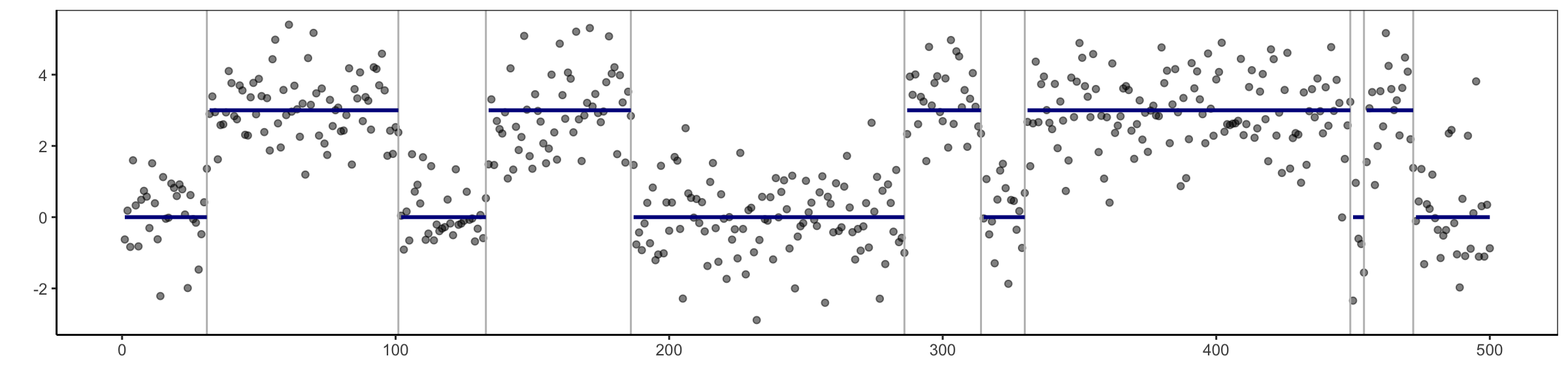}
\end{subfigure}
\begin{subfigure}[b]{0.3\textwidth}
 \caption{}
\includegraphics[width=\linewidth]{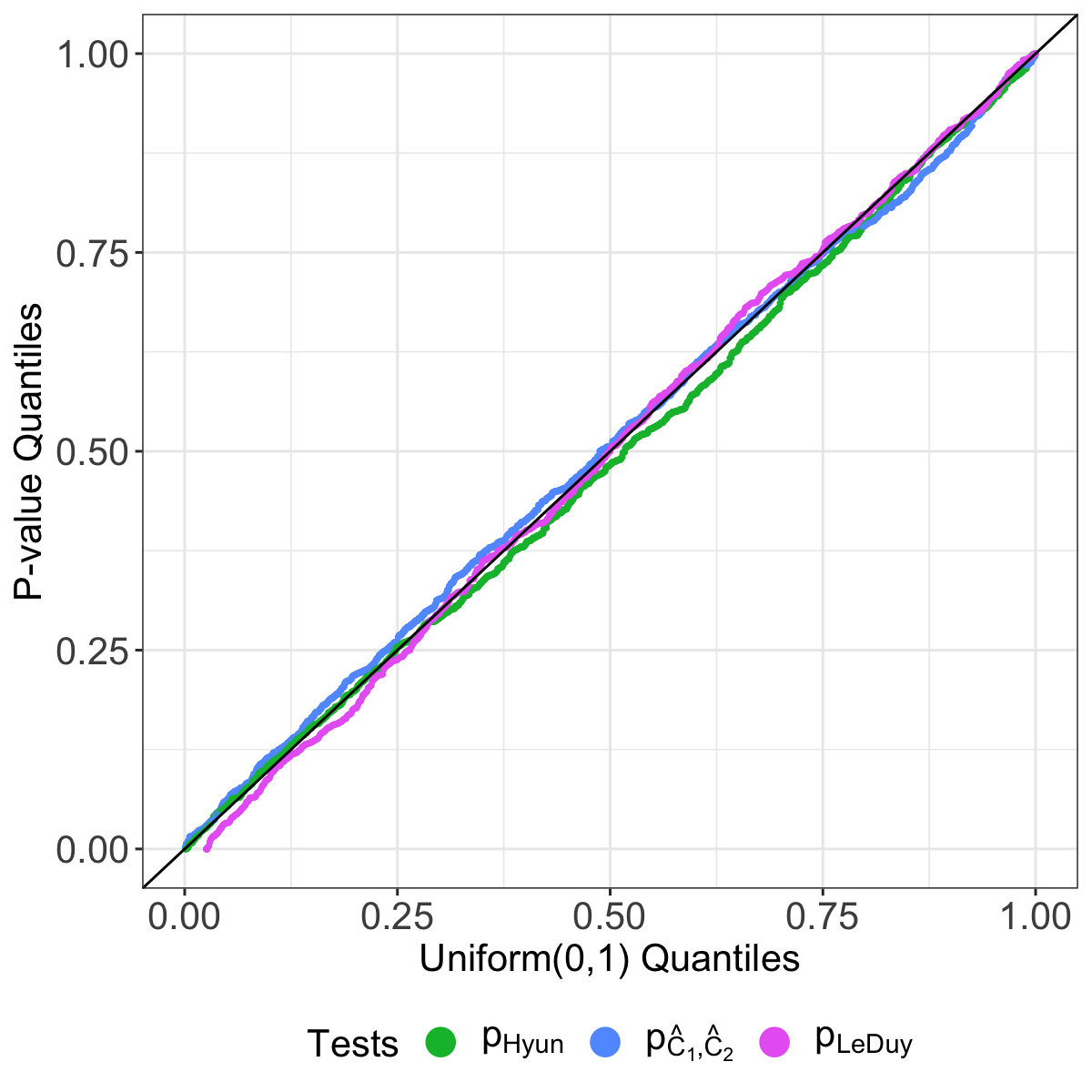}
\end{subfigure}
\begin{subfigure}[b]{0.3\textwidth}
 \caption{}
\includegraphics[width=\linewidth]{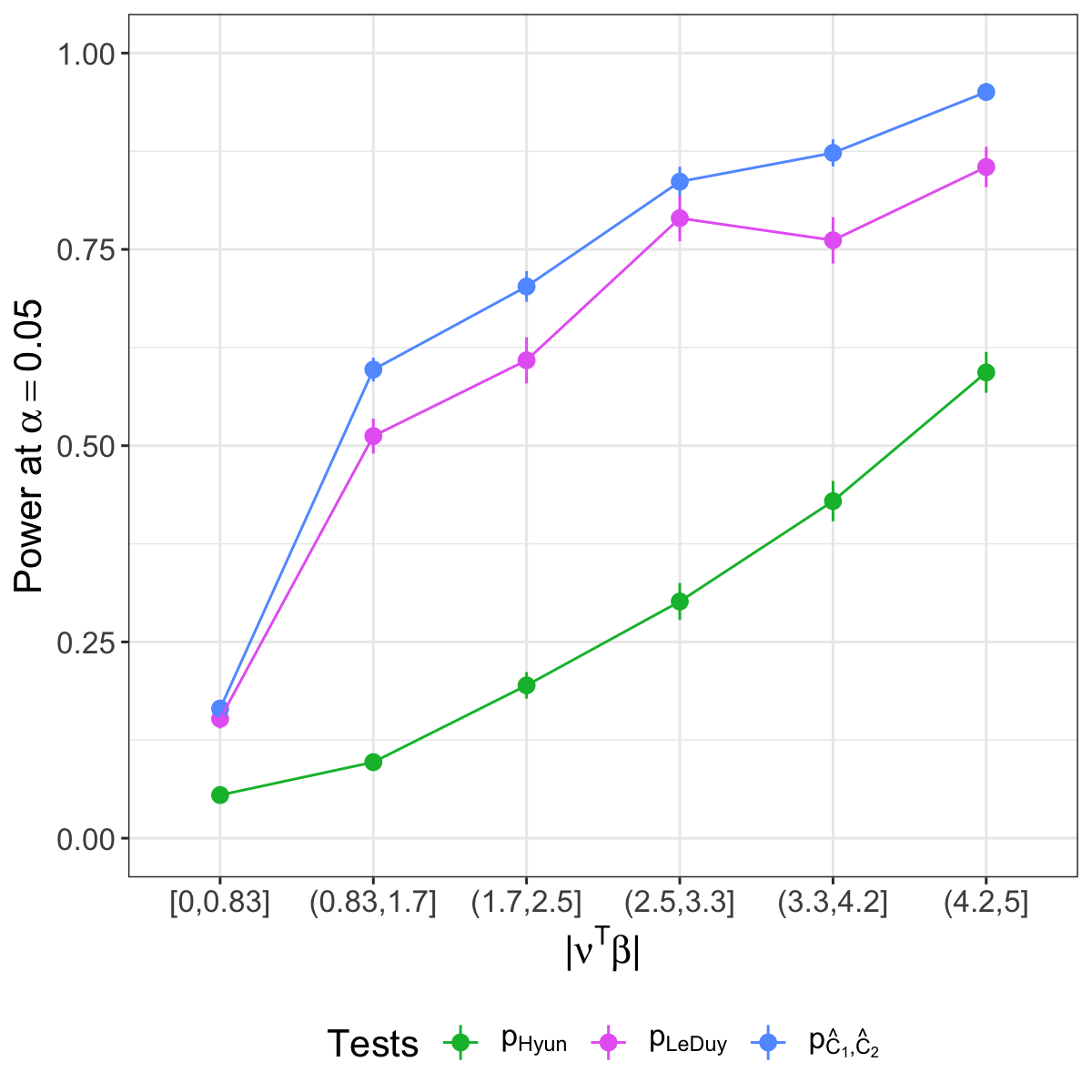}
\end{subfigure}
\begin{subfigure}[b]{0.3\textwidth}
 \caption{}
\includegraphics[width=\linewidth]{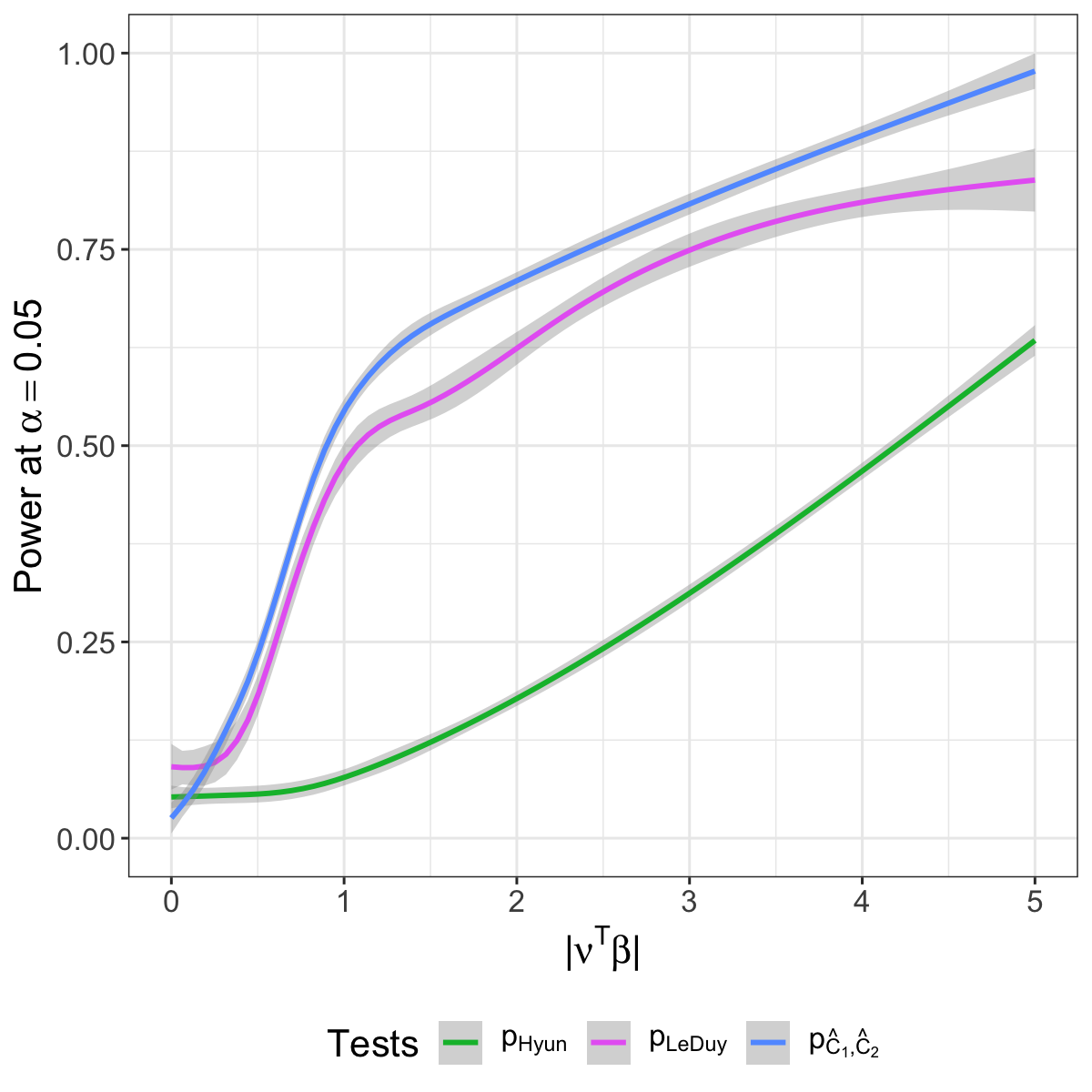}
\end{subfigure}
\caption{\textit{(a): }One realization of $y$ generated according to \eqref{sim:complex_1d_eq} with $\delta=3$ and $\sigma=1$ (grey dots), along with the true signal $\beta$ (blue curve). \textit{(b): } When $\delta=0$, tests based on $\pHyun$ in \eqref{eq:hyun_pval}, $\pB$ in \eqref{eq:pB}, and $\pPP$ in \eqref{eq:PP_pval} control the selective Type I error in \eqref{eq:selective_type_1}. \textit{(c): } The power of the tests based on $\pHyun$, $\pB$, and $\pPP$ increases as a function of the effect size $|\nu^\top\beta|$. For a given bin of $|\nu^\top\beta|$, the test based on $\pB$ has the highest power, followed by the test based on $\pPP$, and finally the test based on $\pHyun$. \textit{(d): } Same as (c), but the power of the three tests are estimated using the \texttt{gam} function in the \texttt{R} package \texttt{mgcv} \protect\citepappendix{wood_2017} instead of binning.}
\label{fig:comparison_PP}
\end{figure}

The signal $\beta\in \mathbb{R}^{500}$ is piecewise constant with 10 changepoints $\{\tau_1,\ldots,\tau_{10}\}$ (or equivalently, 11 piecewise constant segments), and the values of $\beta$ alternate between 0 and $\delta$ after each changepoint:
\begin{align}
 \label{sim:complex_1d_eq}
Y_j \overset{ind.}{\sim} \mathcal{N}(\beta_j,\sigma^2), \quad  \beta_j = \sum_{i=0}^{11}\delta \times 1_{(i \text{ odd, }\tau_i \leq j \leq \tau_{i+1})}, \quad j = 1,\ldots, 500,
\end{align}
where $\tau_0 \equiv 0$ and $\tau_{11} \equiv 500$. Figure~\ref{fig:comparison_PP}(a) displays an instance of \eqref{sim:complex_1d_eq} with $\delta=3$ and $\sigma=1$.

In the simulations that follow, the software accompanying \citetappendix{Le_Duy2021-iy} returned an empty string for $\pPP$ for around 27\% of the hypotheses. Upon inquiring, the authors of \citetappendix{Le_Duy2021-iy} said that this is due to numerical stability issues in identifying the conditioning set in $\pPP$. Consequently, the displayed results for $\pPP$ are based on the subset of the hypotheses for which the authors' software successfully returned a $p$-value.

We first investigate the selective Type I error control by simulating $y_1,\ldots,y_{500}$ from \eqref{sim:complex_1d_eq} with $\delta = 0$ and $\sigma =1$. Therefore, the null hypothesis $H_0:\nu^\top \beta = 0$ holds for all contrast vectors $\nu$, regardless of the pair of piecewise constant segments under investigation. For $\pHyun$ and $\pB$, we solved \eqref{eq:graph_fused_lasso} with $K=10$ steps in the dual path algorithm, which yields exactly 11 piecewise constant segments by the properties of the one-dimensional fused lasso. For $\pPP$, we selected the tuning parameter $\lambda$ so that \eqref{eq:graph_fused_lasso} yields exactly 11 piecewise constant segments on the data. 

Figure~\ref{fig:comparison_PP}(b) displays the observed $p$-value quantiles versus Uniform$(0,1)$ quantiles, aggregated over 1,000 hypothesis tests. We see that all three tests based on $\pHyun$ \eqref{eq:hyun_pval}, $\pB$ \eqref{eq:pB}, and $\pPP$ \eqref{eq:PP_pval} control the selective Type I error as in \eqref{eq:selective_type_1}. 




Next, we show that the test based on $\pB$ has higher power than the test based on $\pPP$, and both tests have higher power than the test based on $\pHyun$. We generated 500 datasets from \eqref{sim:complex_1d_eq} for each of ten evenly-spaced values of $\delta \in [0.5,5]$. For each simulated dataset, we solved \eqref{eq:graph_fused_lasso} with $K=10$ for $\pHyun$, and $\pB$, and chose the tuning parameter $\lambda$ so that \eqref{eq:graph_fused_lasso} yields 10 estimated changepoints for $\pPP$. We rejected the null hypothesis $H_0:\nu^\top \beta = 0$ if the $p$-value was less than $\alpha=0.05$. As in Section~\ref{section:sim_one_d_power}, we consider the power as a function of  $|\nu^\top \beta|$. 

Figure~\ref{fig:comparison_PP}(c) displays the power estimated by first creating six evenly-spaced bins of the observed values of $|\nu^\top\beta|$, and then computing the proportion of simulated datasets for which we rejected $H_0$ within each bin. Alternatively, we could estimate the power as a smooth function of $|\nu^\top\beta|$ using a regression spline (see Figure~\ref{fig:comparison_PP}(d)). In both cases, the test based on $\pB$ has 10--15\% higher power than the test based $\pPP$, and both have substantially higher power than the test based on $\pHyun$.